\documentclass[aps,twocolumn,prc,superscriptaddress,showpacs,nofootinbib,floatfix,amssymb,amsfonts,amsmath]{revtex4-1}

\usepackage{graphicx}
\usepackage{dcolumn}
\usepackage{bm}

\usepackage{amsmath}    
\usepackage{amsfonts}   
\usepackage{amssymb}
\usepackage{graphicx}   
\usepackage{longtable}
\newcommand{\bff}[1]{{\mbox{\boldmath $#1$}}}

\begin{document}
\title{From cluster structures to nuclear molecules: the 
       role of nodal structure of the single-particle 
       wave functions.}

\author{A.\ V.\ Afanasjev}
\affiliation{Department of Physics and Astronomy, Mississippi
State University, MS 39762}

\author{H.\ Abusara}
\affiliation{Department of Physics, Birzeit University, Birzeit, Palestine}

\date{\today}

\begin{abstract}
  The nodal structure of the density distributions of the single-particle 
states occupied in rod-shaped, hyper- and megadeformed structures of 
non-rotating and rotating $N\sim Z$ nuclei has been investigated in detail.
The single-particle states with the Nilsson quantum numbers of the 
$[NN0]1/2$ 
(with $N$ from 0 to 5) and $[N,N-1,1]\Omega$ (with $N$ from 1 to 3 and 
$\Omega=1/2$, 3/2) types are considered. These states are building
blocks of extremely deformed shapes in the nuclei with mass numbers
$A \leq 50$. Because of (near)axial symmetry 
and large elongation of such structures,
the wave functions of the single-particle states occupied are dominated 
by a single basis state in cylindrical basis. This basis state 
defines the nodal structure of the single-particle density distribution.
The nodal structure of the single-particle density distributions
allows to understand in a relatively simple way the necessary conditions for 
$\alpha$-clusterization and the suppression of the $\alpha$-clusterization 
with the increase of mass number. It also explains in a natural way the
coexistence of ellipsoidal mean-field type structures and nuclear molecules
at similar excitation energies and the features of particle-hole excitations
connecting these two types of the structures. Our analysis of the nodal
structure of the single-particle density distributions does not support
the existence of quantum liquid phase for the deformations and nuclei under 
study.
\end{abstract}


\maketitle

\section{Introduction}

  Recent years are characterized by the revival of the interest (both experimental and 
theoretical) to the study of cluster and extremely deformed structures in light nuclei
(see Refs.\ \cite{40Ar.10,K.12,S-34-mol.14,EKNV.12,EKLLM.12,S-34-mol.14,IIIMO.15,KSK.16,ZYLMR.16,RA.16} 
and references quoted therein). Many of these structures are described in terms of 
clusters, the simplest one being the $\alpha$-particle \cite{OFE.06,MKKRHT.06}. Cluster 
or similar type models provide an important insight
into cluster dynamics of nucleus. However, the initial assumptions about clusters 
represent a limitation of this type of models and many shell model 
configurations are beyond of their reach. It is also important 
to remember that the cluster description does not correspond to clearly separated 
$\alpha$-particles, but generates the mean-field states largely by antisymmetrization 
\cite{MKKRHT.06}.

\begin{figure*}[htb]
\includegraphics[angle=0,width=5.5cm]{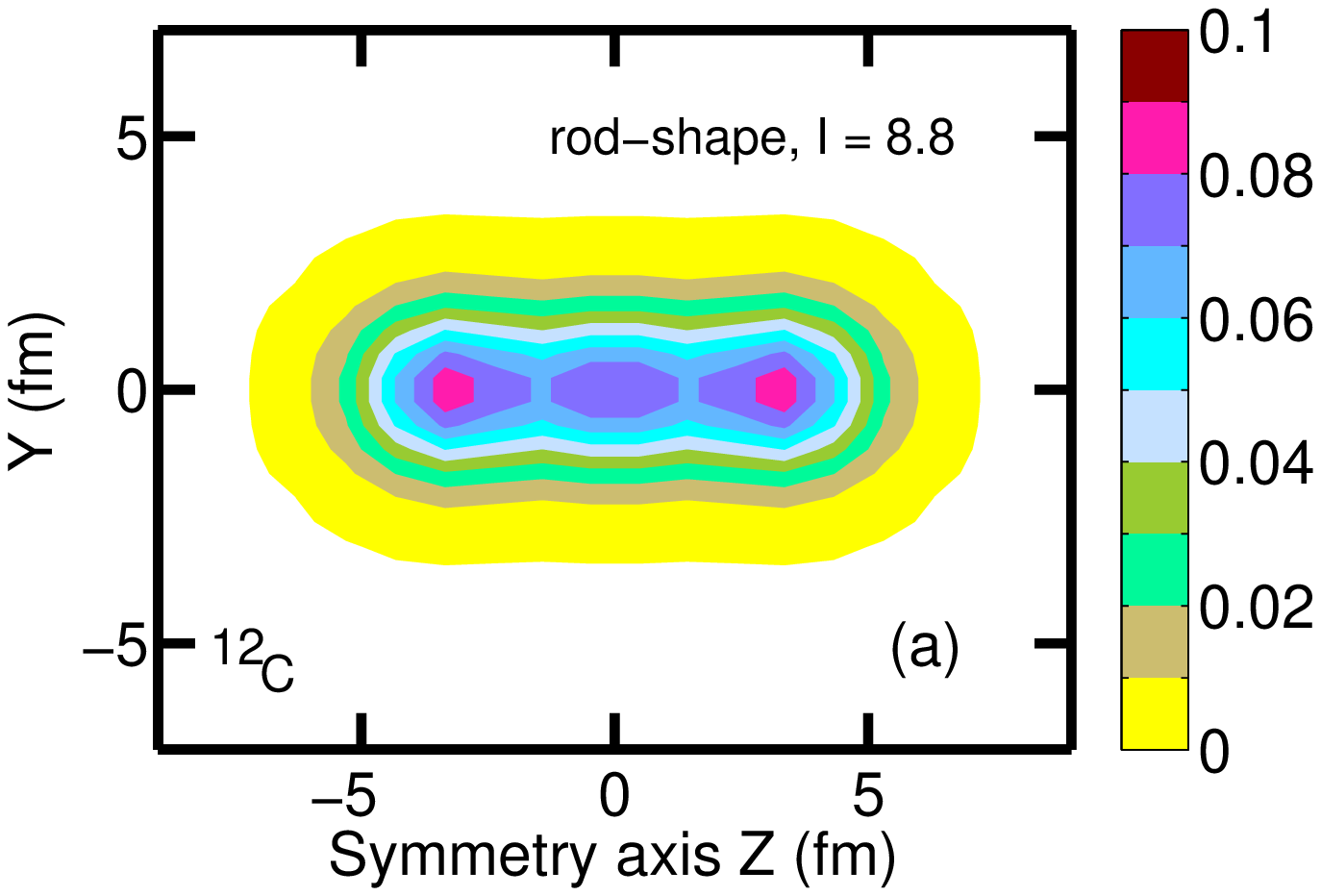}
\includegraphics[angle=0,width=5.5cm]{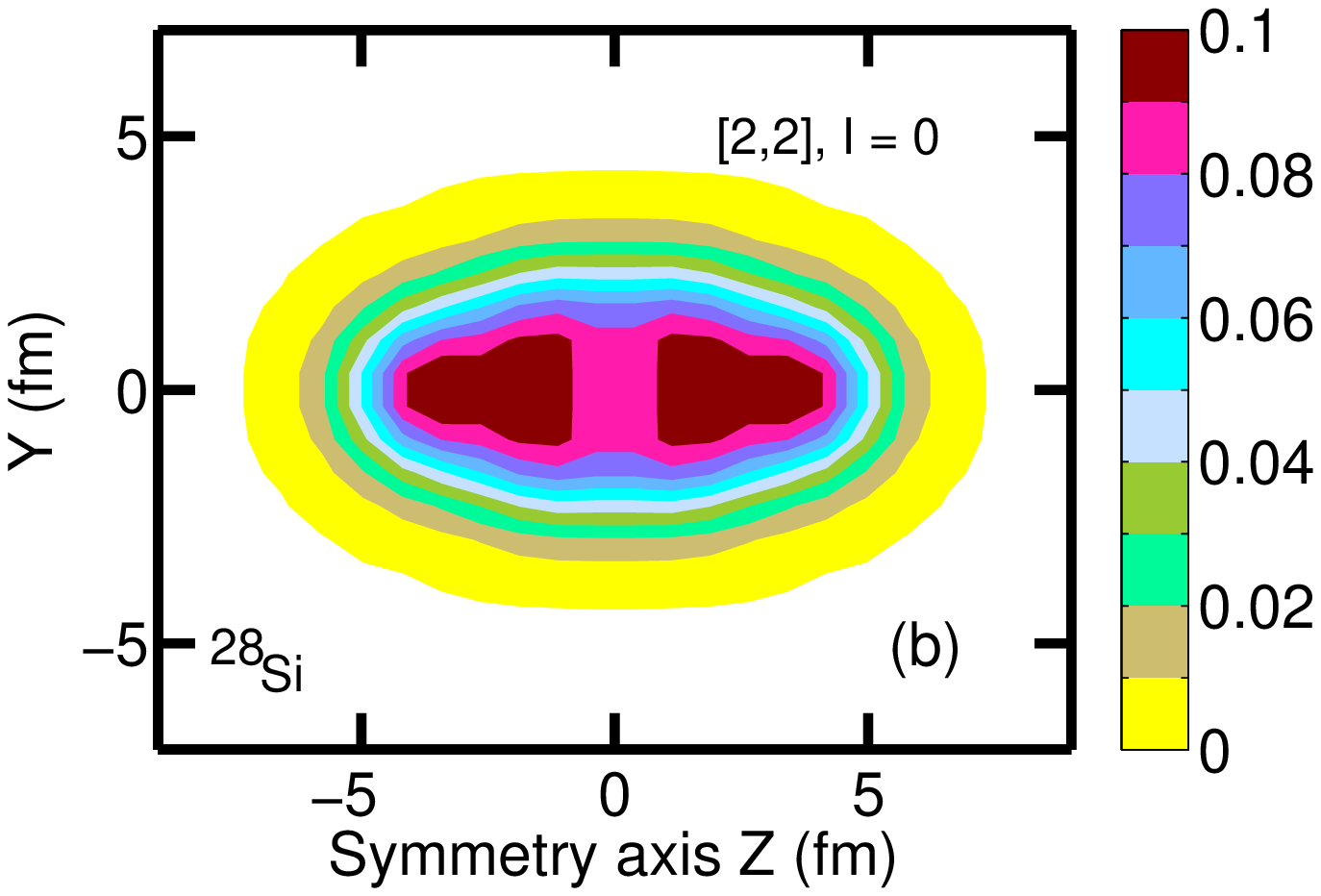}
\includegraphics[angle=0,width=5.5cm]{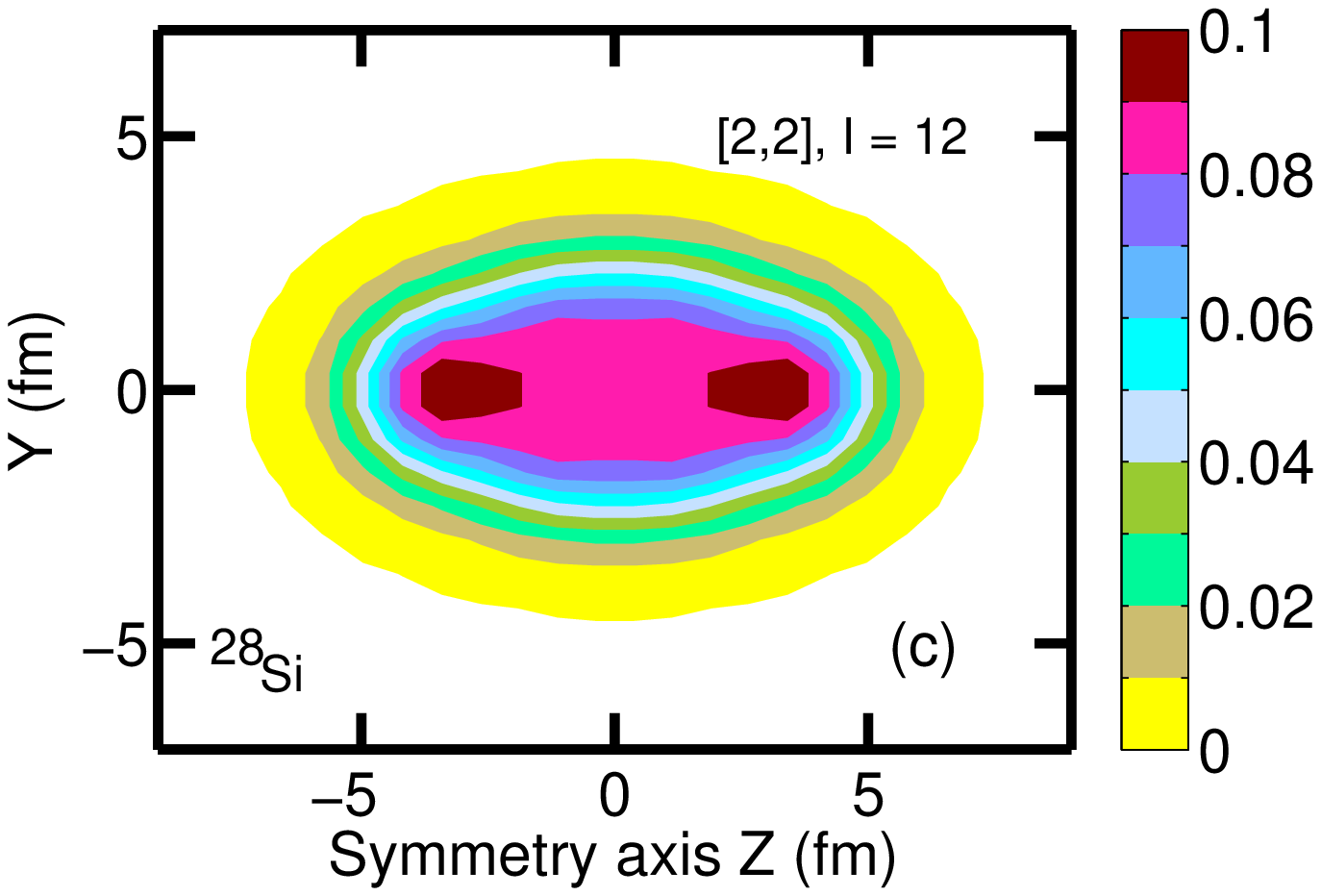}
\includegraphics[angle=0,width=5.5cm]{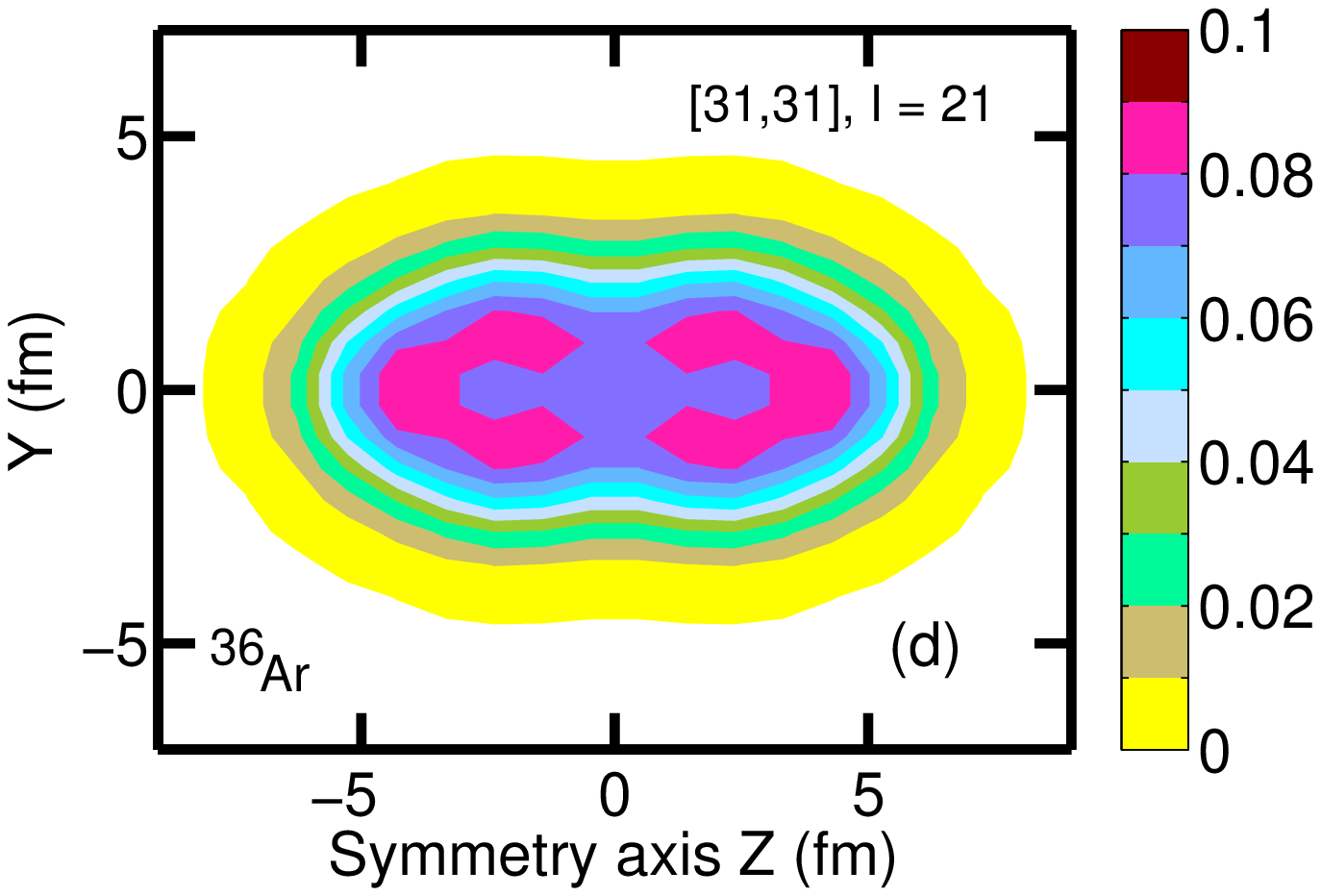}
\includegraphics[angle=0,width=5.5cm]{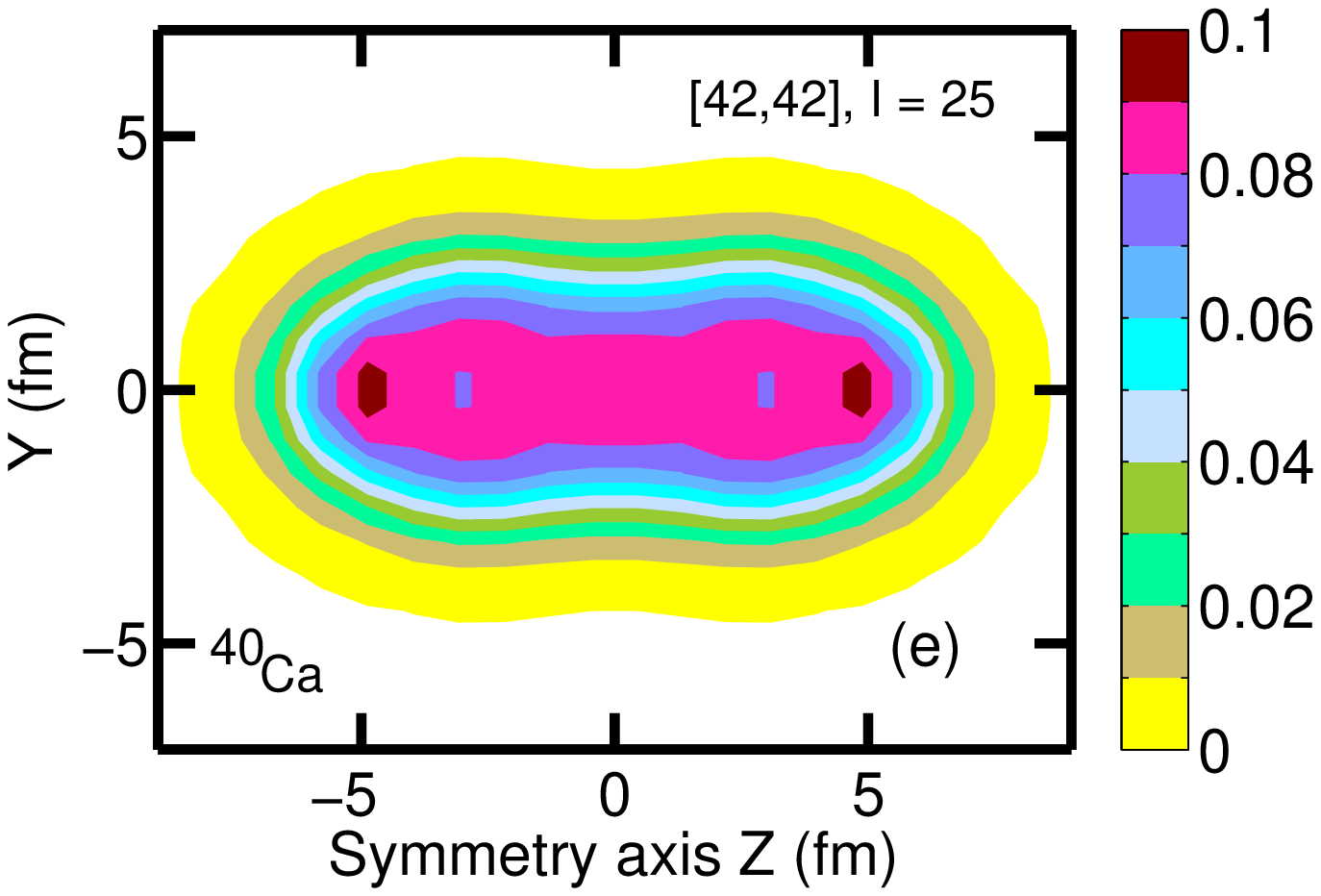}
\includegraphics[angle=0,width=5.5cm]{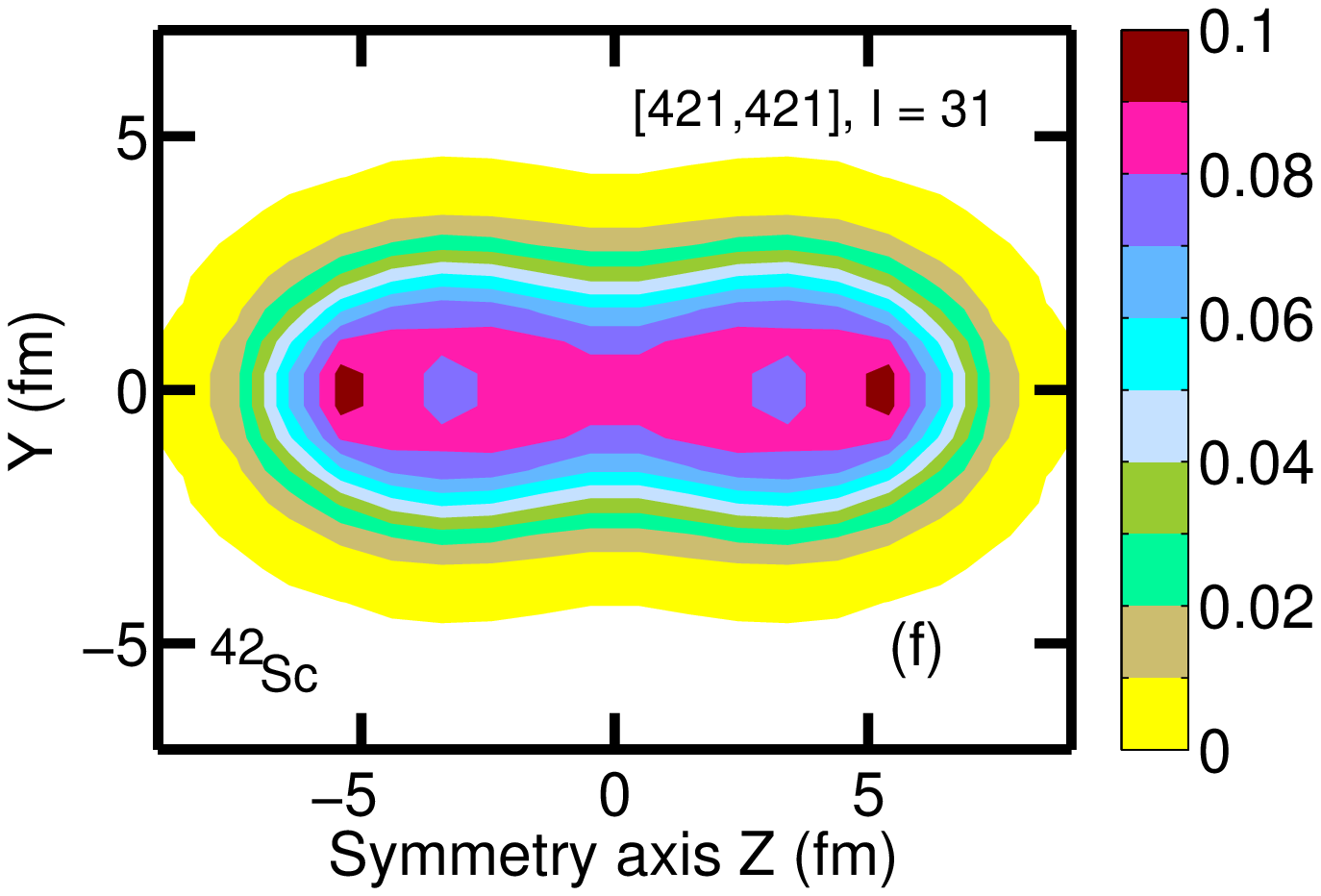}
\caption{(Color online) Total neutron densities [in fm$^{-3}$]
of specified configurations in $^{12}$C, $^{28}$Si, $^{36}$Ar, 
$^{40}$Ca and $^{42}$Sc at indicated spin values. The plotting
of the densities starts with yellow color at $0.001$ fm$^{-3}$.
The densities presented in panels (b-f) are based on the
results of the calculations of Ref.\ \cite{RA.16}.
}
\label{Total-densities}
\end{figure*}

  Alternative way of the description of exotic cluster configurations is 
within the framework of density functional theory (DFT) \cite{EKNV.12,EKNV.14,EKNV.17}. 
This type of models does not assume the existence of cluster structures but 
allows simultaneous treatment of cluster and mean-field-type states 
\cite{ER.04,RMUO.11,EKNV.12,EKNV.14,YIM.14,EKNV.17}. In this framework, the formation 
of clusters proceeds from microscopic single-nucleon degrees of freedom via 
many-body correlations. Let us mention some recent studies of cluster and 
extremely deformed structures in the DFT framework. A linear chain of three  
$\alpha$ clusters, leading to ``rod-shaped'' nucleus and  suggested about 
60 years ago \cite{Mor.56}, was recently studied in the cranked relativistic 
mean field (CRMF) theory in Ref.\ \cite{ZIM.15}; its density distribution 
is presented in Fig.\ \ref{Total-densities}a. This exotic structure (``Hoyle'' 
state) plays a crucial role in the synthesis of $^{12}$C from three $^{4}$He 
nuclei in stars \cite{Apj.54}.  Another example of rod-shaped
nucleus is linear chain configuration of four $\alpha$-clusters in $^{16}$O.
Recently the relationship between the stability of such states  and angular 
momentum was investigated using Skyrme cranked Hartree-Fock (HF) method 
in Ref.\  \cite{IMIO.11} and CRMF approach in Ref.\ \cite{YIM.14}. 
The cranked Skyrme HF framework was employed for the study of the stability 
of rod-shaped  structures in highly-excited states of $^{24}$Mg in Ref.\ \cite{IIIMO.15}.

  However, the phenomenon of clusterization is not limited to the 
$\alpha$-particles. Larger nuclei could play a role of building blocks 
of the clustered configurations. In particular,  the nuclear configurations 
consisting of the $N=Z$ clusters with no or extra few valence nucleons 
could play an important role in the nuclei near the $N=Z$ line 
\cite{OFE.06}. For example, the wave function of the superdeformed (SD) 
[2,2]\footnote{The notation of the configurations is discussed in Sec.\ 
\ref{calc-det}.} band in $^{32}$S contains a significant admixture of 
the molecular $^{16}$O + $^{16}$O structure \cite{MKKRHT.06,MH.04}. 
Extremely deformed structures such as super-, hyper- (HD) and megadeformed 
(MD) configurations as well as nuclear molecules have been systematically 
studied in the rotating $A=28-50$ $N\approx Z$ nuclei in the CRMF 
framework (Refs.\ \cite{RA.16,AR.16}). A number of the configurations 
with cluster structures have been found in these calculations. Fig.\ 
\ref{Total-densities}(b-f) show some examples of such structures with 
different pattern of density distribution. 
The best example of nuclear molecule is the MD [421,421] configuration in 
$^{42}$Sc (Fig.\ \ref{Total-densities}f) followed by the MD [42,42] 
configuration in $^{40}$Ca (Fig.\ \ref{Total-densities}e) and the 
MD [31,31] configuration in $^{36}$Ar (Fig.\ \ref{Total-densities}d). These 
three configurations show pronounced neck. On the other hand, the HD 
[2,2] configuration in $^{28}$Si shows the clusterization at spin zero but 
with less pronounced neck (Fig.\ \ref{Total-densities}b). The rotation 
somewhat hinders these features and suppresses the neck in 
this configuration (see Fig.\ \ref{Total-densities}c and Ref.\ \cite{AR.16}).

  In the DFT framework, the formation of clusters proceeds from microscopic 
single-nucleon degrees of freedom via many-body correlations. Although this 
fact is wildly recognized, in reality very little attention has been paid to 
the detailed study of the role of the single-particle states in clusterization. 
In that context, mostly the impact of the underlying single-particle and shell 
structure on the clusterization has been investigated.  For example, the impact 
of underlying single-particle structure on the transition from spheroidal 
superdeformed configurations to doubly spherical configurations, which are 
analogs of cluster configurations, and the connection between the magic numbers 
at both shapes has been investigated in Ref.\ \cite{AJ.94}. Even less attention has 
been paid to the structure of the single-particle wave functions and related 
single-particle densities and their impact on clusterization in the DFT 
framework.  To our knowledge this has been discussed only in two publications. 
The formation of the total density of $^{8}$Be nucleus from single-particle 
densities of occupied states has been discussed in Ref.\ \cite{EKNV.14}. The 
contribution of the single-particle densities of the [220]1/2 states into 
buildup of the ground state densities of $^{20}$Ne has been presented in Ref.\ 
\cite{EKNV.12}.

  To address this gap in our knowledge of clusterization we perform 
systematic investigation of the densities of the single-particle states 
and their nodal structure in clustered and extremely deformed configurations 
of the $N=Z$ $^{12}$C, $^{28}$Si and $^{40}$Ca nuclei. The selection of these 
nuclei is dictated by several factors. First, typical single-particle 
orbitals, which play a role at hyper- and megadeformation in the 
$N\approx Z$ nuclei with $Z=6-24$, have to be considered. Second, the 
impact of reasonable changes of the nucleonic potential in axial and 
radial directions on the nodal structure of single-particle density 
distributions have to be investigated. Since nucleonic potential depends
on total nucleonic density, this is achieved by considering nuclei which 
differ substantially in that respect. Note that the single-particle 
densities bear a clear fingerprint of underlying single-particle wave 
functions. Third, such choice of nuclei allows to see how significant 
is the impact of the lowering of the position of the single-particle orbital 
in nucleonic potential on the single-particle density distributions.
In this paper we consider rotating and non-rotating nuclei and define 
which single-particle states favor the $\alpha-$clusterization, which states 
suppress this type of clusterization and which particle-hole excitations 
are important for the creation of nuclear molecules.

  The difficulty in investigating cluster and extremely deformed states 
at spin zero is that they are generally unbound and lie at high excitation 
energies at low spins \cite{OFE.06,J.clust.16}. Moreover, they are either 
formed on the shoulder or in very shallow minima of potential energy 
surfaces \cite{EKNV.14,AA.08}; thus, they are inherently unstable. The 
high density of nucleonic configurations at these energies and possible 
mixing among them is another factor hindering their experimental observation. 
As shown in Ref.\ \cite{RA.16}, the rotation could help to overcome these 
obstacles. This is because extremely deformed configurations are favored 
by rotation at high spins (Refs.\ \cite{DPSD.04,AA.08,RA.16}) so that only 
such states could be populated above some specific spin values in the mass
region of interest \cite{RA.16}.

  In the present paper the analysis is performed within the framework 
of covariant density functional theory (CDFT) \cite{VALR.05}. It provides 
a fully self-consistent description of many nuclear phenomena. The CDFT 
well describes the experimental proton density distributions in spherical 
nuclei \cite{GRT.90,GAABet.05}, the deformations of superdeformed nuclei 
\cite{AKR.96,CRHB,A60,Ca40-PRL.01} and charge radii \cite{GTM.05,AARR.14,AA.16} 
across the  nuclear chart.  In addition, it provides good description of 
the changes in deformation and charge radii with the change of the 
configuration/particle number. These two facts strongly suggest that the CDFT 
properly reproduces the single-particle density contributions on which 
these observables depend\footnote{Note that at present there is no 
experimental technique which would allow to measure the density distribution 
of specific single-particle orbital (and thus its localization).}. This is 
an important factor in the context of the present investigation since the 
total density distribution (which 
could point either to $\alpha$-clusterization or nuclear molecule) is built 
as a sum of the single-particle density distributions of occupied single-particle 
states. Note also that covariant (relativistic) energy density functionals 
(CEDFs) show more pronounced clusterization of the density distribution as 
compared with non-relativistic ones because of deeper single-nucleon potentials 
\cite{EKNV.12}.

  The studies of the single-particle densities in the present paper are 
also related to the general problem of the nucleon localization in finite
nuclei. This problem was considered in Refs.\ \cite{RMUO.11,EKNV.13,ZSN.16}, 
however, only total nucleon densities were used in the discussion of the
localization. It was described in terms of different parameters 
reflecting different aspects of nuclear many-body problem. The 
$\alpha$ parameter representing the ratio of the spatial dispersion of the 
wave function to the average internucleon distance has been introduced in 
Ref.\ \cite{EKNV.13}. This parameter generally increases with the number 
of nucleons. Based on that, it was concluded that cluster 
states are more easily formed in light nuclei and that the transition 
from localized clusters to quantum liquid state occurs for nuclei with 
$A\approx 30$. An alternative localization measure has been employed
in Refs.\ \cite{RMUO.11,ZSN.16}. It is defined as a conditional probability
of finding a nucleon within a distance $\delta$ from a given nucleon at 
point {\bf r} with the same spin and isospin. This measure has been applied 
both to light and very heavy nuclei. 

   The paper is organized as follows. Section \ref{calc-det} describes 
the details of the calculations. The basic features of the nodal structure 
of the single-particle wave function (and thus of its density distribution)
in the case of extremely elongated  prolate shapes are discussed in Sec.\ 
\ref{nodal-wf}. Secs.\ \ref{12C-rod}, \ref{28Si-sect} and \ref{Sect-40Ca} 
analyse the densities and their nodal structures obtained in the CDFT
calculations for the single-particle states occupied in rod-shaped structure 
of $^{12}$C, hyperdeformed band of $^{28}$Si and megadeformed  structure of 
$^{40}$Ca.  Sec.\ \ref{sect-general} summarizes the general features of the 
nodal structure of the single-particle density distributions and analyse 
how they affect the $\alpha$-clusterization and the formation of nuclear 
molecules.  Finally, Sec. \ref{concl} summarizes the results of our 
work.

\section{The details of the calculations}
\label{calc-det}

  The calculations are  performed in the cranked relativistic mean field
(CRMF) framework \cite{AKR.96,VALR.05} using the NL3* CEDF \cite{NL3*}. Note 
that one-dimensional rotation is along the $x$-axis in this framework. The 
pairing is neglected in the calculations since it has very little impact 
on the configurations of interest \cite{RA.16}. The CRMF equations are solved 
in the basis of an anisotropic three-dimensional harmonic oscillator in 
Cartesian coordinates,
for details see Refs.\ \cite{AKR.96,KR.89}. The truncation of basis is performed 
in such a way that all states belonging to the major shells up to
$N_F=14$ fermionic shells for the Dirac spinors and up to $N_B=20$ bosonic 
shells for the meson fields are taken into account. This truncation scheme provides
sufficient numerical accuracy (see Ref.\ \cite{AA.08} for details).

 The calculated configurations are labeled by shorthand 
[$n_{1}$($n_{2}$)($n_{3}$),$p_{1}$($p_{2}$)($p_{3}$)] labels, 
where $n_{1}$, $n_{2}$ and $n_{3}$ are the number of neutrons 
in the $N=3$, 4 and 5 intruder/hyperintruder/megaintruder orbitals 
and $p_{1}$, $p_{2}$ and $p_{3}$ are the number of protons in the 
$N=3$, 4 and 5 intruder/hyperintruder/megaintruder orbitals.  
If some of these orbitals are not occupied, the respective 
numbers are omitted in the configuration labels.

  To give a full 3-dimensional representation of the single-particle density 
distributions, they are plotted in the figures below in the $xz$ and $yz$ planes 
at the positions of the Gauss-Hermite integration points in the $y$ and $x$ 
directions closest to zero. The density cross-section in the $xy$ plane is
taken at the Gauss-Hermite integration point in the $z$-coordinate which gives 
the largest density. The numerical values of these $x$ and $y$ coordinates are 
given in figure captions, while the value of the $z$ coordinate is shown in 
middle panels of the figures which present single-particle density distributions.
Note that some graphical results of the calculations are provided in
the Supplemental Material with this article as Ref.\ \cite{Suppl-node}.

\section{The nodal structure of the single-particle wave function}
\label{nodal-wf}

 Considering that the structures under investigation are characterized
by the extreme prolate deformation and near axial symmetry (see Ref.\ 
\cite{RA.16}), the expansion
of the wave functions of the single-particle states in terms of quantum 
numbers specific for asymptotic Nilsson quantum numbers (see Sec. 8.2 of
Ref.\ \cite{NilRag-book}) is the most appropriate. Thus, the wave function 
$\Psi_{[Nn_z\Lambda]\Omega}$ of the single-particle state denoted by the 
Nilsson quantum number\footnote{We use here the 
standard notation in which the single-particle states are labeled by 
the asymptotic quantum numbers $[Nn_z\Lambda]\Omega$ (Nilsson quantum 
numbers) of the dominant component of the wave function.} 
$[Nn_z\Lambda]\Omega$ is expanded into 
the basis states $|N'n'_z\Lambda'\Omega'>$ by 
\begin{eqnarray}
\Psi_{[Nn_z\Lambda]\Omega} = \sum_{N',n'_z,\Lambda',\Omega'} c_{N'n'_z\Lambda'\Omega'} |N'n'_z\Lambda'\Omega'>
\label{wave-funct}
\end{eqnarray}
Here, the basis states are characterized by principal quantum number $N'$, 
the number $n'_z$ of nodes in the axial direction 
($z$-direction)  and the projections of orbital ($\Lambda'$) and total 
($\Omega'$) single-particle angular momenta on the axis of symmetry. The 
sum in Eq.\ (\ref{wave-funct}) runs over all allowable combinations of the 
quantum numbers $N', n'_z, \Lambda'$ and $\Omega'$.

 Since the single-particle density $\rho_{[Nn_z\Lambda]\Omega}$ of the 
Nilsson state $[Nn_z\Lambda]\Omega$ is given by
\begin{eqnarray}
\rho_{[Nn_z\Lambda]\Omega} = \sum_{N',n'_z,\Lambda',\Omega'} c^2_{N'n'_z\Lambda'\Omega'} 
<N'n'_z\Lambda'\Omega'|N'n'_z\Lambda'\Omega'> \nonumber \\
\end{eqnarray}
the weights  $c^2_{N'n'_z\Lambda'\Omega'}$ define the contributions of the 
basic states $|N'n'_z\Lambda'\Omega'>$ into the single-particle density.

The nodal structure of the single-particle wave function $\Psi_{[Nn_z\Lambda]\Omega}$
and thus of its density distribution is defined by the spatial part of the wave 
function. The spatial parts of the basis states are expressed in cylindrical 
$(r,\phi,z)$ coordinates in the following way (see Ref.\ \cite{N.55} and Sec. 
8.2 of Ref.\ \cite{NilRag-book})
\begin{eqnarray}
|N'n'_z\Lambda'>  \sim  
                        H_{n'_z}\left(\frac{z}{b_z}\right) 
                         L_{n'_r}^{|\Lambda'|}\left(\frac{r^2}{b^2_\bot}\right) 
                         \left(\frac{r}{b_\bot}\right)^{|\Lambda'|}
\nonumber  \\
e^{-\frac{1}{2}\left(\frac{z^2}{b_z^2} + \frac{r^2}{b_\bot^2}\right)} 
e^{i\Lambda' \phi} 
\label{wf-spatial}
\end{eqnarray}
where $H$ and $L$ are the Hermite and associated Laguerre polynomials. The $b_z$ and 
$b_{\bot}$ are oscillator lenghts in axial and radial directions 
and $n'_r$ is the number of radial nodes. The condition $N=n'_z+2n'_r +|\Lambda'|$ 
defines the possible combinations of the quantum numbers $n'_r$, $n'_z$ and
$\Lambda'$ and thus the nodal structure of the density distribution of the basis 
state.

  The nodal structure in the axial direction of the wave function (Eq. (\ref{wf-spatial})) 
and thus of related single-particle density distribution  is defined 
by the zeros of the Hermite polynomials. In general, the nodal structure of
density distribution in radial direction is defined by zeros in associated 
Laguerra polynomials and in the $\left(\frac{r}{b_\bot}\right)^{|\Lambda'|}$
term. It turns that all basis states (see Table \ref{table-wf}), providing the 
most important contributions to the wavefunctions of the single-particle states 
of interest, have $n_r=0$ for which $L_0^{|\Lambda'|}\left(\frac{r^2}{b^2_\bot}\right)=1$. 
With a single exception these states also have only two possible values of 
$\Lambda'$, namely, 0 and 1. Thus, for these states the node at the axis of 
symmetry $(r=0)$ emerges only at $\Lambda'=1$ from the 
$\left(\frac{r}{b_\bot}\right)^{|\Lambda'|}$ term. As a result, when considering
the pattern of the density distribution only two types of the
basis states are important, namely, the $|N,N,0>$ and $|N,N-1,1>$ states. The 
density distributions of the basis states with $|NN0>$ will be axially symmetric 
with the maximum of density located at the axis of symmetry. The basis states with  
$|N,N-1,1>$ structure will have a zero density at the axis of symmetry.

\begin{figure*}[htb]
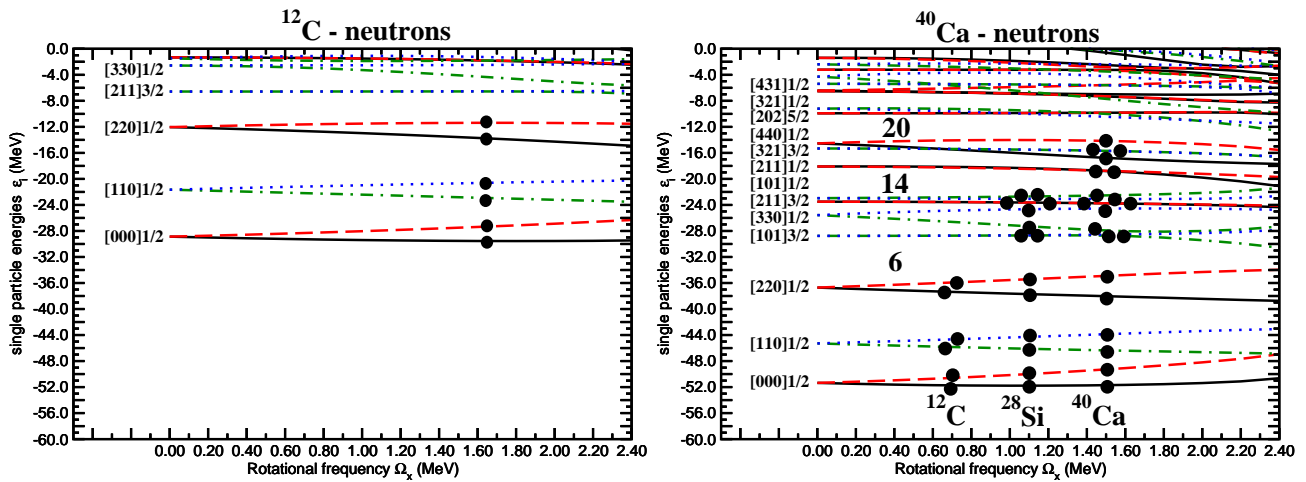

\includegraphics[angle=0,width=8.5cm]{fig-2-a.eps}
\includegraphics[angle=0,width=8.5cm]{fig-2-b.eps}
\caption{(Color online) Neutron single-particle energies (routhians) in the 
self-consistent rotating potential as a function of the rotational frequency 
$\Omega_x$. They are given along the deformation path of the rod-shaped 
structure in $^{12}$C and yrast MD configuration [42,42]  
in $^{40}$Ca. Long-dashed, solid, dot-dashed and dotted lines indicate  
$(\pi=+, r=+i)$, $(\pi=+, r=-i)$, $(\pi=-, r=+i)$ and $(\pi=-, r=-i)$ orbitals, 
respectively.  At $\Omega_x=0.0$ MeV, the single-particle orbitals are labeled by 
the asymptotic quantum numbers $[Nn_z\Lambda]\Omega$ (Nilsson quantum numbers) of 
the dominant component of the wave function. Solid circles indicate the 
occupied orbitals. Large shell gaps are indicated in the right panel.
}
\label{Fig-routh}
\end{figure*}

 However, as follows from Eq.\ (\ref{wave-funct}) the basis states are 
expected to be mixed in the structure of the single-particle wave function. 
As a result, the single-particle wave functions may not have a well 
pronounced nodal structure specific for basis states. However, a number 
of factors leads to the reduction of such mixing in extremely deformed 
structures of light nuclei. First, the quality of asymptotic quantum
numbers is improving with the increase of the elongation of the nuclear
system (see Sec. 8.2 in Ref.\ \cite{NilRag-book}).  In addition, such 
mixing depends on the energy distance between the basis states and the 
number of possible counterparts with which appreciable mixing could 
take place. At the bottom of nucleonic potential, these energy distances 
are large and the number of counterparts is quite limited (see Fig.\ 
\ref{Fig-routh}). This leads to appreciable suppression of the 
mixing.

This is illustrated in Table \ref{table-wf} in which the weights $c^2_{N'n'_z\Lambda'\Omega'}$ 
of the three largest components of the wave functions of the single-particle states occupied in 
the megadeformed [42,42] configuration of $^{40}$Ca are presented. One can see that the 
single-particle wave functions are dominated by a single very large component which in turn will 
define the spatial distribution of the single-particle density. This domination is especially 
pronounced at no rotation and for the single-particle orbitals located at the bottom of the 
nucleonic potential. Note that the [440]1/2 and [211]1/2 orbitals with $r=-i$ strongly 
interact in substantial frequency range near $\Omega_x \sim 1.8$ MeV (see  Fig.\ 
\ref{Fig-routh}b). This leads to an increase of the fragmentation of the wave function of these 
two states at $\Omega_x=1.8$ MeV (Table \ref{table-wf}). However, such interaction 
is absent in the [440]1/2 and [211]1/2 orbitals with $r=+i$. As a result, their wave functions 
are substantially less fragmented (Table \ref{table-wf}).

It is interesting to compare the level of the fragmentation of the single-particle 
wave functions presented in Table \ref{table-wf} with the ones obtained in the 
calculations for other nuclei. Unfortunately, as a rule the studies within the cranking 
models based on different frameworks do not show detailed information on the structure 
of the single-particle wave functions. To our knowledge, there are only two exceptions 
which provide some hints on the fragmentation of the  single-particle wave functions 
and their evolution with rotational frequency. The superdeformed bands in $^{32}$S and 
neighboring odd-mass nuclei have been studied in cranking Skyrme Hartree-Fock approach 
in Ref.\ \cite{MDD.00}. The analysis of routhian diagrams presented in this manuscript  
indicate the change of  the dominant components of the single-particle wavefunctions with 
rotational frequency for appreciable number of routhians which do not undergo unpaired 
band crossing. On the contrary, the rotation does not change the dominant components 
of the single-particle wave functions of the above mentioned type of the routhians shown
in Table \ref{table-wf}. The substantial fragmentation of the single-particle wave functions 
has also been seen in the CRMF calculations of the yrast hyperdeformed configuration in 
$^{124}$Xe \cite{AA.08}. Thus, the level of the fragmentation of the single-particle wave 
functions shown in Table \ref{table-wf} is on average substantially lower than the one 
seen in these two examples.

\begin{table*}[ht]
\begin{center}
\caption{The squared amplitudes $c^2_{N'n'_z\Lambda'\Omega'}$ of three largest 
components of the wave functions of the single-particle states occupied in the 
megadeformed [42,42] configuration of $^{40}$Ca. The states are shown from the bottom
of nucleonic potential in the same sequence as they appear in the
routhian diagram of this configuration (see Fig.\ \ref{wave-funct}b).
The results are shown at no rotation ($\Omega_x=0.0$ MeV) and at rotational
frequency $\Omega_x=1.8$ MeV which corresponds to spin $I=25\hbar$. If not 
indicated otherwise, the orbitals have signature $r=-i$. Note that the [550]1/2
state is not occupied in this configuration.
\label{table-wf}
}
\begin{tabular}{|c|c|c|} \hline 
  State    &  $\Omega_x$ [MeV] &             Wave function                     \\ \hline
$[000]1/2$ &     0.0      &   92.7\%$|000,1/2>$ + 6.7\%$|220,1/2>$ + 0.4\%$|200,1/2>$   \\
           &     1.8      &   90.4\%$|000,1/2>$ + 8.7\%$|220,1/2>$ + 0.3\%$|200,1/2>$   \\
$[110]1/2$ &     0.0      &   89.6\%$|110,1/2>$ + 9.6\%$|330,1/2>$ + 0.4\%$|101,1/2>$   \\
           &     1.8      &   87.3\%$|110,1/2>$ + 9.6\%$|330,1/2>$ + 1.3\%$|101,3/2>$   \\
$[220]1/2$ &     0.0      &   85.1\%$|220,1/2>$ + 6.5\%$|000,1/2>$ + 6.2\%$|440,1/2>$   \\
           &     1.8      &   73.2\%$|220,1/2>$ + 8.9\%$|000,1/2>$ + 6.9\%$|440,1/2>$   \\
$[101]3/2$ &     0.0      &   92.9\%$|101,3/2>$ + 6.7\%$|321,3/2>$ + 0.3\%$|301,3/2>$   \\
           &     1.8      &   82.9\%$|101,3/2>$ + 9.4\%$|321,3/2>$ + 3.4\%$|101,1/2>$   \\
$[330]1/2$ &     0.0      &   75.7\%$|330,1/2>$ + 8.5\%$|110,1/2>$ + 6.8\%$|321,1/2>$   \\
           &     1.8      &   65.5\%$|330,1/2>$ + 8.9\%$|101,1/2>$ + 5.8\%$|101,3/2>$   \\
$[211]3/2$ &     0.0      &   92.5\%$|211,3/2>$ + 6.9\%$|431,3/2>$ + 3.6\%$|202,3/2>$   \\
           &     1.8      &   78.4\%$|211,3/2>$ + 7.2\%$|431,3/2>$ + 6.1\%$|220,1/2>$   \\
$[101]1/2$ &     0.0      &   90.2\%$|101,1/2>$ + 5.0\%$|330,1/2>$ + 4.4\%$|321,1/2>$   \\
           &     1.8      &   76.7\%$|101,1/2>$ + 7.2\%$|321,1/2>$ + 6.5\%$|330,1/2>$   \\
$[211]1/2$ &     0.0      &   81.4\%$|211,1/2>$ + 9.6\%$|431,1/2>$ + 5.2\%$|440,1/2>$   \\
           &     1.8      &   65.3\%$|211,1/2>$ + 22.1\%$|440,1/2>$ + 5.4\%$431,3/2>$  \\
$[211]1/2$ &     0.0      &   81.4\%$|211,1/2>$ + 9.6\%$|431,1/2>$ + 5.2\%$|440,1/2>$   \\
 $(r=+i)$  &     1.8      &   82.9\%$|211,1/2>$ + 6.8\%$|431,1/2>$ + 3.2\%$|211,3/2>$  \\
$[321]3/2$ &     0.0      &   85.7\%$|321,3/2>$ + 6.1\%$|101,3/2>$ + 6.0\%$|541,1/2>$   \\
           &     1.8      &   57.6\%$|321,3/2>$ + 9.1\%$|330,1/2>$ + 7.1\%$|312,5/2>$    \\
$[440]1/2$ &     0.0      &   77.4\%$|440,1/2>$ + 10.7\%$|211,1/2>$ + 4.1\%$|660,1/2>$   \\
           &     1.8      &   36.0\%$|440,1/2>$ + 20.3\%$|431,1/2>$ + 15.3\%$|211,1/2>$  \\
$[440]1/2$ &     0.0      &   77.4\%$|440,1/2>$ + 10.7\%$|211,1/2>$ + 4.1\%$|660,1/2>$   \\
 $(r=+i)$  &     1.8      &   63.3\%$|440,1/2>$ + 15.2\%$|431,3/2>$ + 6.0\%$|220,1/2>$  \\ \hline \hline
$[550]1/2$  &     0.0     &   79.3\%$|550,1/2>$ + 6.8\%$|541,1/2>$ + 6.7\%$|770,1/2>$    \\
            &     1.8     &   46.9\%$|550,1/2>$ + 23.9\%$|541,3/2>$ + 5.9\%$|770,1/2>$   \\ \hline
\end{tabular}
\end{center}
\end{table*}

\section{Rod-shape structure in $^{12}$C} 
\label{12C-rod}

  One of the examples of the cluster structures is the linear chain of 
three $\alpha$-particles in $^{12}$C \cite{OFE.06}. These rod-shape 
structures in rotating $^{12}$C and neighboring nuclei have been 
investigates in the framework of cranked relativistic mean field 
theory  in Ref.\ \cite{ZIM.15}.  The total neutron density distribution 
for this configuration in $^{12}$C is presented in Fig.\ \ref{Total-densities}a 
and its routhian diagram is shown in Fig.\ \ref{Fig-routh}a. The proton and neutron 
single-particle states with structure [000]1/2, [110]1/2 and [220]1/2 
of both signatures are occupied in this configuration. Note that proton 
routhians are very similar to neutron ones; however, they are less bound 
(by roughly 8 MeV) because of the Coulomb interaction.

\begin{figure*}[htb]
\includegraphics[angle=0,width=5.5cm]{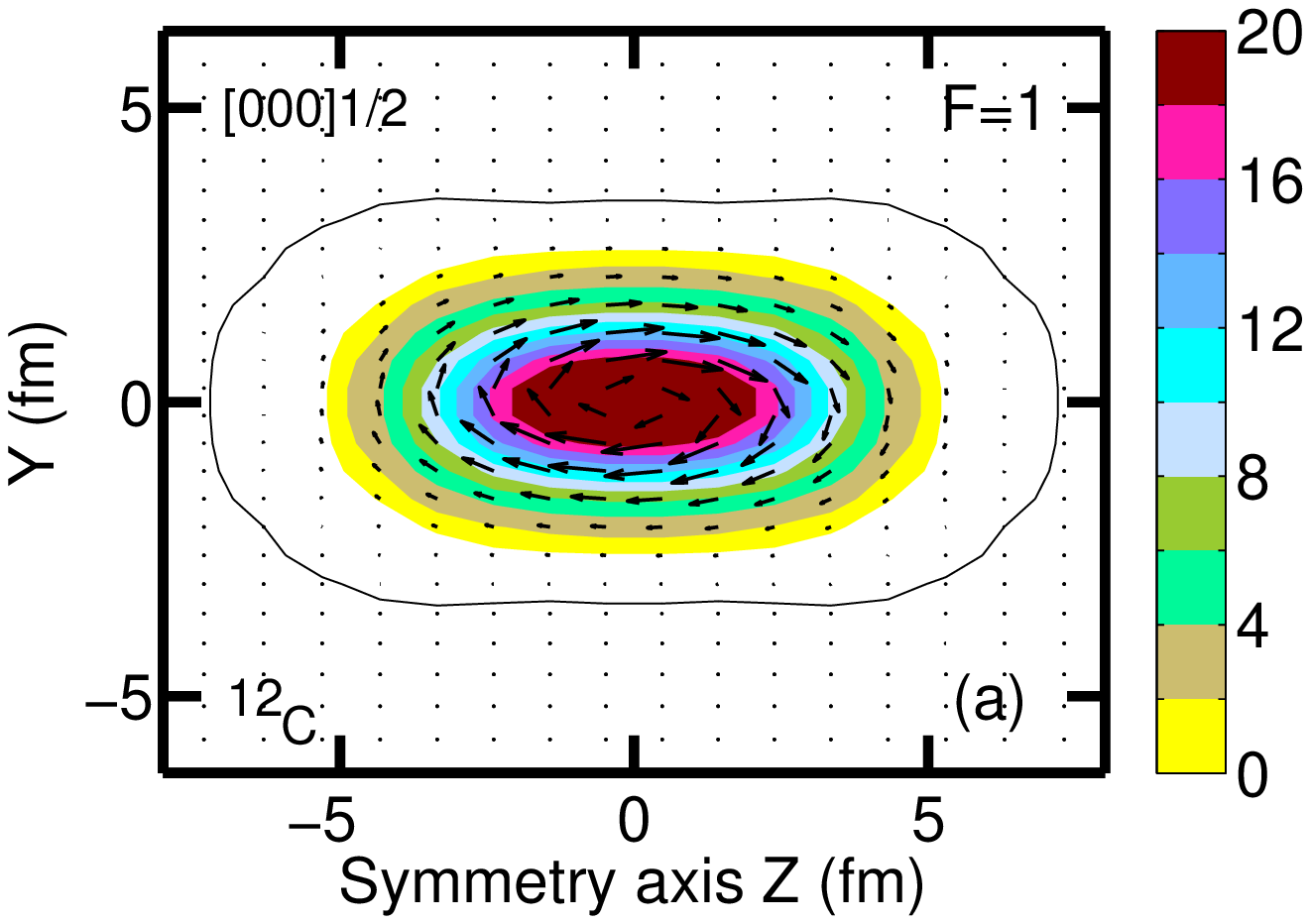}
\includegraphics[angle=0,width=5.5cm]{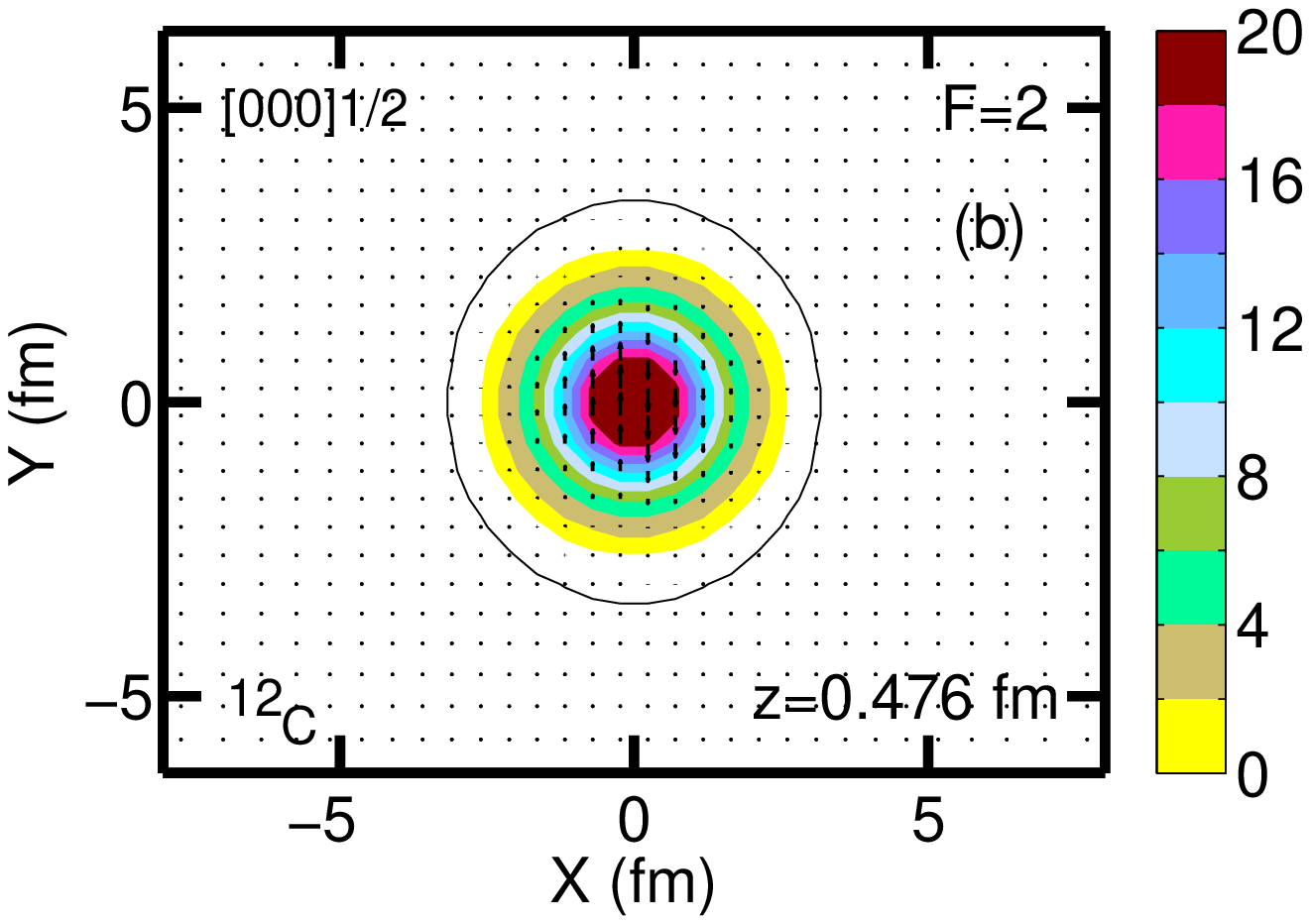}
\includegraphics[angle=0,width=5.5cm]{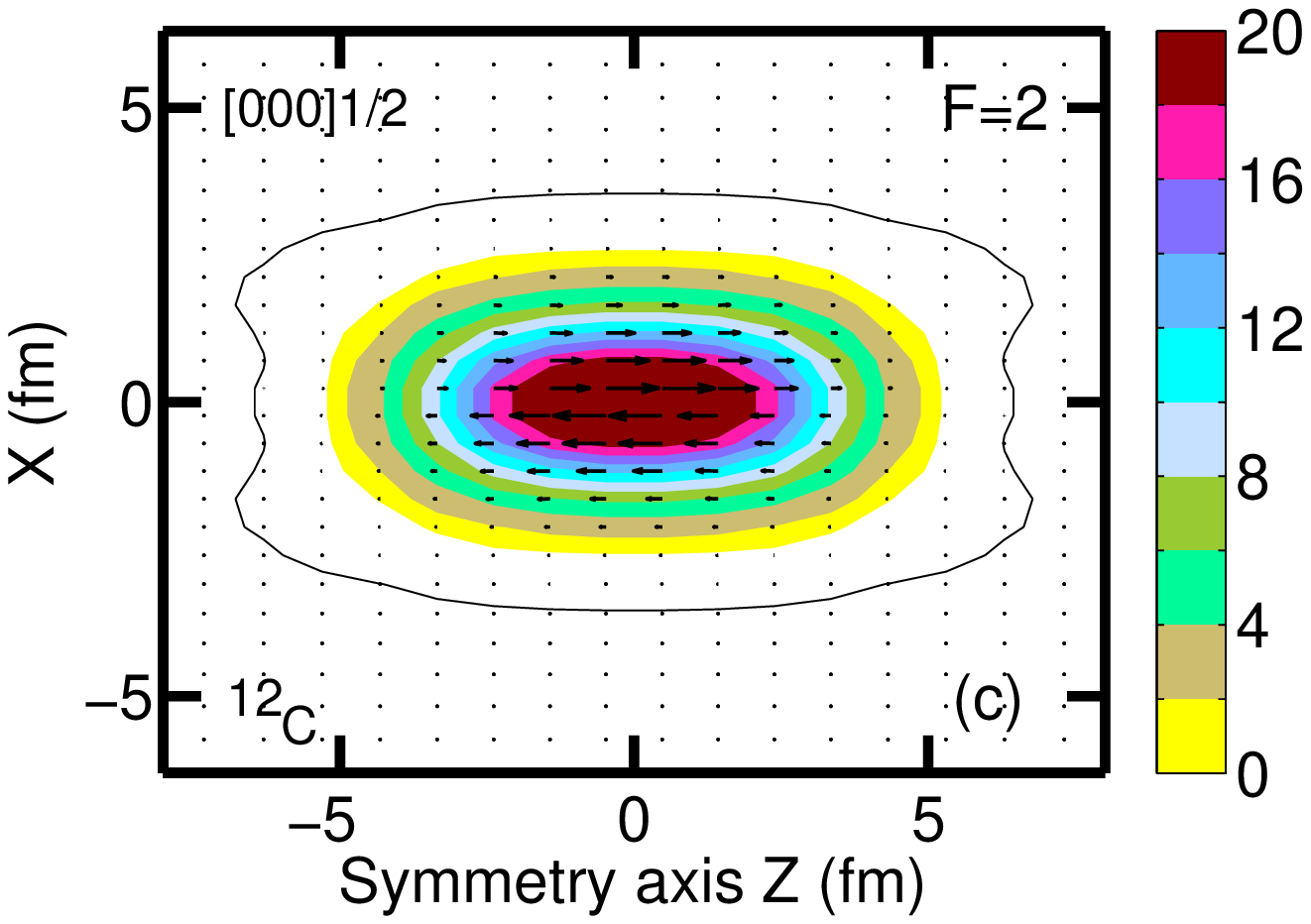}
\includegraphics[angle=0,width=5.5cm]{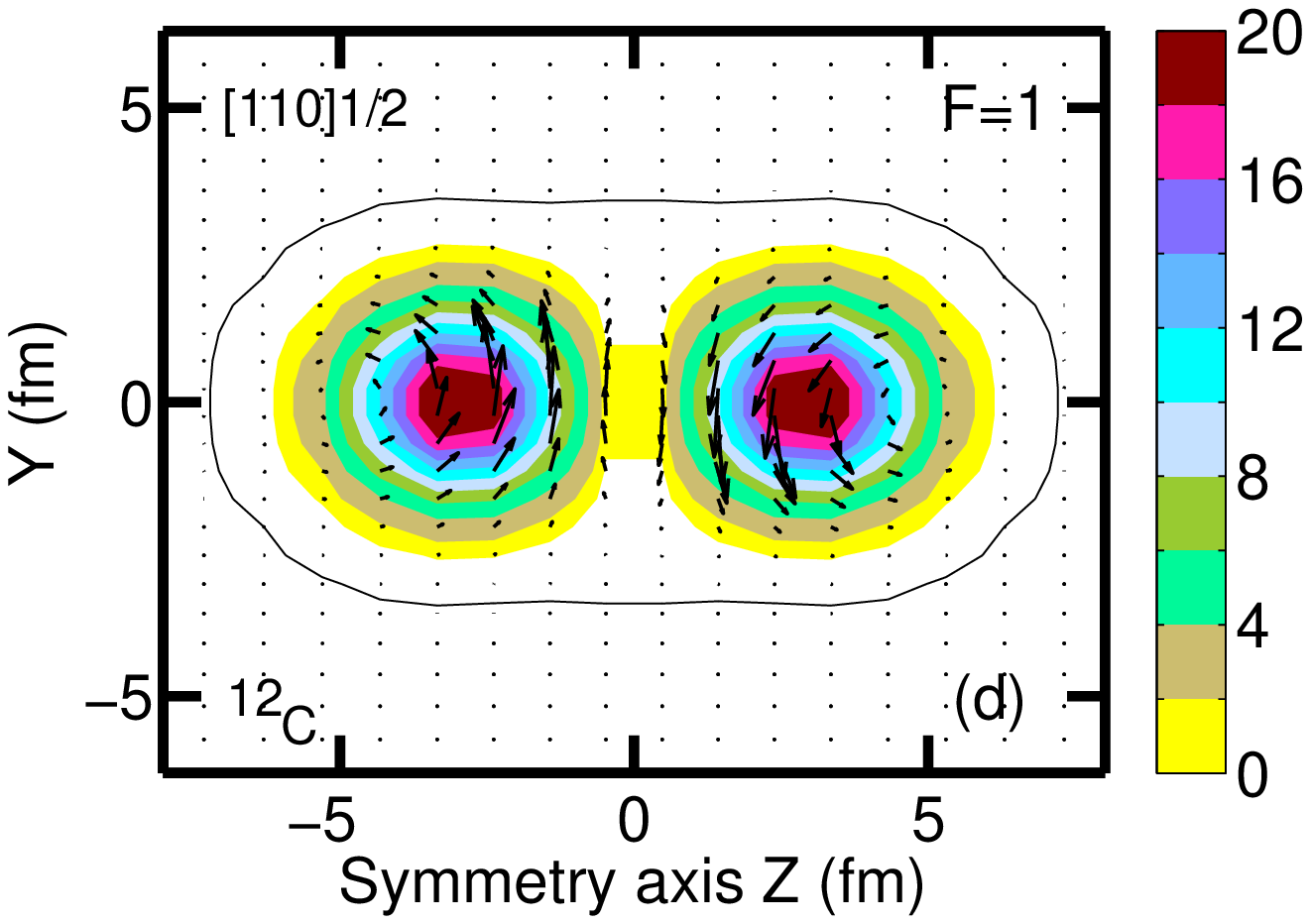}
\includegraphics[angle=0,width=5.5cm]{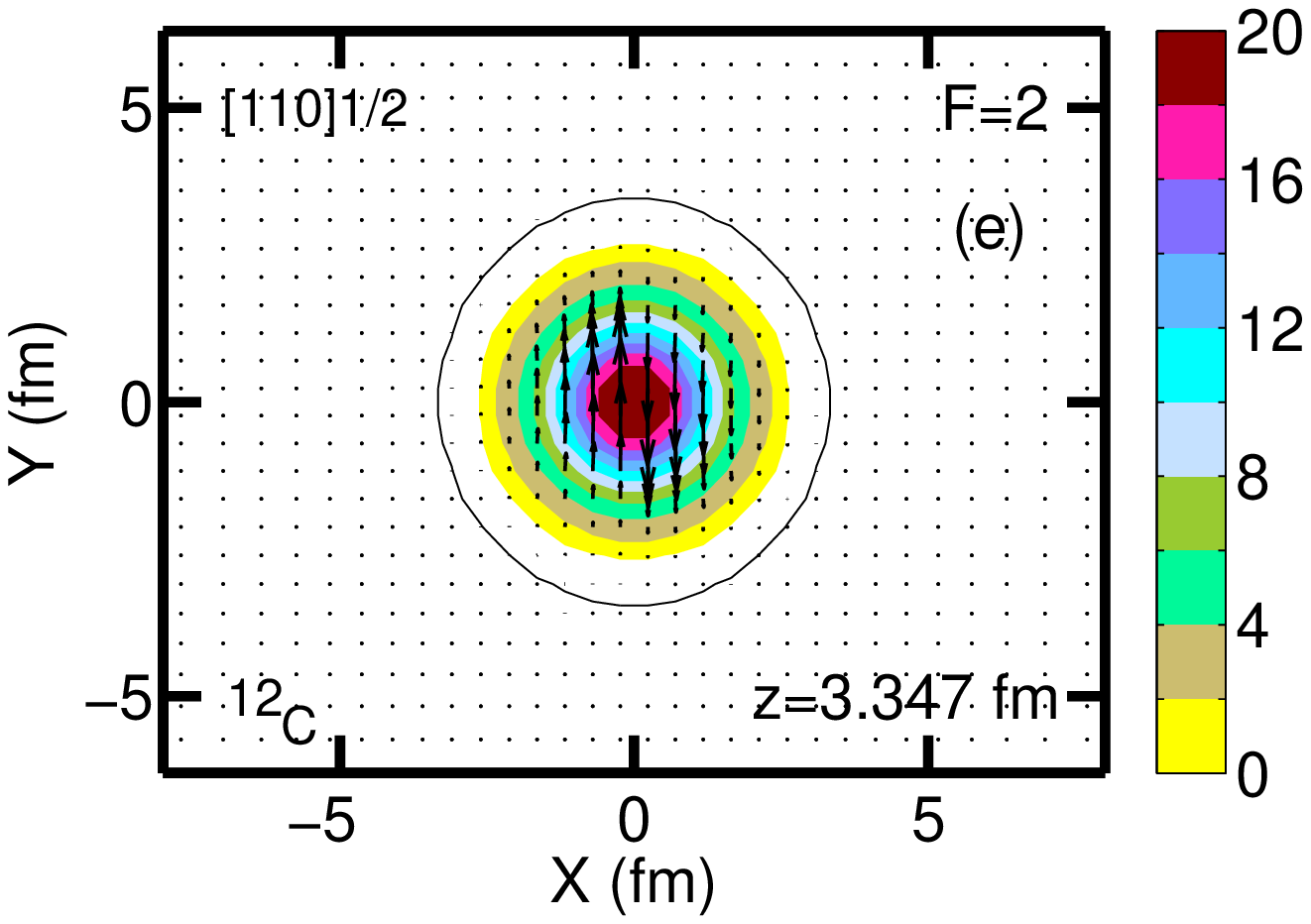}
\includegraphics[angle=0,width=5.5cm]{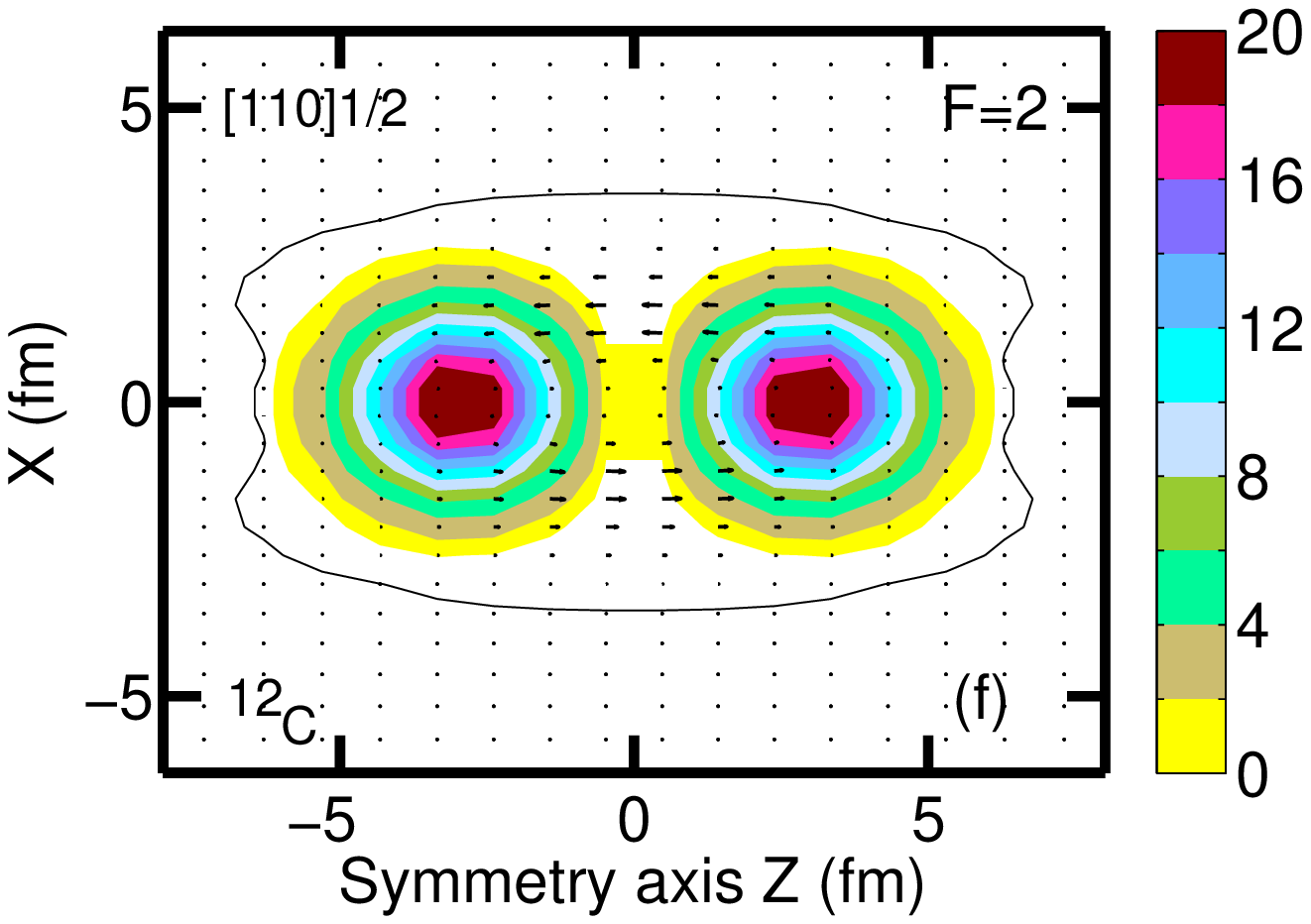}
\includegraphics[angle=0,width=5.5cm]{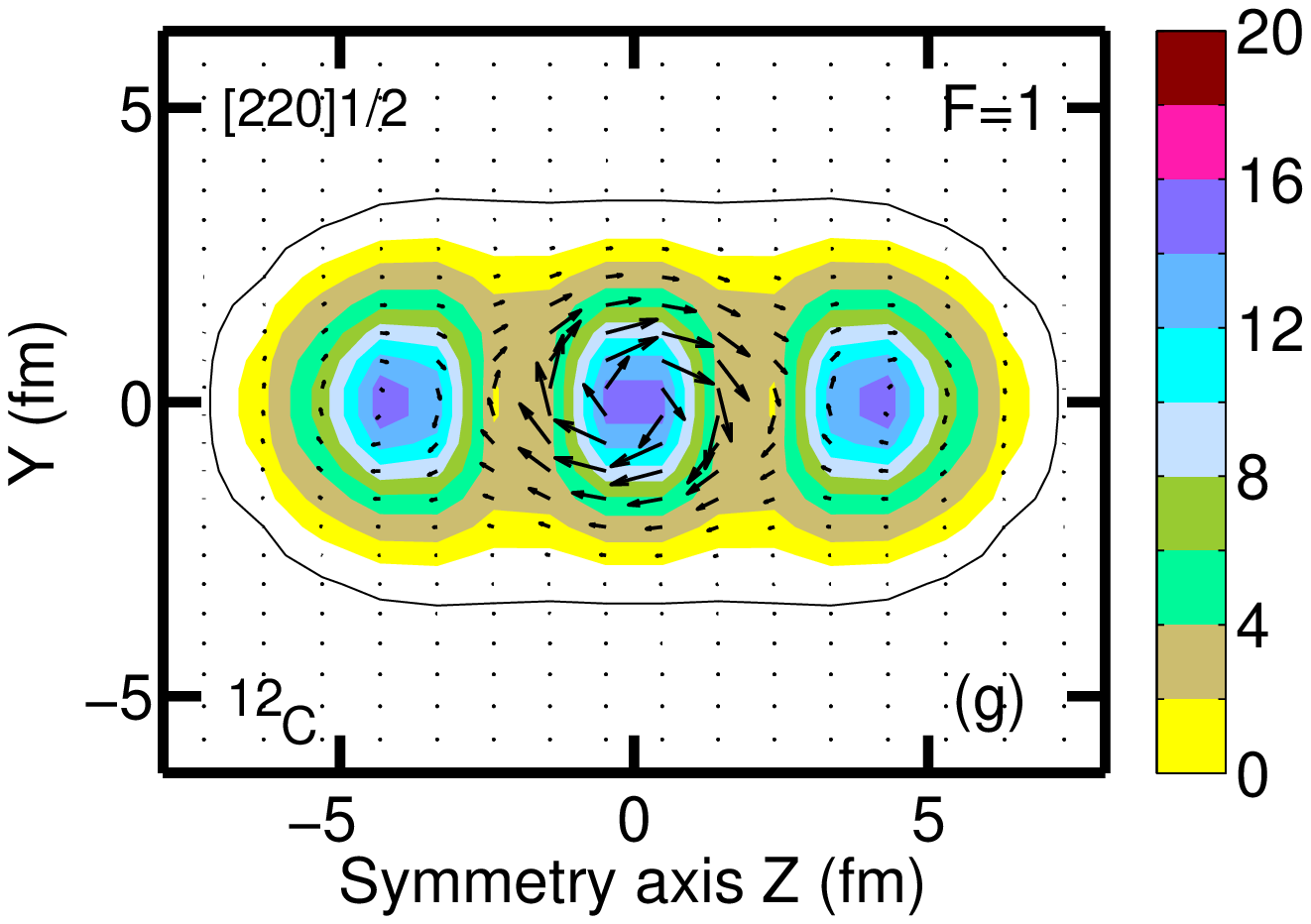}
\includegraphics[angle=0,width=5.5cm]{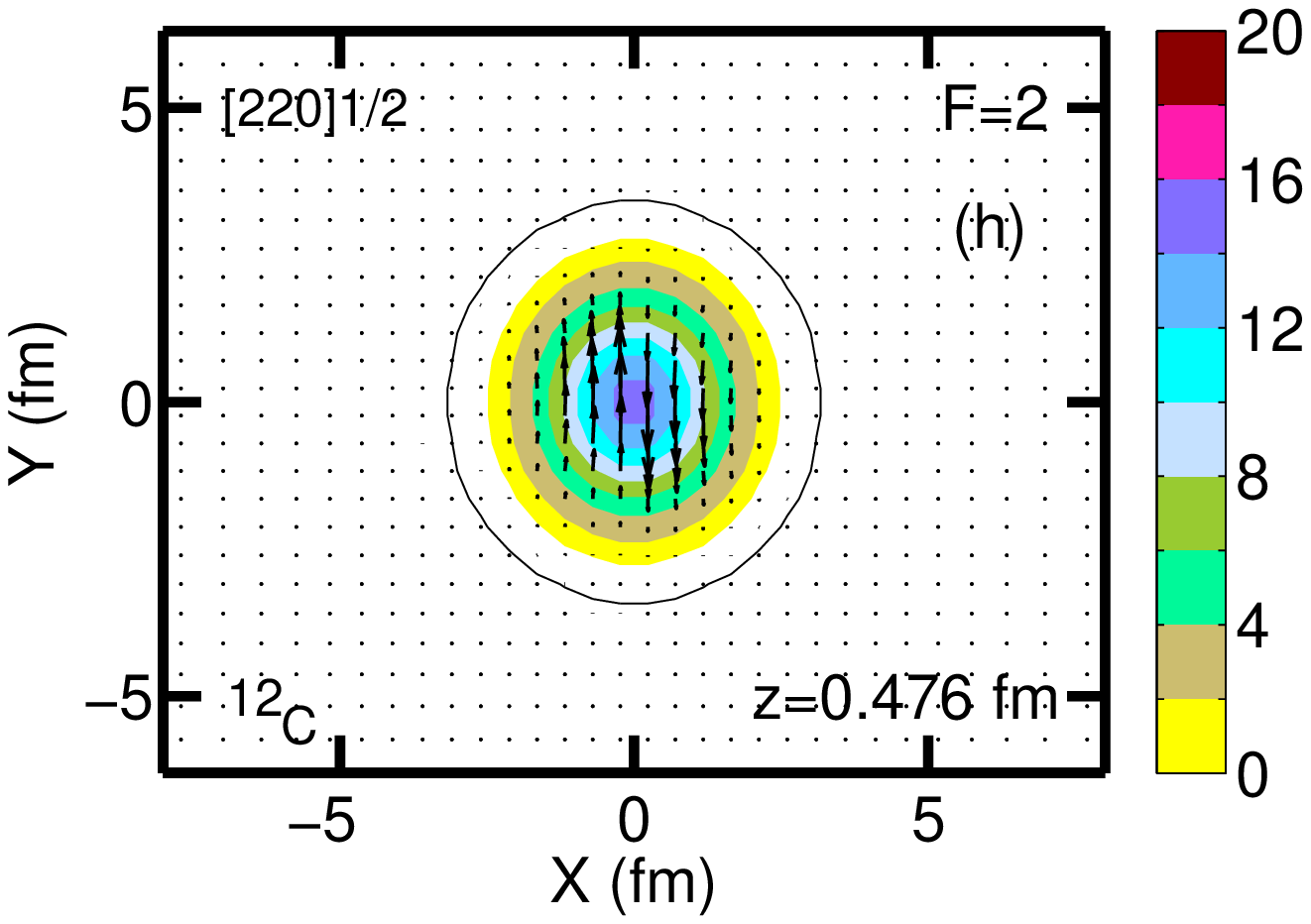}
\includegraphics[angle=0,width=5.5cm]{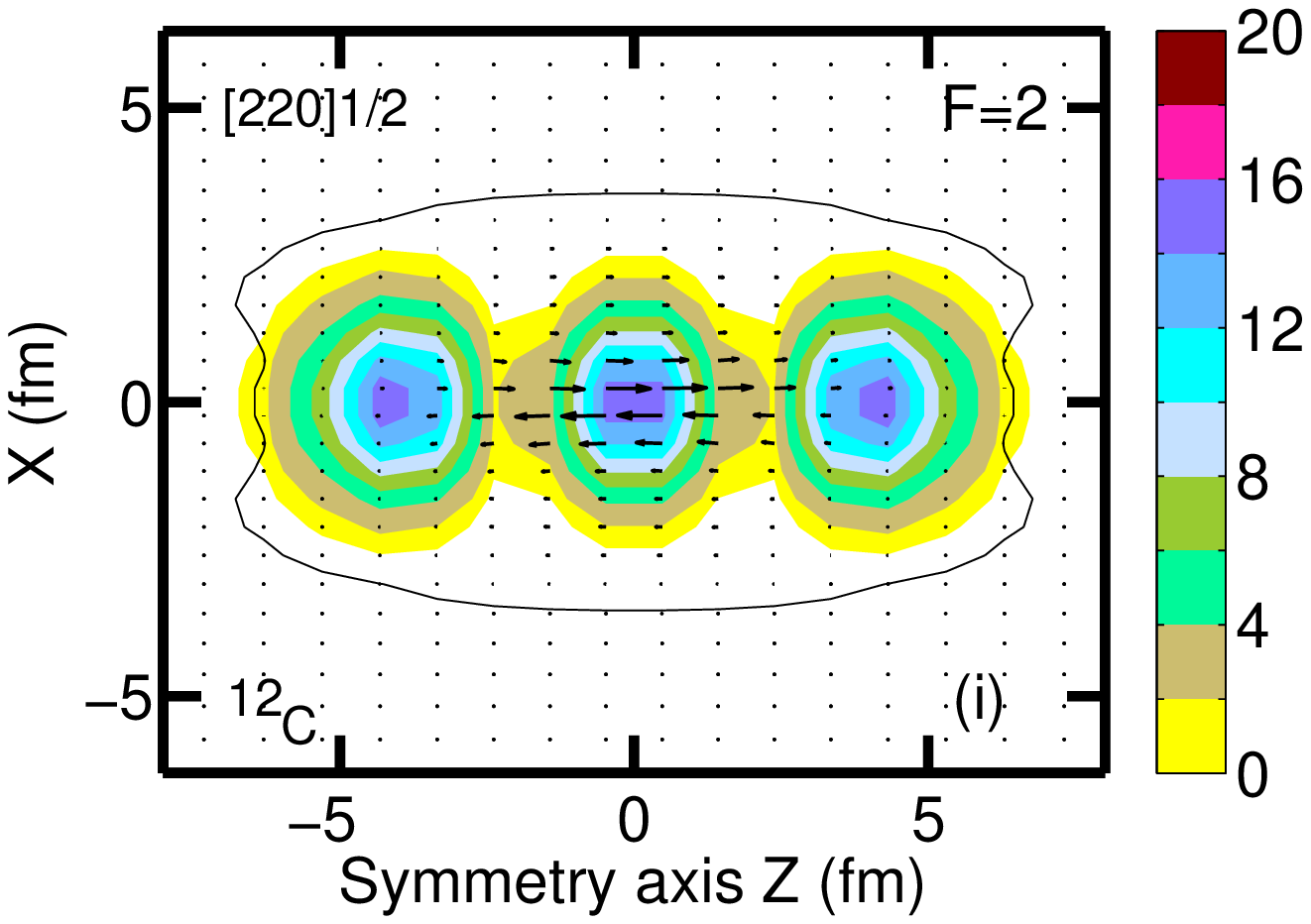}
\caption{(Color online) The single-neutron density distributions due to 
the occupation of the indicated Nilsson states with signature $r=-i$ in the 
rod-shape configuration of $^{12}$C. The colormap shows the densities as 
multiplies of $0.001$ fm$^{-3}$. The plotting of the densities starts with 
yellow color at $0.001$ fm$^{-3}$. The results of the calculations are shown 
at rotational frequency $\Omega_x=3.2$ MeV which corresponds to spin $I=8.77 
\hbar$. For each state, the cross-sections in the $yz$ and $xz$ planes are 
plotted at $x=0.234$ fm and $y=0.234$ fm, respectively. 
The shape and size of the nucleus are indicated by black solid line which 
corresponds to total neutron density line of $\rho=0.001$ fm$^{-3}$. In addition, 
the current distributions {\bff j}$^n$({\bff r}) produced by these states are 
shown by arrows. The currents in panels (a),(d) and (g) are plotted at arbitrary 
units for better visualization. The currents in other panels are normalized to 
the currents in above mentioned panels by using factor F.
}
\label{C12-NL3*}
\end{figure*}

 Single-particle density distributions of these states are shown in Fig.\ 
\ref{C12-NL3*}. One can see that they are almost axially symmetric (see density 
cross-sections in the $xy$-plane which is perpendicular to the symmetry 
axis $z$ ). The density distributions for opposite signatures of the 
specific orbital are almost the same. The same is also true for single-particle 
density distributions for the proton and neutron states with the same 
structure. Thus, it is sufficient to consider only neutron states with 
signature $r=+i$ as it is done in Fig.\ \ref{C12-NL3*}.

 The density distributions of these states are almost axially symmetric 
with the maximum of density inside of each cluster located at the axis 
of symmetry. This is because their wave functions do not have nodes in 
radial direction. However, they show different nodal structure along 
the axis of symmetry since $n_z$ is changing from 0 in the [000]1/2 
state via 1 in the [110]1/2 state to 2 in the [220]1/2 state.

  The density distribution of the [000]1/2 state, which is emerging 
from spherical $1s_{1/2}$ subshell, is the ellipsoid of revolution with 
the maximum of the density located at the center of nucleus. The density 
distribution of the [110]1/2 state, emerging from the spherical $1p_{1/2}$ 
subshell, is formed by two spheroids located symmetrically with respect of 
$z=0$. The [220]1/2 orbital, emerging from spherical $1d_{5/2}$ subshell, 
shows three spheroidal clusters in density distribution; one of them is 
located at the center of nucleus and two others symmetrically with respect 
of it.  Among these states, the highest localization of the wave function 
is seen in the [000]1/2 state. With increasing principal quantum number 
$N$ (and the number of density clusters) the localization of the wave 
function and the maximum density in the center of the density cluster 
decreases.

\begin{figure*}[htb]
\includegraphics[angle=0,width=5.5cm]{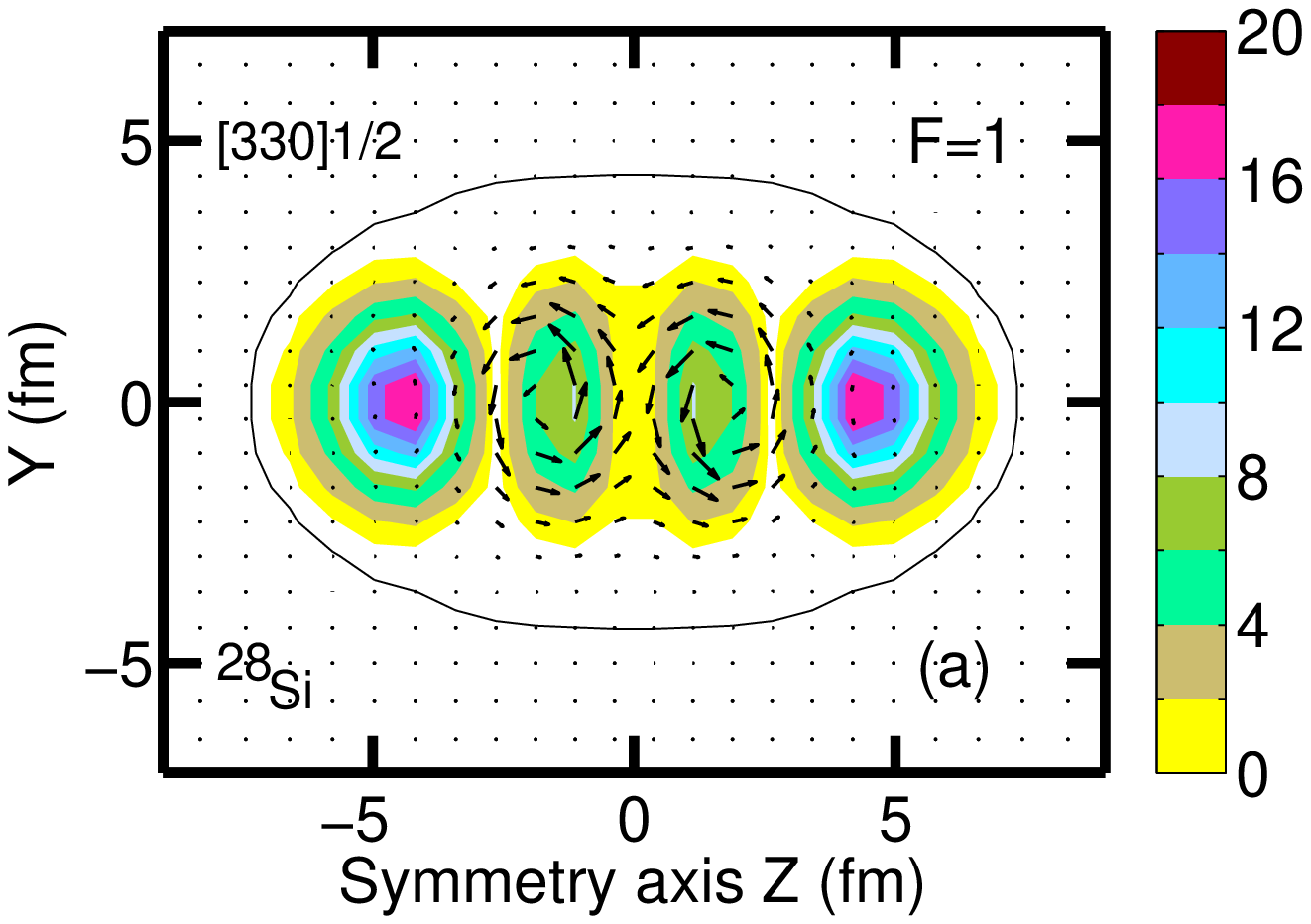}
\includegraphics[angle=0,width=5.5cm]{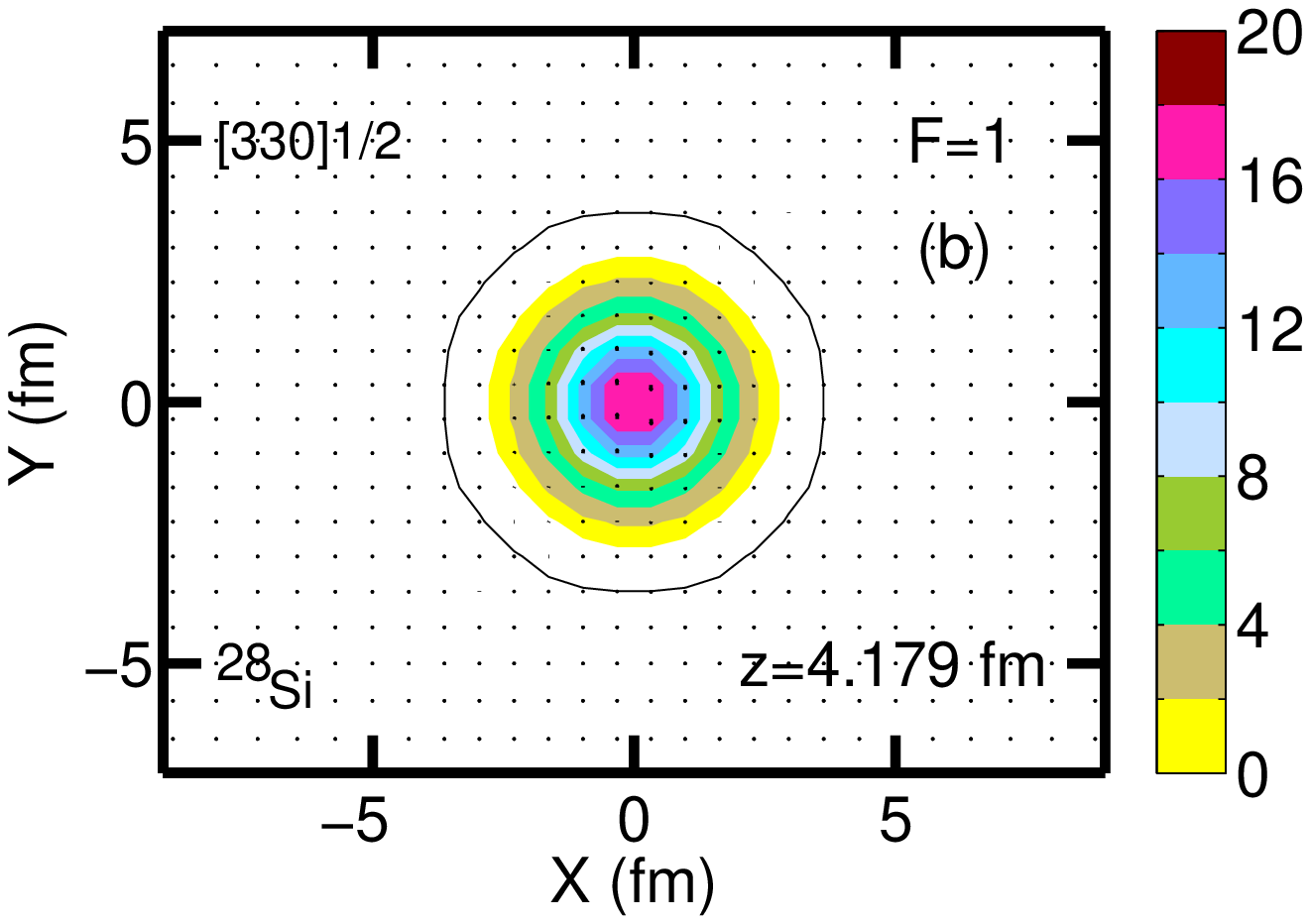}
\includegraphics[angle=0,width=5.5cm]{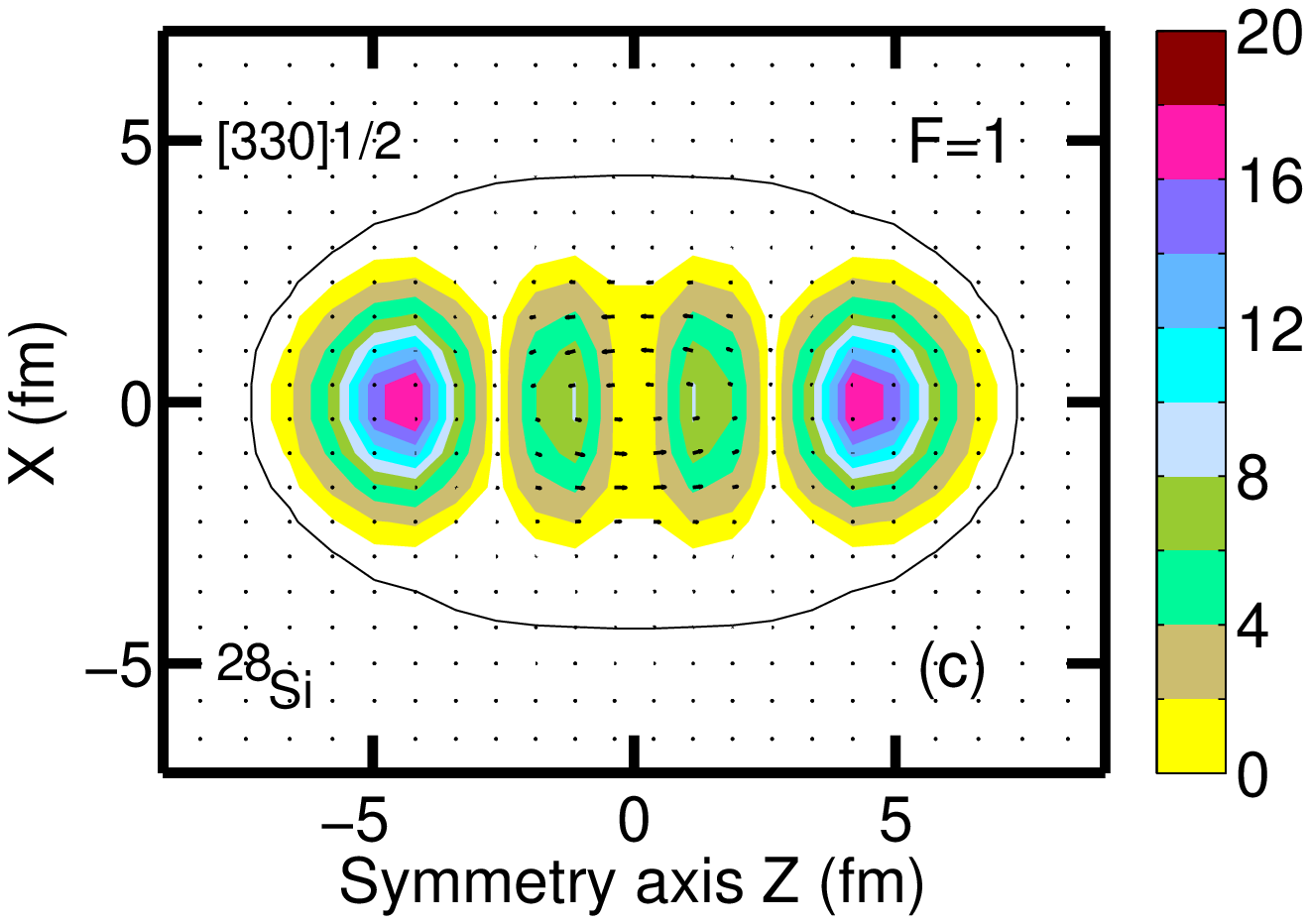}
\includegraphics[angle=0,width=5.5cm]{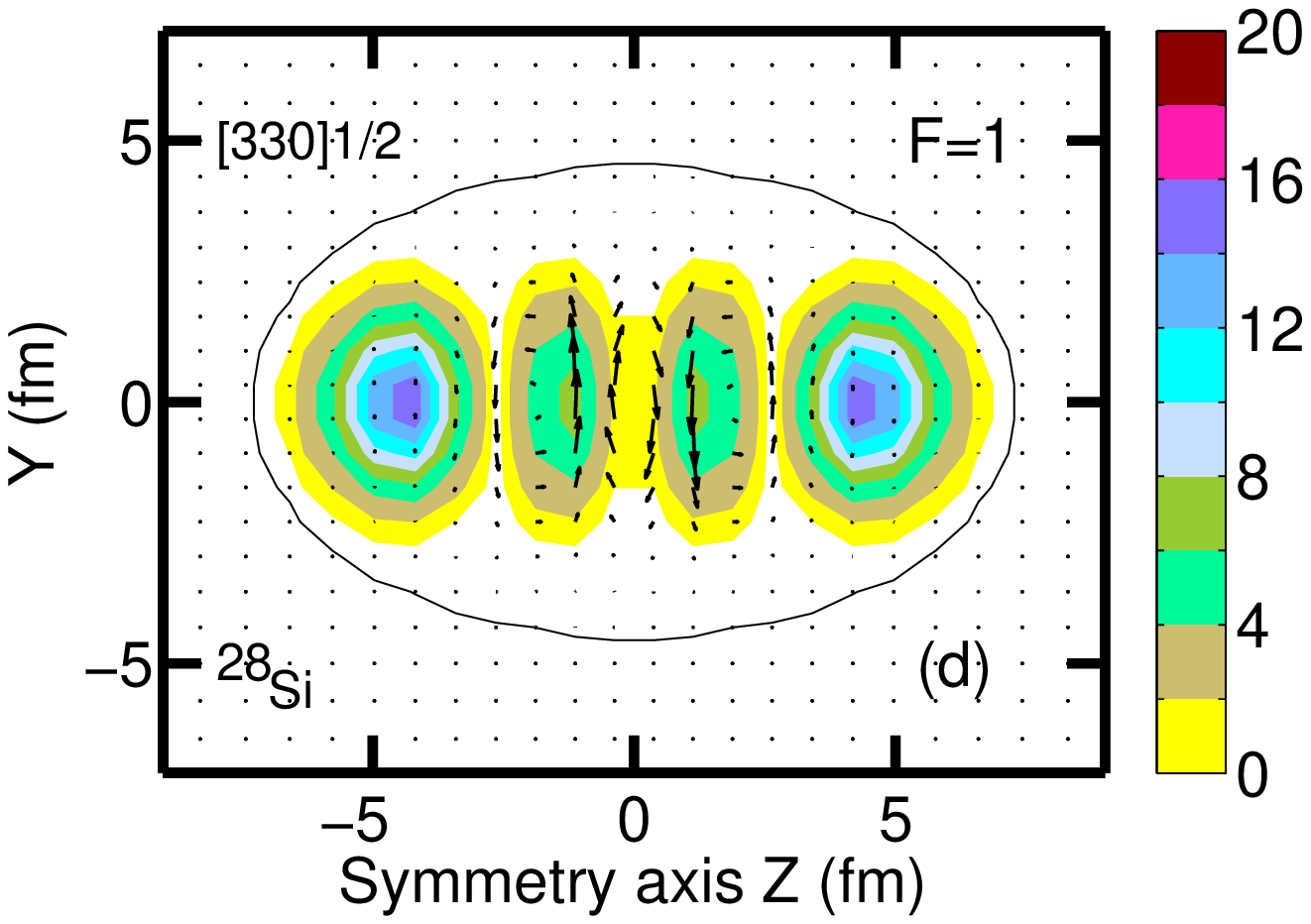}
\includegraphics[angle=0,width=5.5cm]{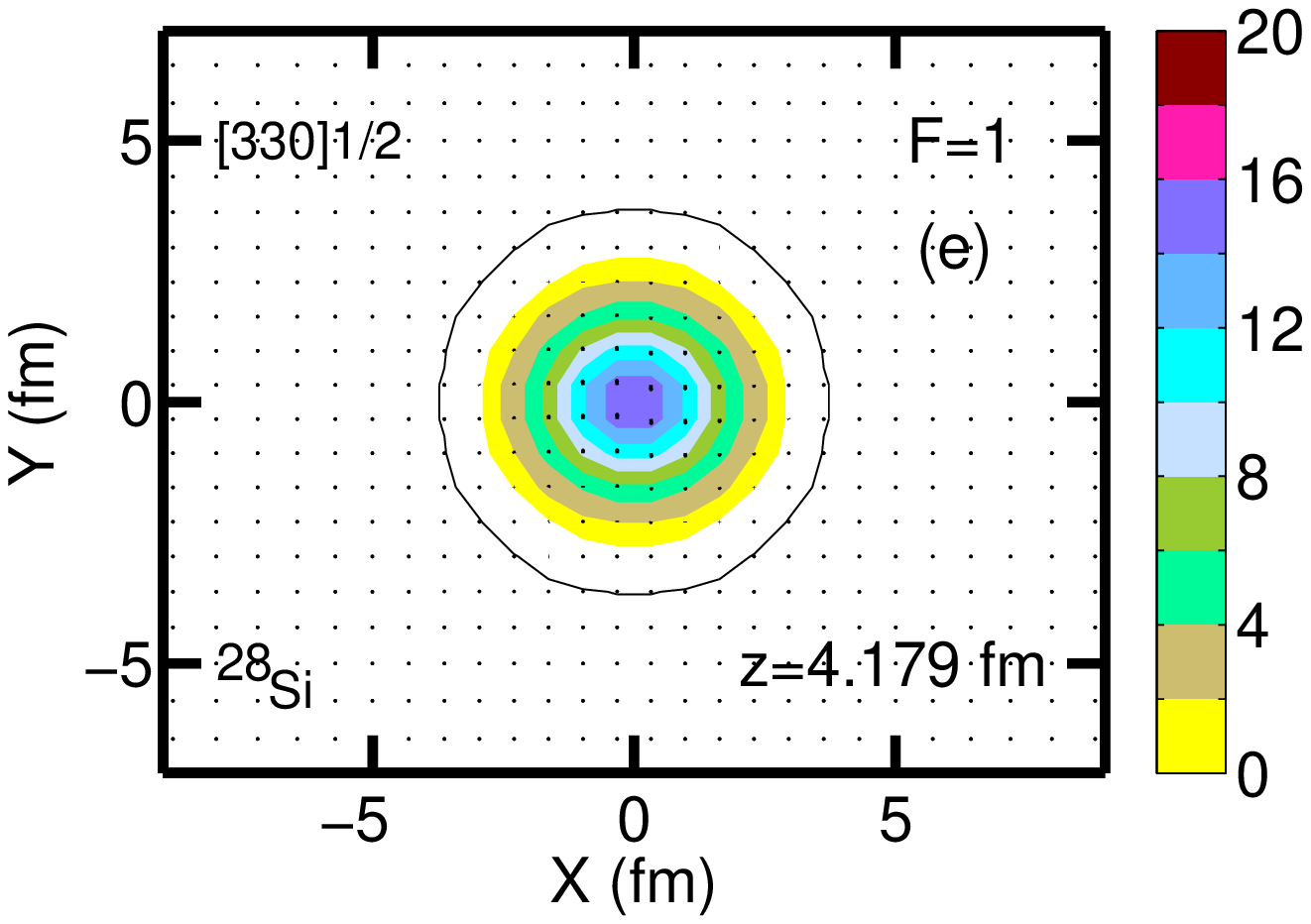}
\includegraphics[angle=0,width=5.5cm]{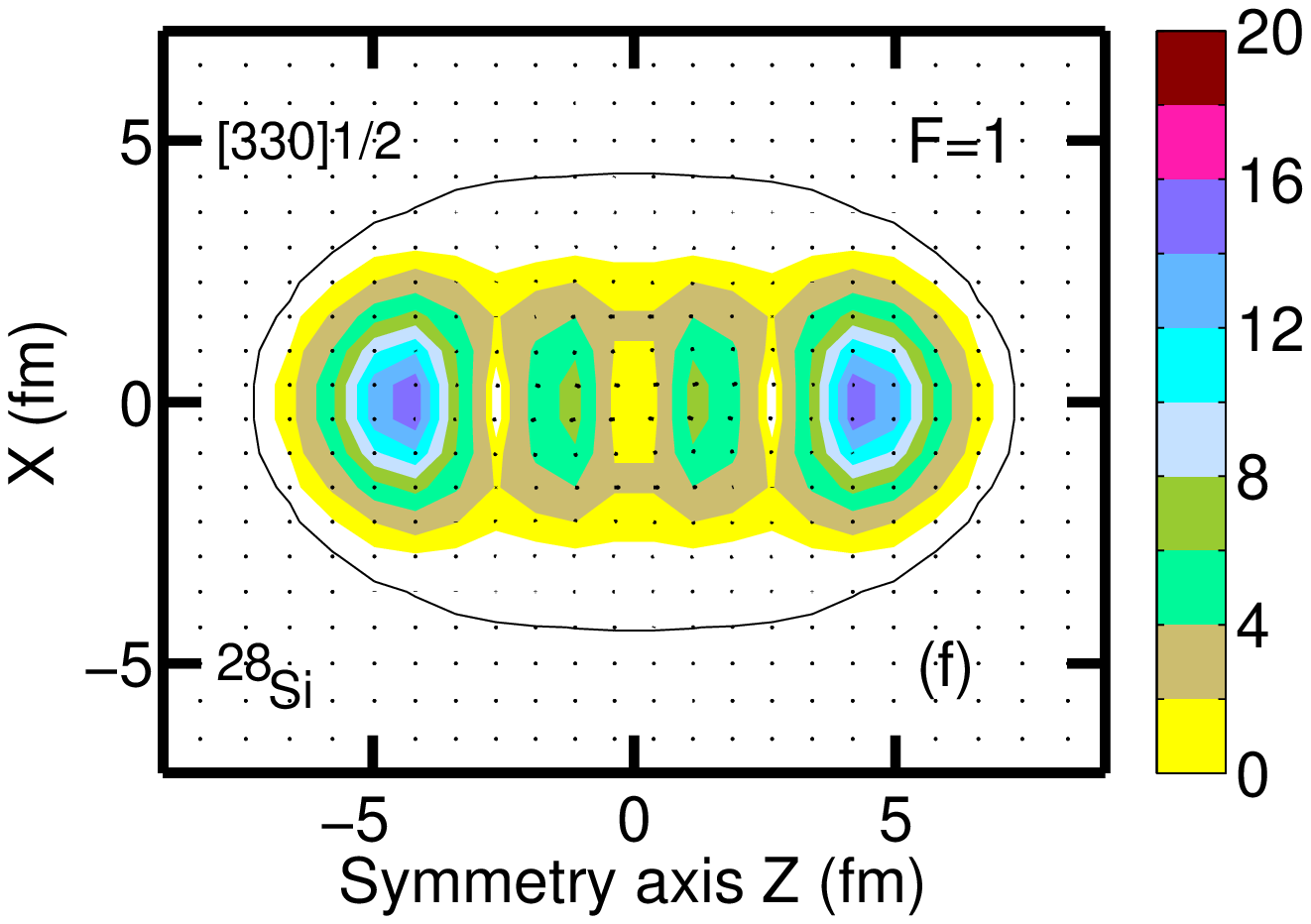}
\caption{(Color online) The same as Fig.\ \ref{C12-NL3*} but for the [330]1/2$(r=-i)$ 
orbital in the HD [2,2] configuration of $^{28}$Si. The densities in the $yz$ and 
$xz$ planes are taken at $x=0.326$ fm and $y=0.326$ fm, respectively. Top and 
bottom panels show the results at $\Omega_x=0.0$ MeV and $\Omega_x=1.8$ MeV, 
respectively.}
\label{Si28-33012}
\end{figure*}

 The asymptotic Nilsson labels are quite good approximate quantum numbers
at the extreme deformations of interest (see also the discussion in
Sec. 8.2 of Ref.\ \cite{NilRag-book}). For example, the lowest state 
in the routhian diagram has the structure 96.0\%$|000,1/2>$ + 3\%$|200,1/2>$ 
+ 0.5$\%|220,12>$ + ... at rotational frequency $\Omega_x=0.0$ MeV. Here and 
below we show only three largest squared components (in the format 
$c^2_{N'n'_z\Lambda'\Omega'}$\%$|N'n'_z\Lambda,'\Omega'>$) of the single-particle 
wave function.
The rotation only somewhat modifies the structure  of its wave function which 
at $\Omega_x=3.2$ MeV has the structure 92.5\%$|000,1/2>$ 
+ 3.0\%$|200,1/2>$ + 2.8\%$|220,1/2>$ + ... for the $r=-i$ branch.
The same is true for other states of interest. The lowest negative parity state 
has the structure 94.6\%$|110,1/2>$ + 4.0\%$|310,1/2>$ + 0.4\%$|101,1/2>$ + ... and 
87.1\%$|110,1/2>$ + 3.4\%$[310,12>$ + 3.1\%$|101,3/2>$ + ... (for the $r=-i$ branch) at 
$\Omega_x=0.0$ and $\Omega_x=3.2$ MeV, respectively. The lowest $N=2$ state has the 
structure 89.7\%$|220,1/2>$ + 5.8\%$|420,1/2>$ + 1.5\%$|211,1/2>$ + ... and 
72.3\%$|220,1/2>$ + 8.9\%$|211,1/2>$ + 6.6\%$|211,3/2>$ +... (for the $r=-i$ branch)
at $\Omega_x=0.0$ and $\Omega_x=3.2$ MeV, respectively. Two general trends in the 
structure of the wave function are clearly seen on these examples. These are the 
increase of the fragmentation of the wave function (with related decrease of the 
dominant component of the wave function) with the raise of the position of the 
single-particle state with respect of the lowest state in the nucleonic potential 
and with increasing rotation of the nucleus. The first effect brings the state of 
interest into the region of increased density of the single-particle states and thus 
to the region where the interactions of the states are more abundant. The second 
is a consequence of the Coriolis interaction.

\section{The hyperdeformed [2,2] configuration in $^{28}$Si} 
\label{28Si-sect}
 
  Next we consider the hyperdeformed [2,2] configuration in $^{28}$Si. 
The structure of this nucleus has been studied in detail in Ref.\ 
\cite{AR.16} where it was shown that this configuration is calculated 
at relatively low excitation energy at spins above $10\hbar$. Its total 
neutron density distribution is shown at spins $I=0\hbar$ and $I=12\hbar$ 
in Fig.\ \ref{Total-densities}(b-c). The HD [2,2] configuration shows 
clear signatures of clusterization which are especially pronounced at 
$I = 0\hbar$ (Fig.\ \ref{Total-densities}b). Although the rotation 
somewhat hinders these signatures (Fig.\ \ref{Total-densities}c), they 
are still present at $I=12\hbar$.

\begin{figure*}[htb]
\includegraphics[angle=0,width=5.5cm]{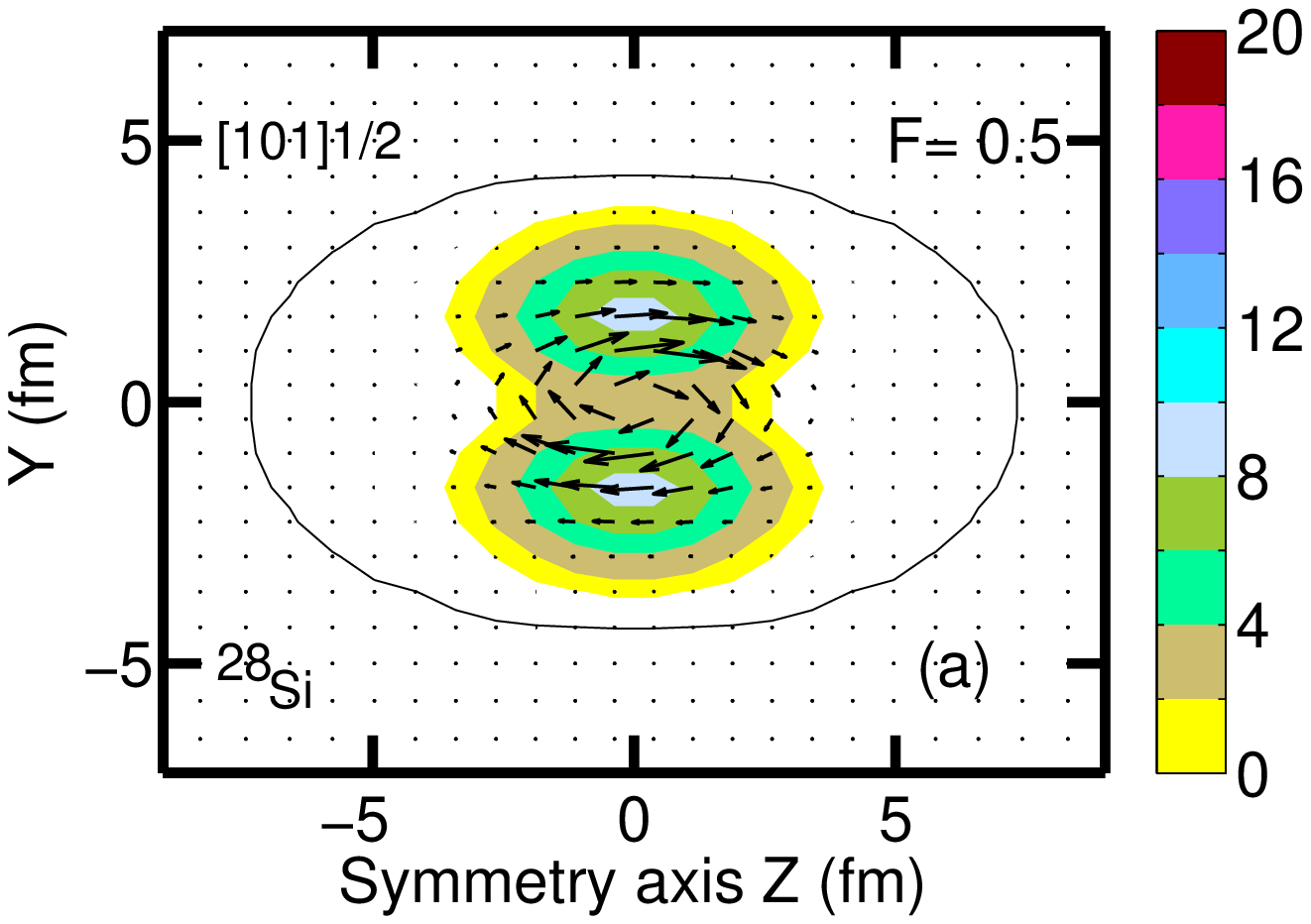}
\includegraphics[angle=0,width=5.5cm]{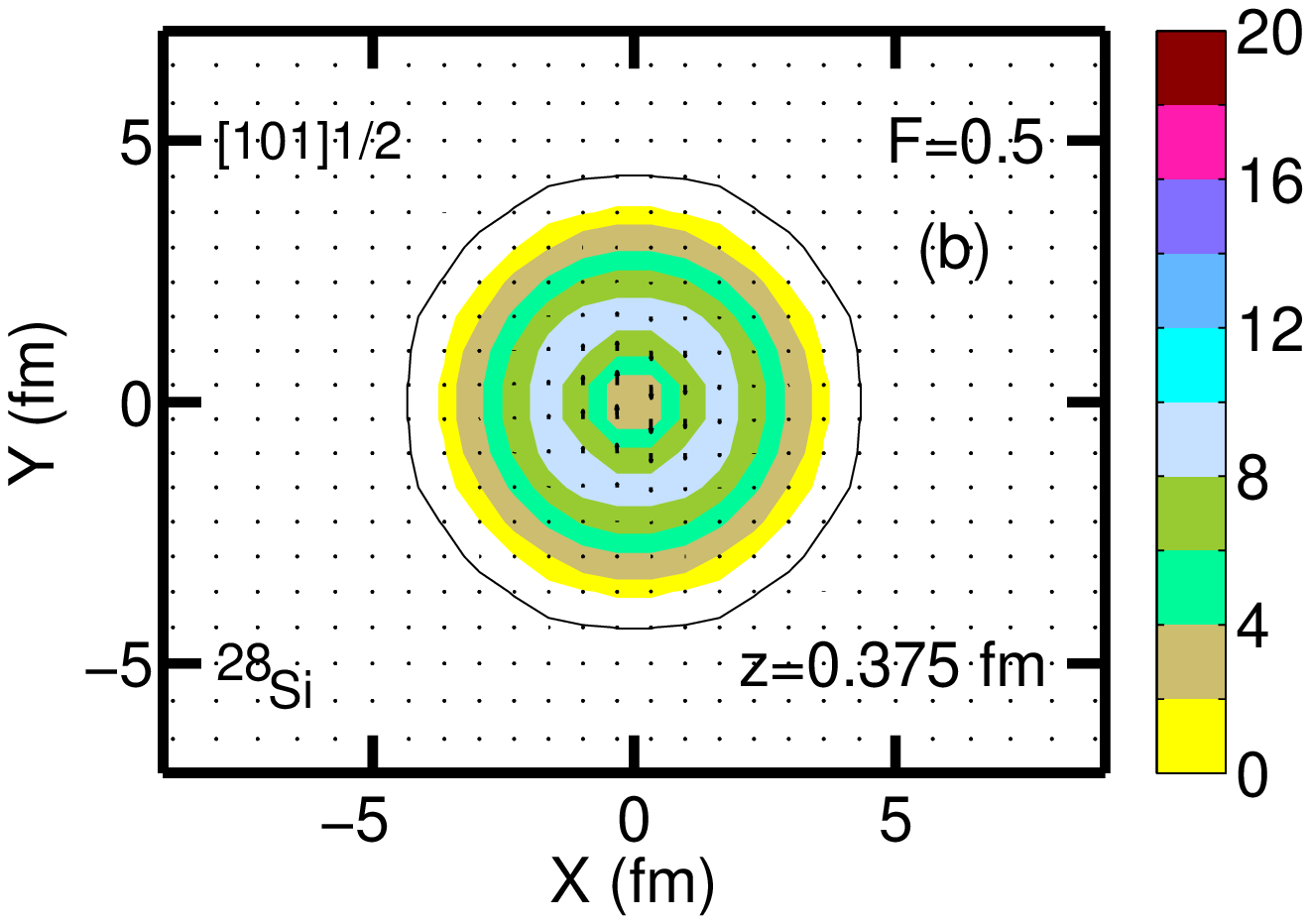}
\includegraphics[angle=0,width=5.5cm]{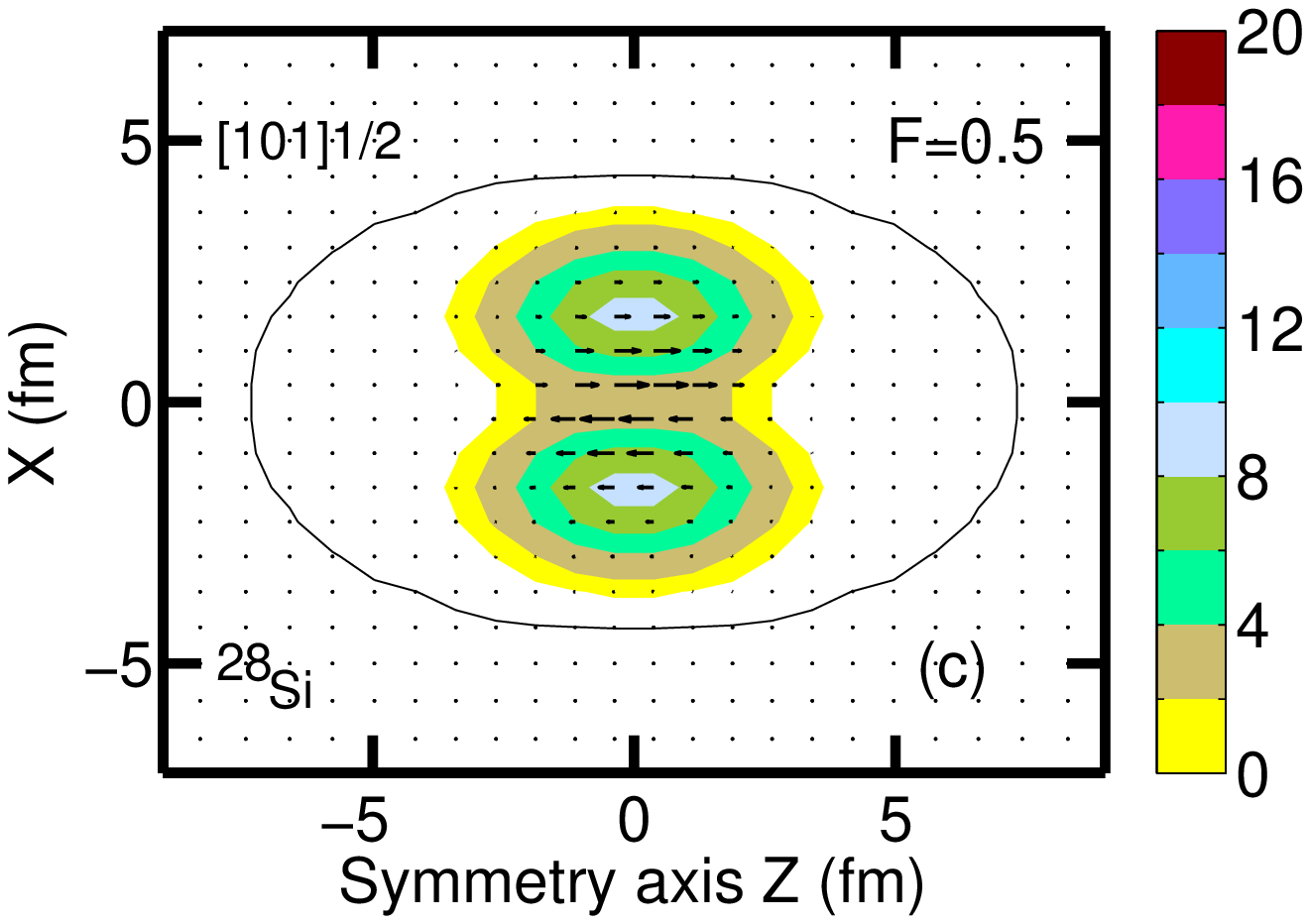}
\includegraphics[angle=0,width=5.5cm]{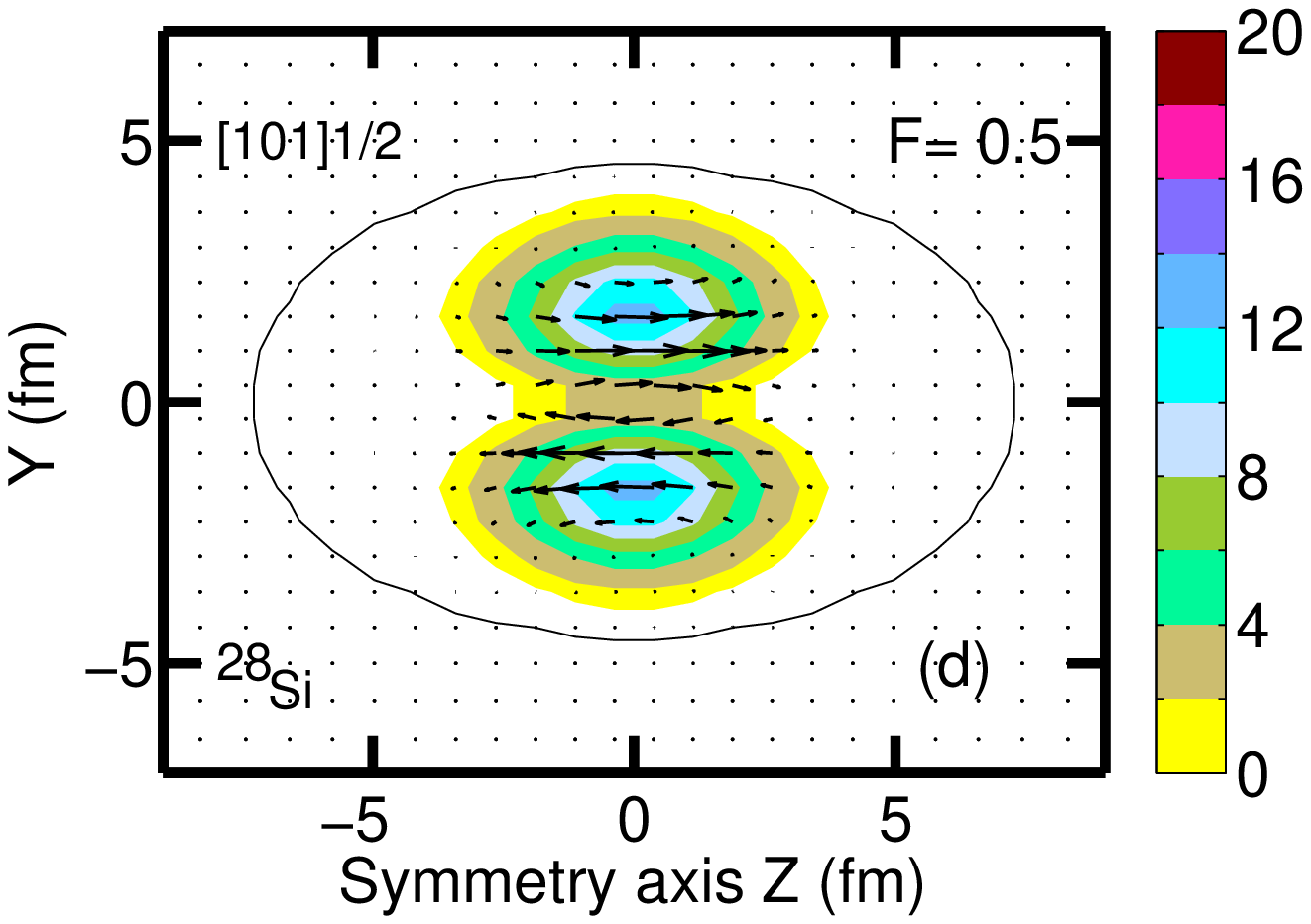}
\includegraphics[angle=0,width=5.5cm]{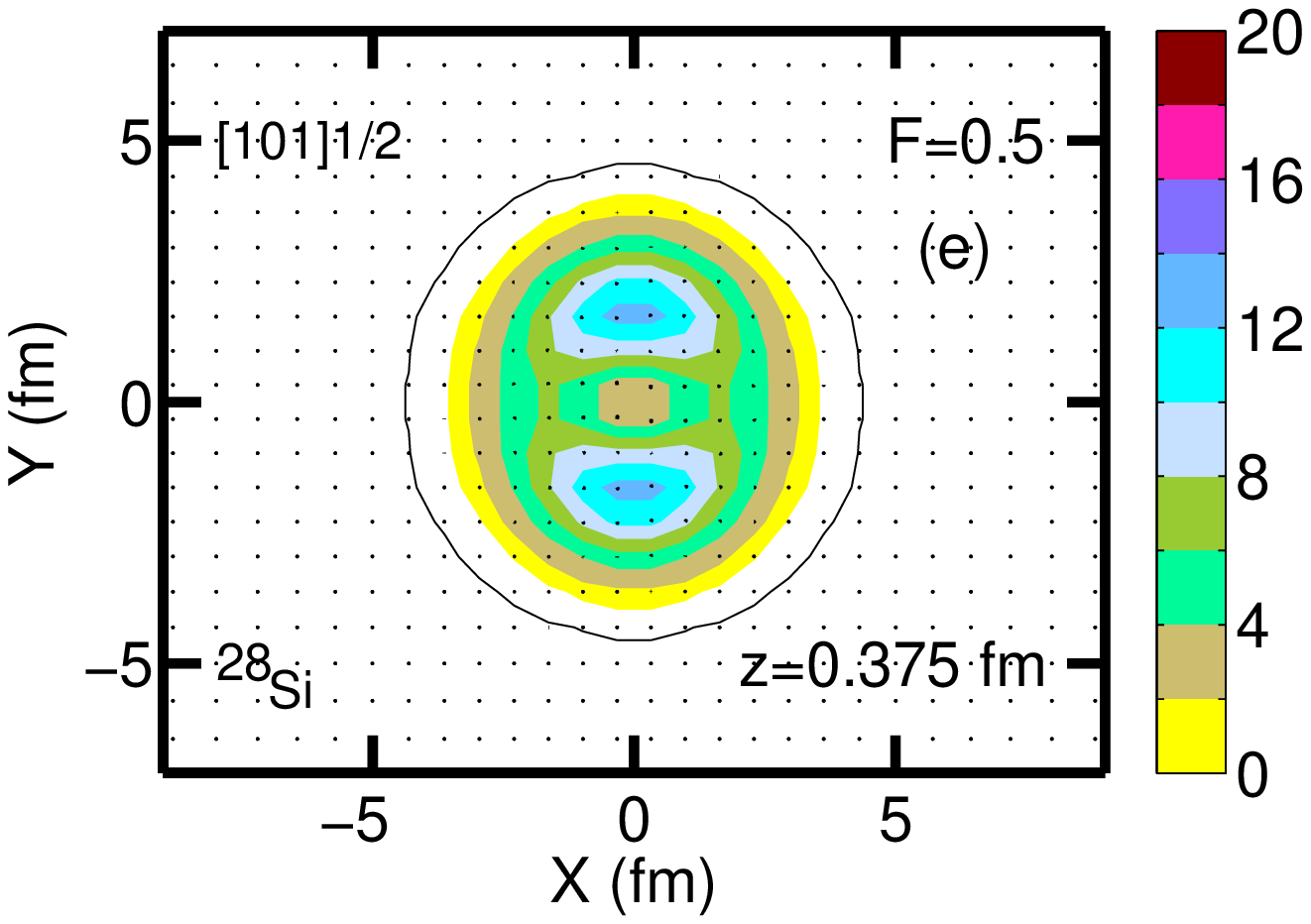}
\includegraphics[angle=0,width=5.5cm]{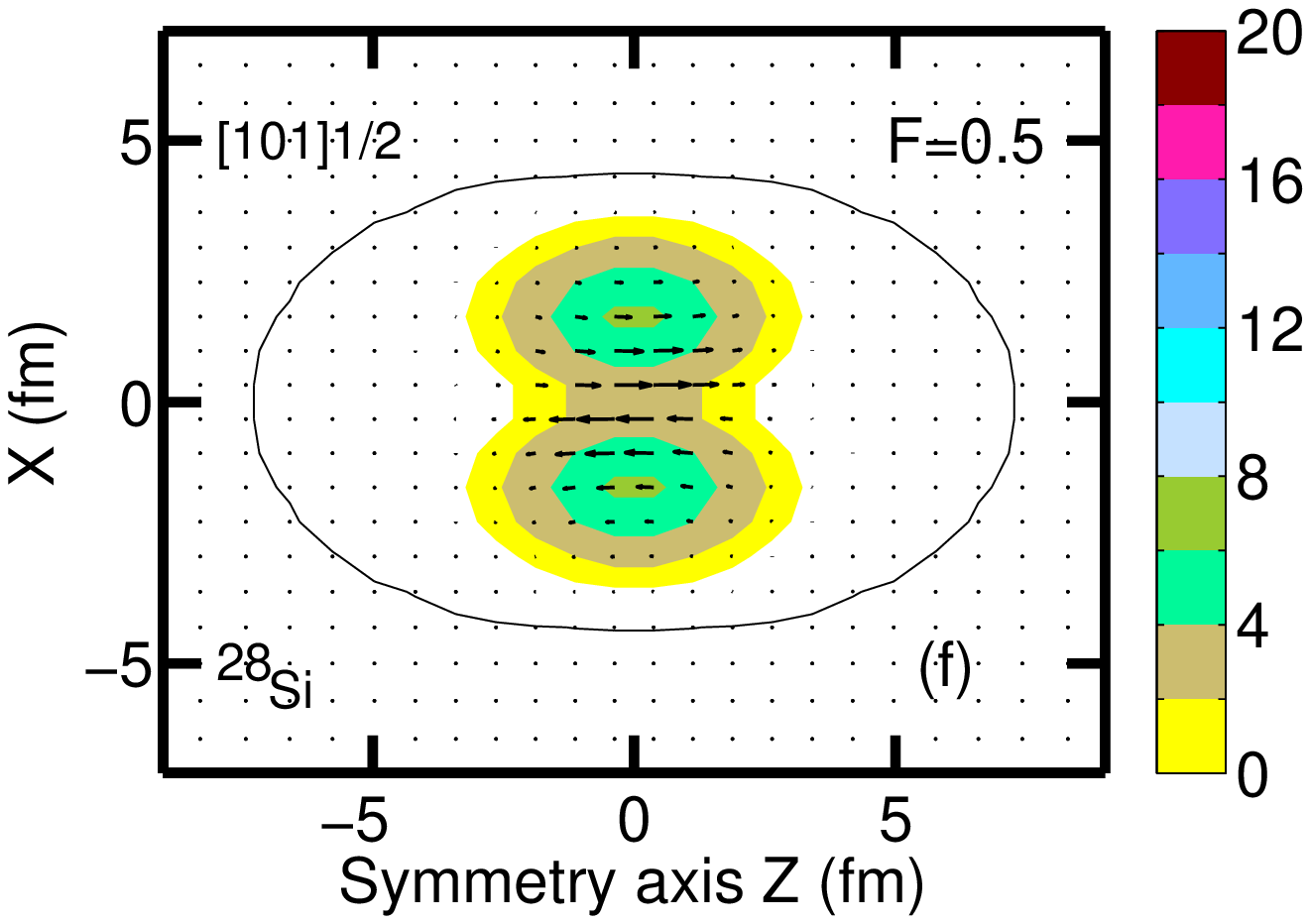}
\includegraphics[angle=0,width=5.5cm]{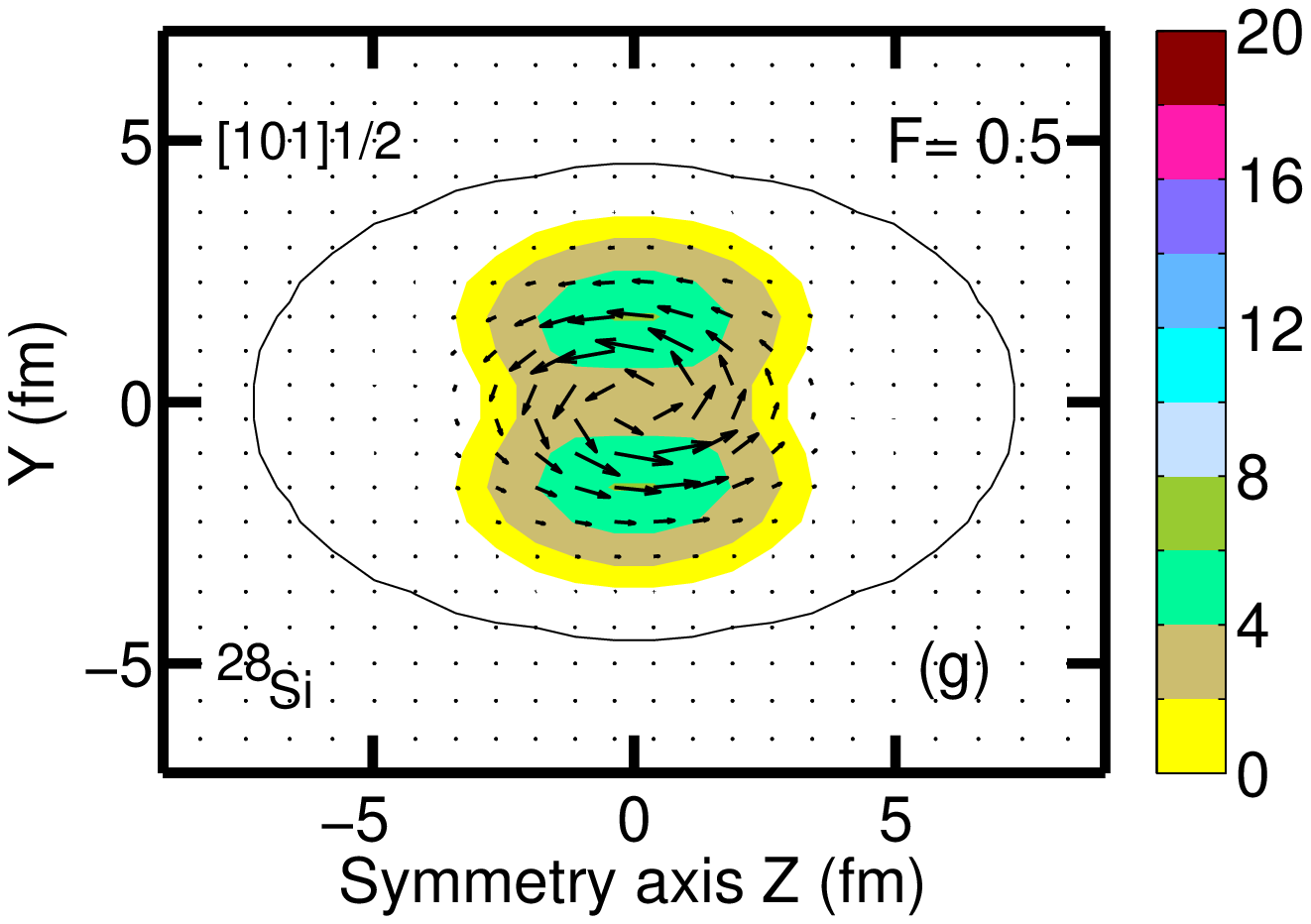}
\includegraphics[angle=0,width=5.5cm]{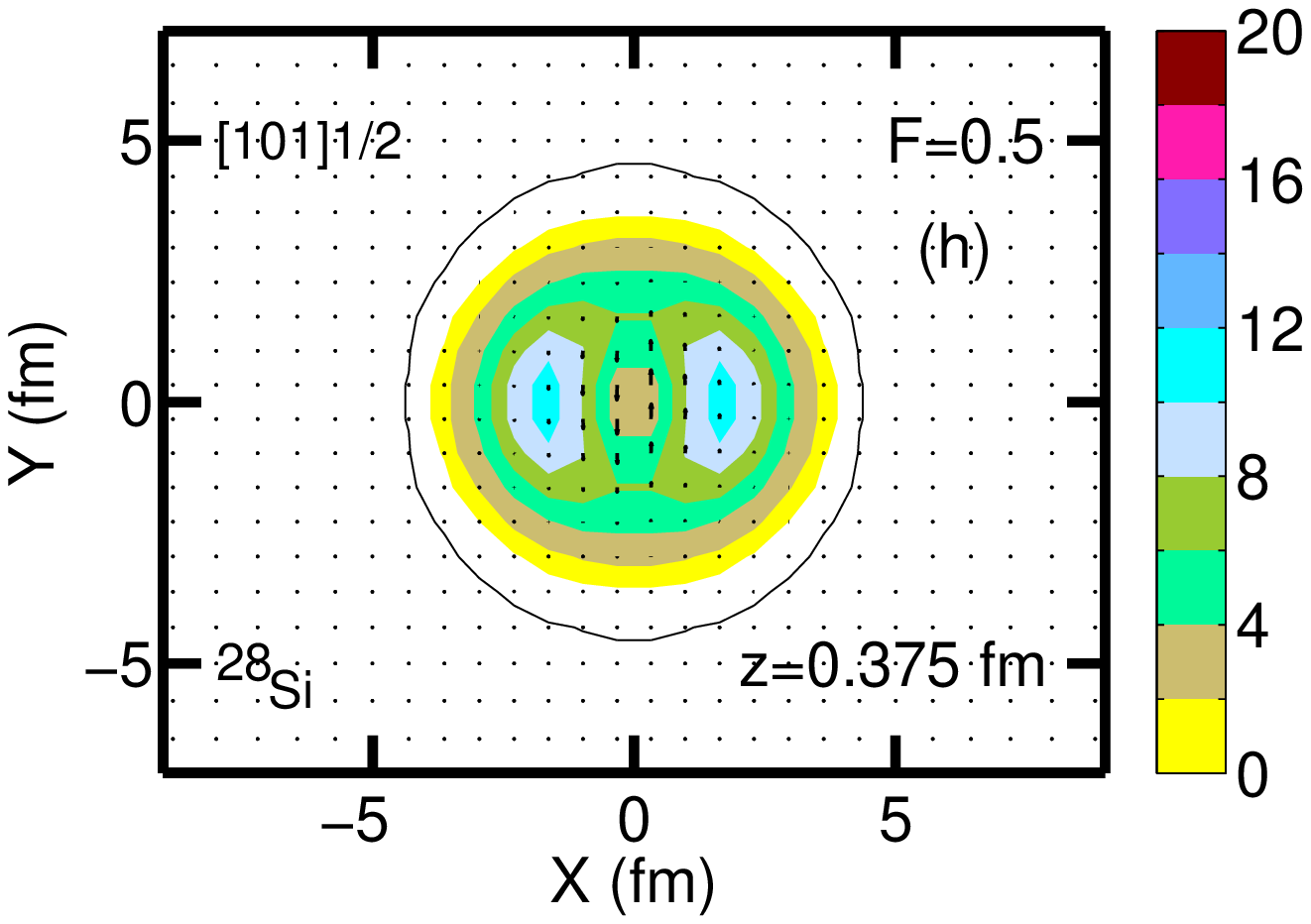}
\includegraphics[angle=0,width=5.5cm]{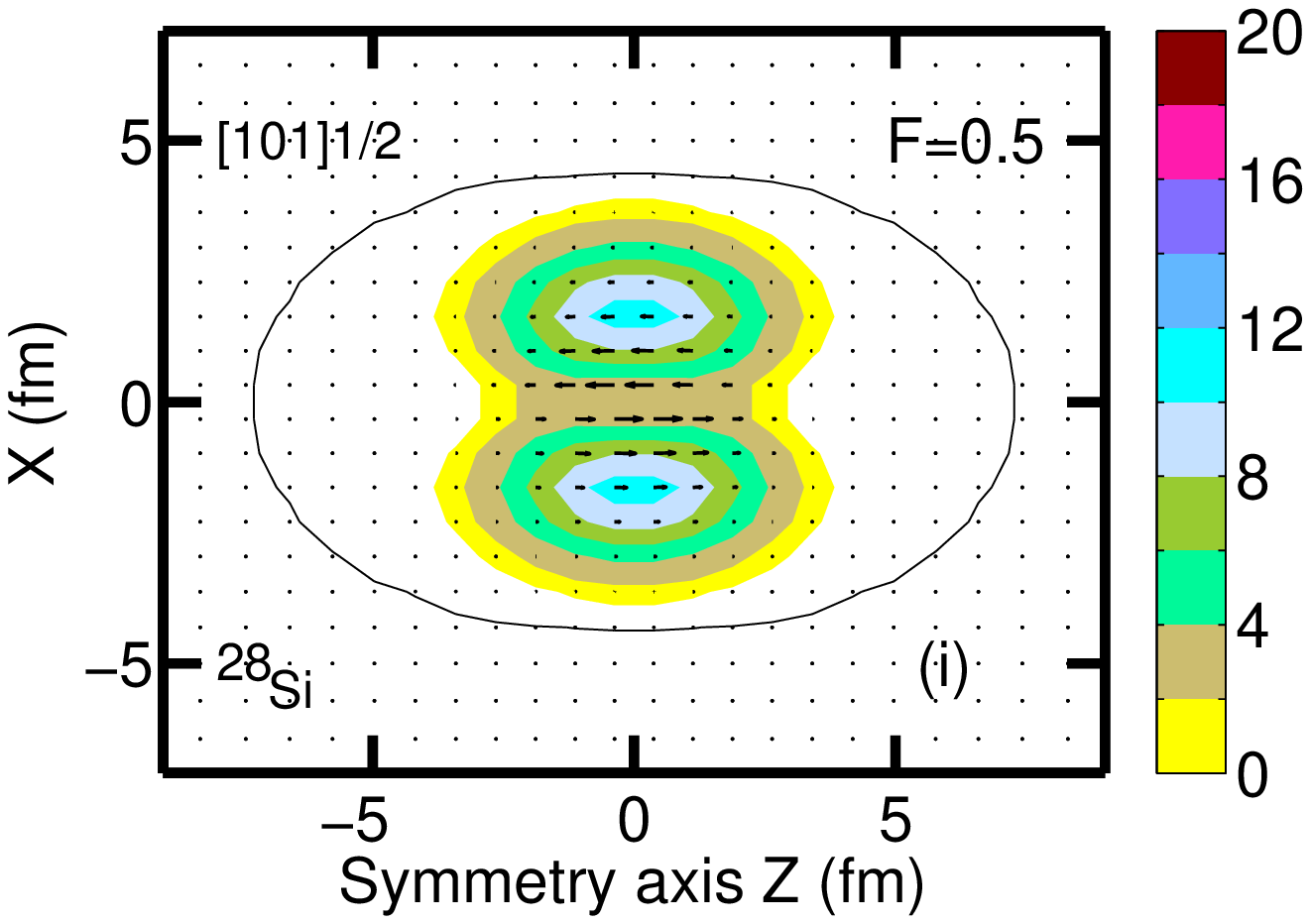}
\caption{(Color online) The same as Fig.\ \ref{Si28-33012} but for the 
[101]1/2$(r=\pm i)$ orbitals. The top panels show the results for the $(r=-i)$ 
orbital at $\Omega_x=0.0$ MeV. Because the density distributions in both 
signatures of the [101]3/2 orbital are identical at $\Omega_x=0.0$ MeV, the 
results for the [101]3/2$(r=+i)$ orbital are not shown here. Middle and bottom 
panels show the densities and currents at $\Omega_x=1.8$ MeV for the [
101]1/2$(r=-i)$ and  [101]1/2$(r=+i)$ orbitals, respectively. 
}
\label{Si28-10112}
\end{figure*}

 The length of the rod-shape structure in $^{12}$C and the HD [2,2]
configuration in $^{28}$Si is almost the same in axial ($z$-) direction.
However, the density  of the HD [2,2] configuration is broader in the
radial direction as compared with the one seen in rod-shape structure
of $^{12}$C. As follows from the discussion below, these differences 
are traced back mainly to the density distributions of the orbitals by 
which these two configurations differ.

   The [000]1/2, [110]1/2 and [220]1/2 states are occupied both in the
HD configuration of $^{28}$Si and rod-shape structure of $^{12}$C. Their 
density  distributions and nodal structure are very similar in both 
nuclei (compare Fig.\ \ref{C12-NL3*} with Figs. 1, 2 and 3 in the Supplemental 
Material \cite{Suppl-node}). However, the density distributions of these 
states in the HD [2,2] 
configuration of $^{28}$Si are slightly less elongated in axial direction 
and somewhat more stretched out in radial direction as compared with the ones 
in rod-shape structure of $^{12}$C. This is a consequence of two facts. 
First, the difference in total density distributions (rod-shape in $^{12}$C 
versus ellipsoid like in $^{28}$Si, Figs. \ref{Total-densities}(a-c)) 
affects the nucleonic potential. The sizes of the nuclei and thus of nucleonic 
potentials have also an impact. The charge radii of $^{12}$C and $^{28}$Si
in the ground state are $\sim 2.8$ fm and $\sim 3.15$ fm, respectively (see 
Fig.\ 23 in Ref.\ \cite{AARR.14}). However, these differences are compensated 
to a degree by the fact that these states are located deeper in the nucleonic 
potential of $^{28}$Si (at single-particle energies $\varepsilon_i =-50.22, -40.20$ 
and -26.56 MeV at $\Omega_x=0.0$ MeV for the [000]1/2, [110]1/2 and [220]1/2 
states, respectively) as compared with the ones in $^{12}$C (Fig.\ \ref{Fig-routh}a) 
which leads to their smaller effective radius as compared with the radius of 
the ground state.

\begin{figure*}[htb]
\includegraphics[angle=0,width=5.5cm]{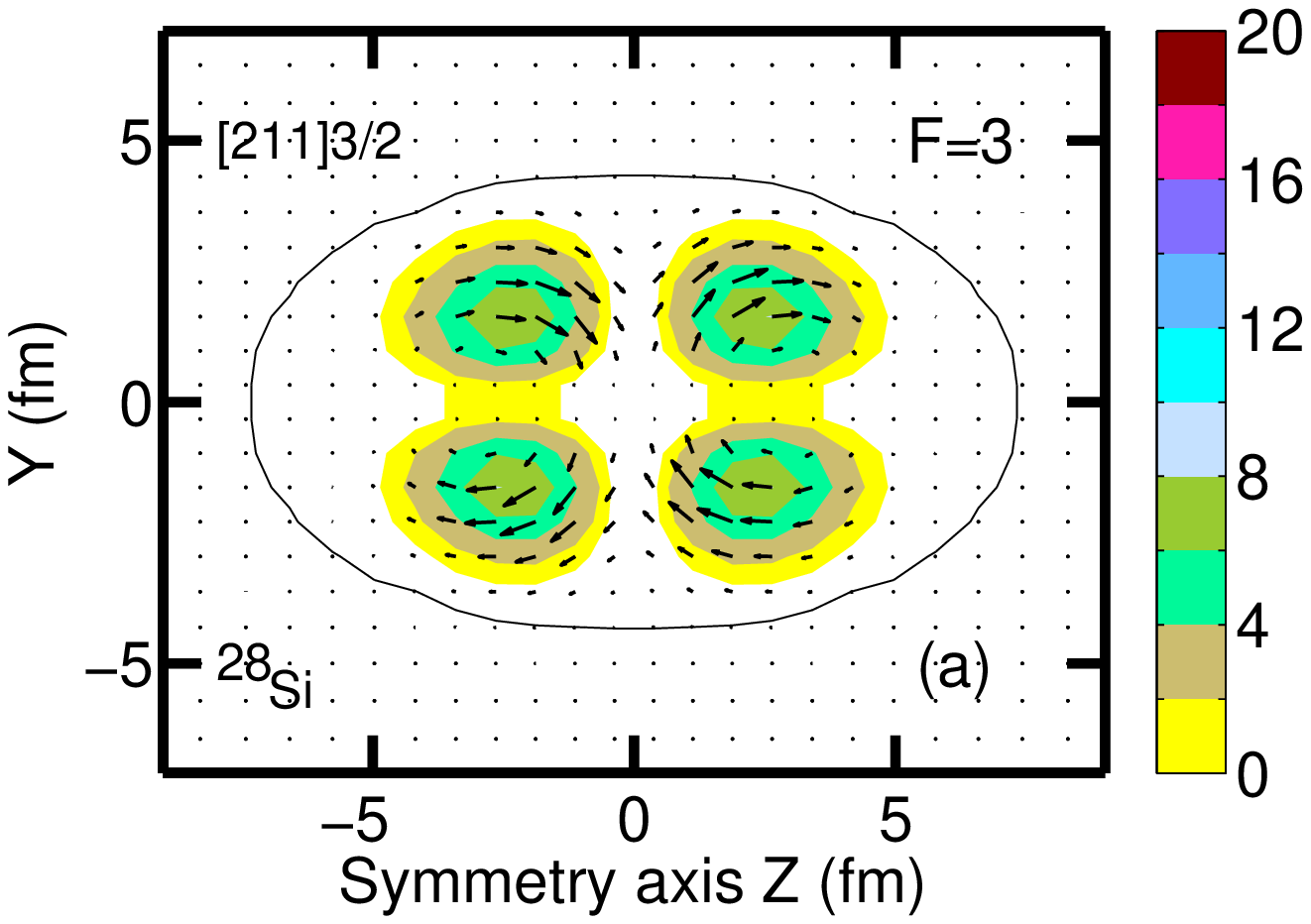}
\includegraphics[angle=0,width=5.5cm]{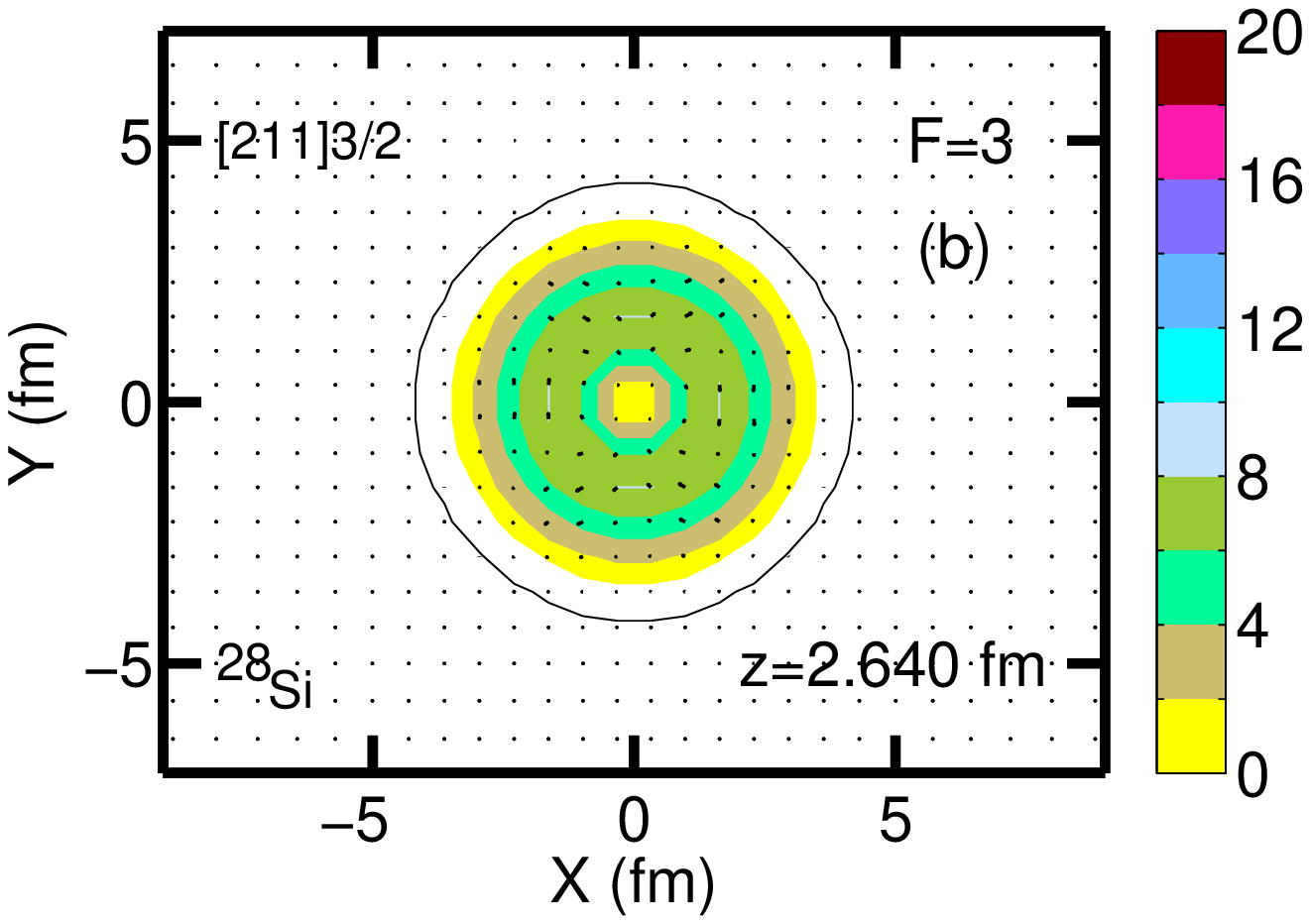}
\includegraphics[angle=0,width=5.5cm]{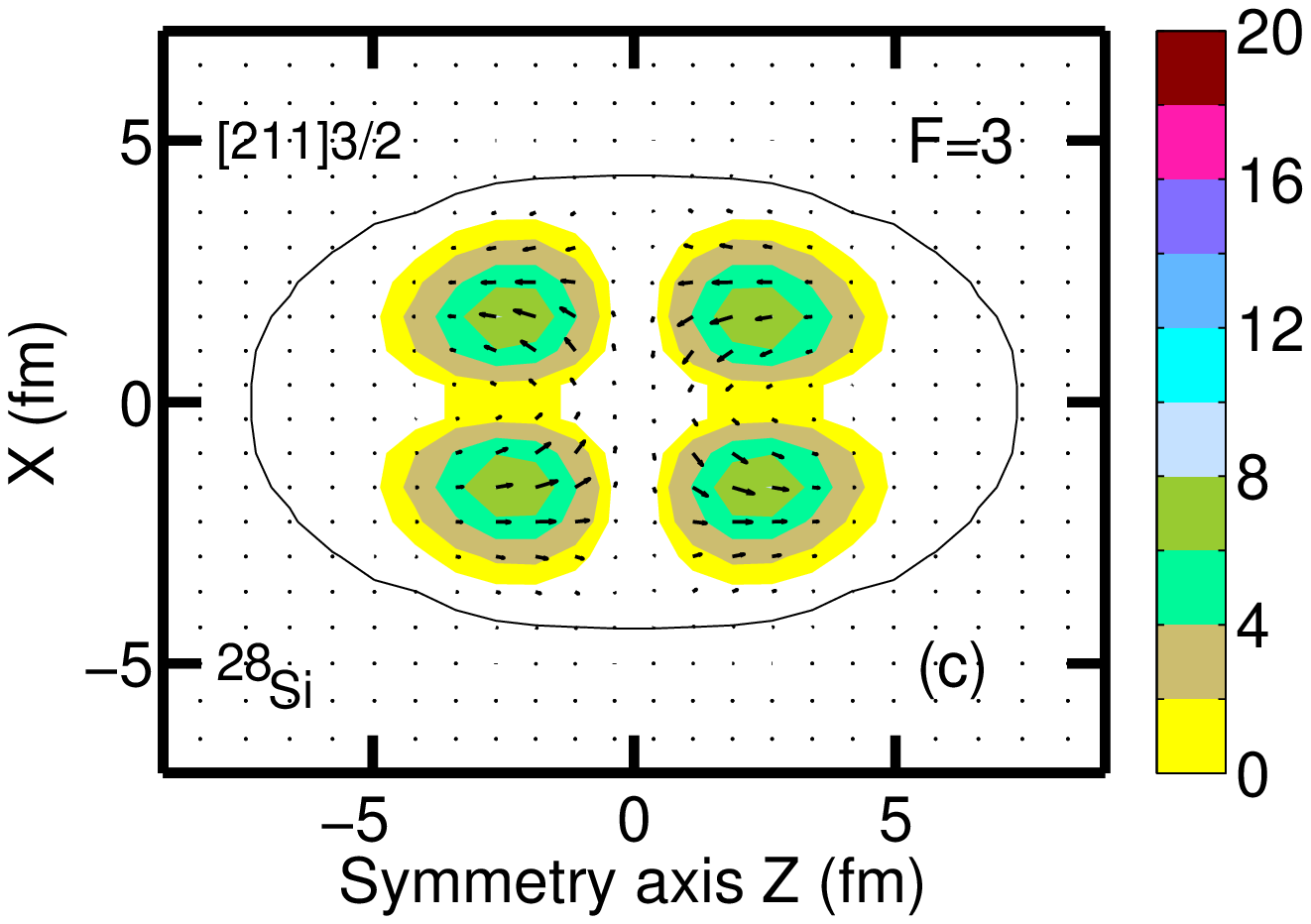}
\includegraphics[angle=0,width=5.5cm]{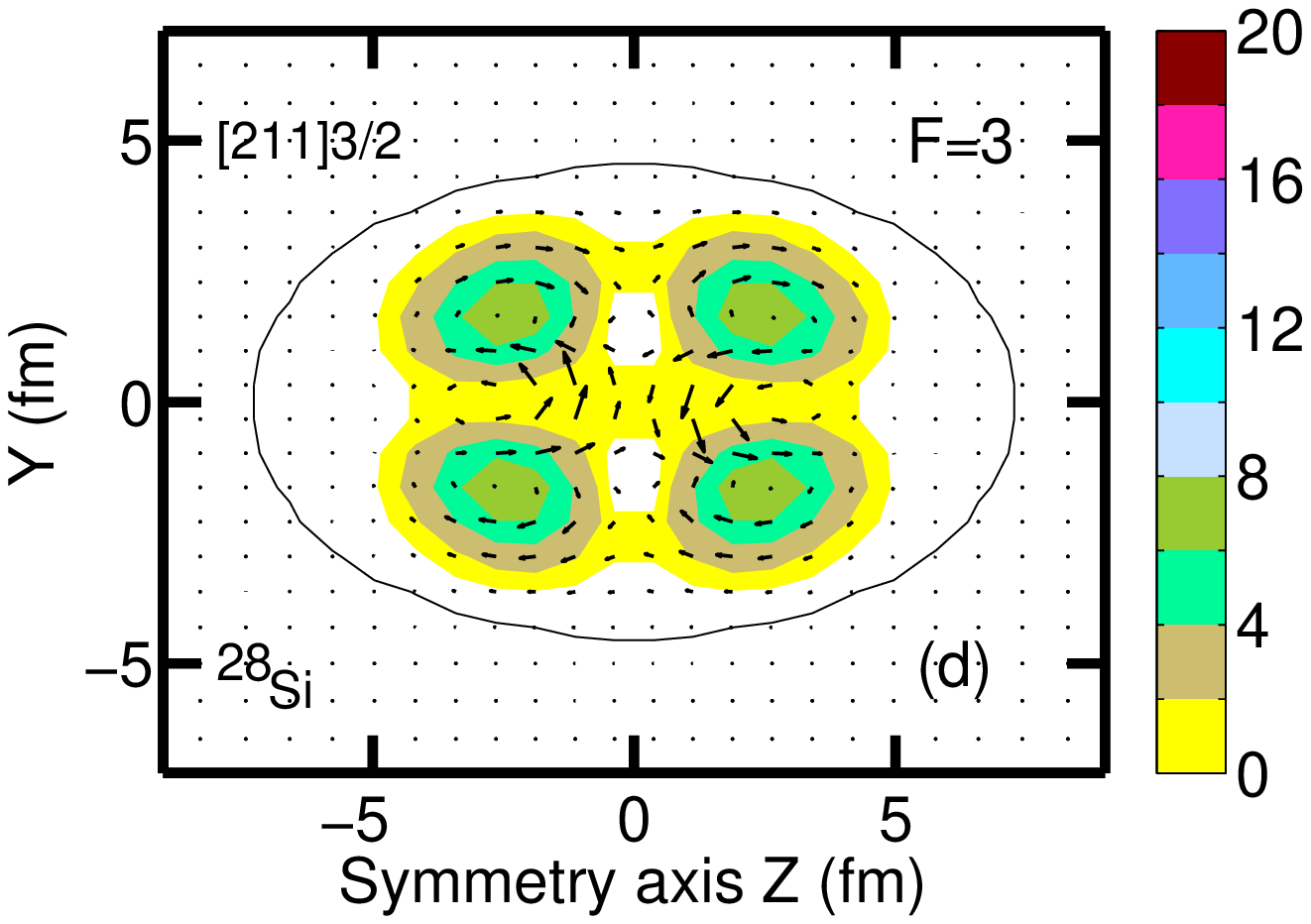}
\includegraphics[angle=0,width=5.5cm]{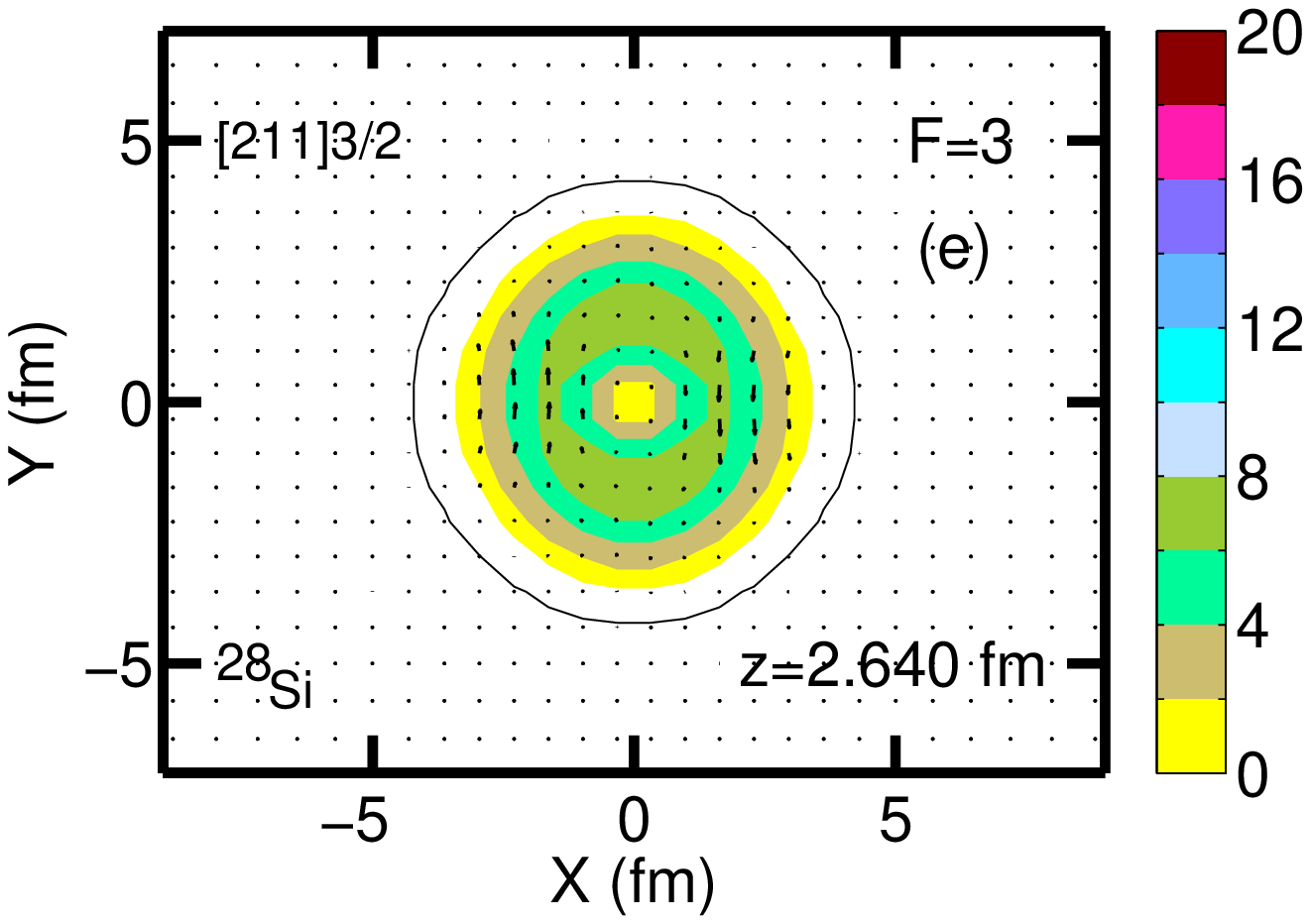}
\includegraphics[angle=0,width=5.5cm]{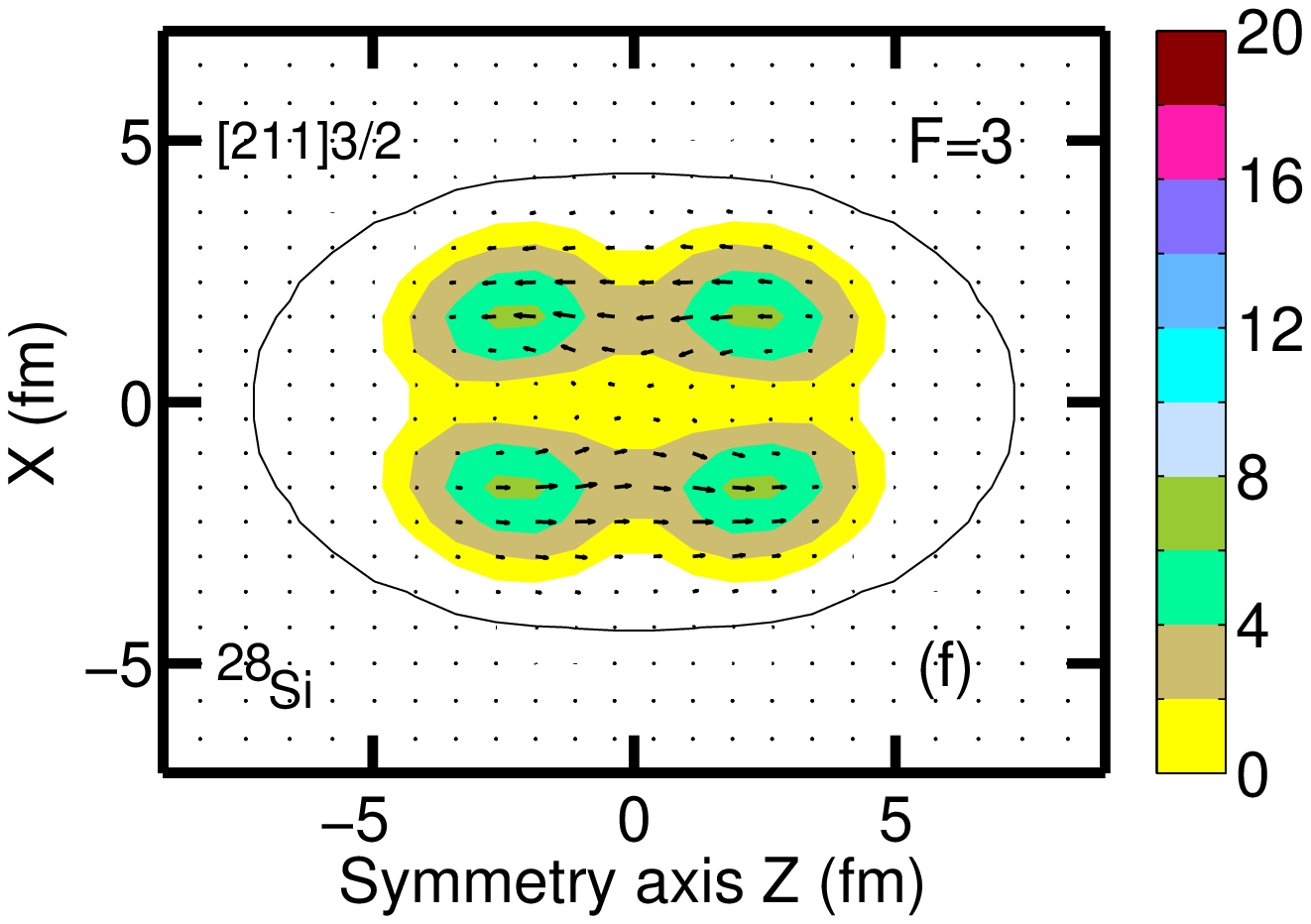}
\caption{(Color online) The same as Fig.\ \ref{Si28-33012} but for the 
[211]3/2$(r=-i)$ orbital.}
\label{Si28-21132}
\end{figure*}

  In addition,  the Nilsson [101]3/2, [330]1/2, [211]3/2 and [101]1/2 orbitals 
are occupied in this configuration (see Fig.\ \ref{Fig-routh}b). Their density 
distributions are presented in Figs. \ref{Si28-33012}, \ref{Si28-10112} and 
\ref{Si28-21132} as well as in Fig. 4 of the Supplemental Material
(Ref.\ \cite{Suppl-node}) at no rotation 
($\Omega_x=0.0$ MeV) and at rotational frequency $\Omega_x=1.8$ corresponding 
to spin $I\sim 12 \hbar$. The density distributions of these single-particle 
orbitals are characterized by different nodal structure. 

The density distribution of the [330]1/2 orbital is similar in structure to 
the one seen for the $[NN0]1/2$ orbitals in $^{12}$C and $^{28}$Si: there are 
four density clusters located along the axis of symmetry with the maximum of 
the density in each cluster located at the axis of symmetry.  The largest 
clusters with the highest density in the center are located in the polar 
region.  

 The orbitals with the Nilsson labels [101]3/2, [101]1/2 and [211]3/2 belong
to the group of the states the wavefunctions of which are dominated by the basis 
states of the $|N,N-1,1>$ type (see Table \ref{table-wf}). These basis states 
produce zero density at the axis of symmetry (see Sec.\ \ref{nodal-wf}). 
Significant reduction of the density on approaching the axis of symmetry is 
seen in the density distributions of these Nilsson states (see Figs.\ \ref{Si28-10112} 
and \ref{Si28-21132} in the manuscript and Fig. 4 in the Supplemental Material
\cite{Suppl-node}). However, 
not always we see zero density at or close to the axis of symmetry. This is due to 
two reasons. First, there are the contributions into the wave functions emerging 
from the basis states of the $|NN0>$ type (see Table \ref{table-wf}) which build 
the density at the axis of symmetry. Second, because of calculational features 
the plots are made at the cross-sections which are located slightly off the axis 
of symmetry. 

 The [101]1/2 (Figs.\ \ref{Si28-10112}(a-c)) and [101]3/2 (Figs. 4(a-c) in the Supplemental 
Material \cite{Suppl-node}) Nilsson states show very similar density distributions 
of doughnut type in 
which the maximum of the density is located in the equatorial plane. These two states 
at spin zero differ only in the orientation of the single-particle spin along the symmetry 
axis which has only moderate impact on the density distribution. As a result, their 
densities are similar at spin zero (compare Figs.\ \ref{Si28-10112}(a-c) with Figs. 4(a-c) 
in the Supplemental Material \cite{Suppl-node}); minor 
differences are due to different single-particle 
energies and different projections $\Omega$ of the total single-particle  angular 
momentum on the axis of  symmetry  which leads to the interaction of the single-particle 
states within the groups with different $\Omega$ (see Table \ref{table-wf}).

  The wave function of the [211]3/2 Nilsson state is dominated by the $|211,3/2>$ 
basis state the density of which has one node in radial direction and two nodes 
in axial direction. As a result, the density distribution of this Nilsson state
is the combination of two circular axially symmetric density rings located 
symmetrically with respect of equatorial plane (Figs.\ \ref{Si28-21132}(a-c)).

\begin{figure*}[htb]
\includegraphics[angle=0,width=5.5cm]{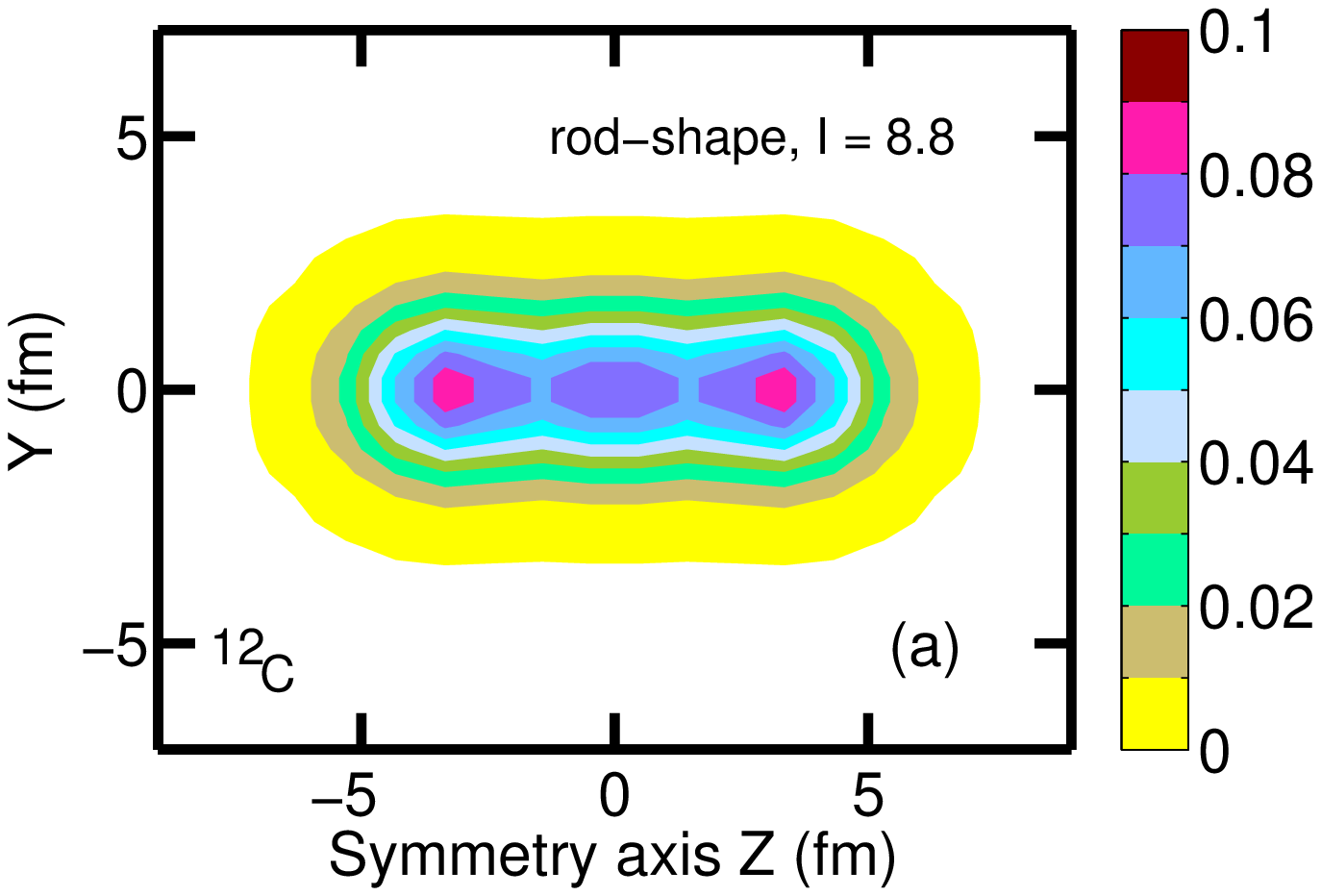}
\includegraphics[angle=0,width=5.5cm]{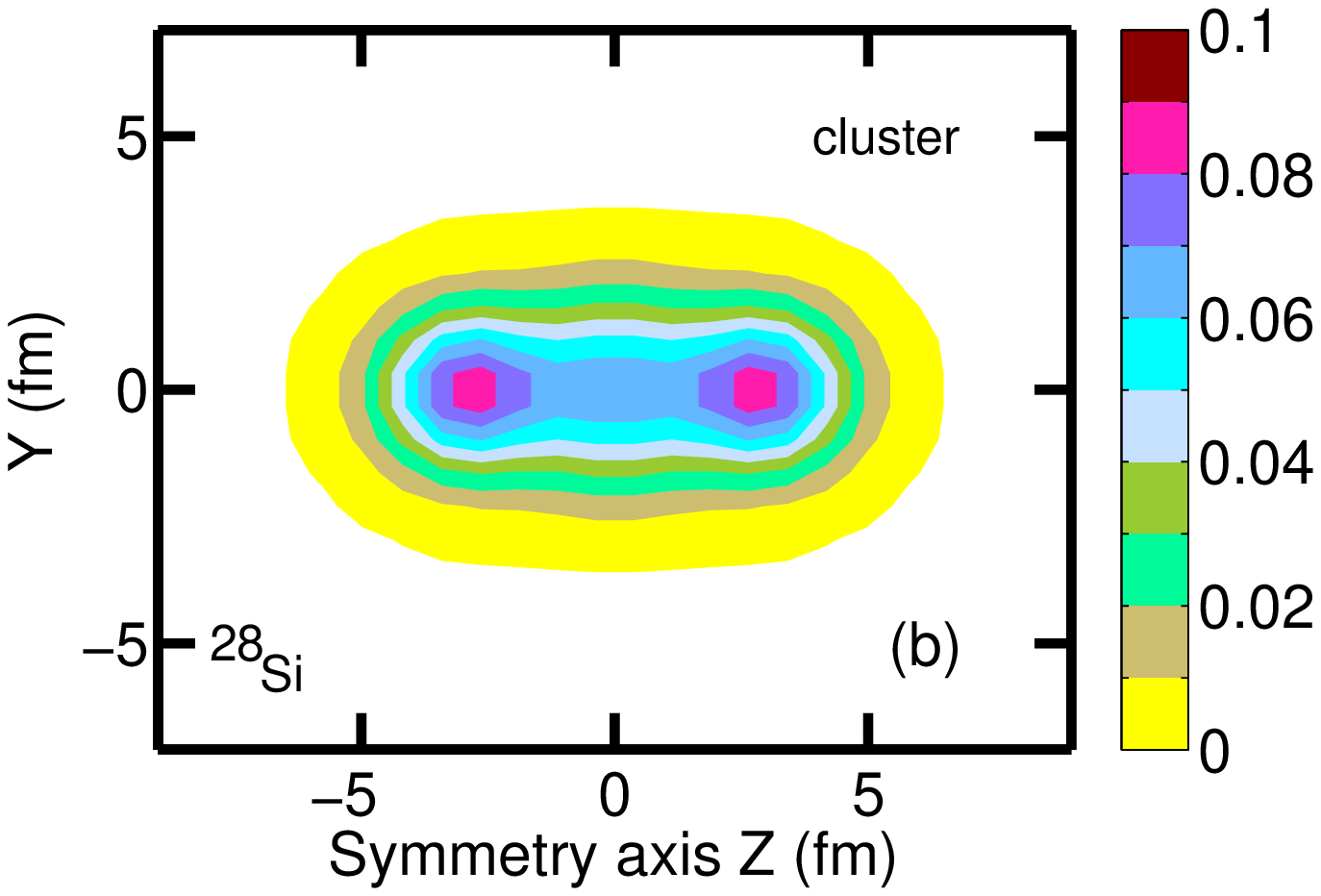}
\includegraphics[angle=0,width=5.5cm]{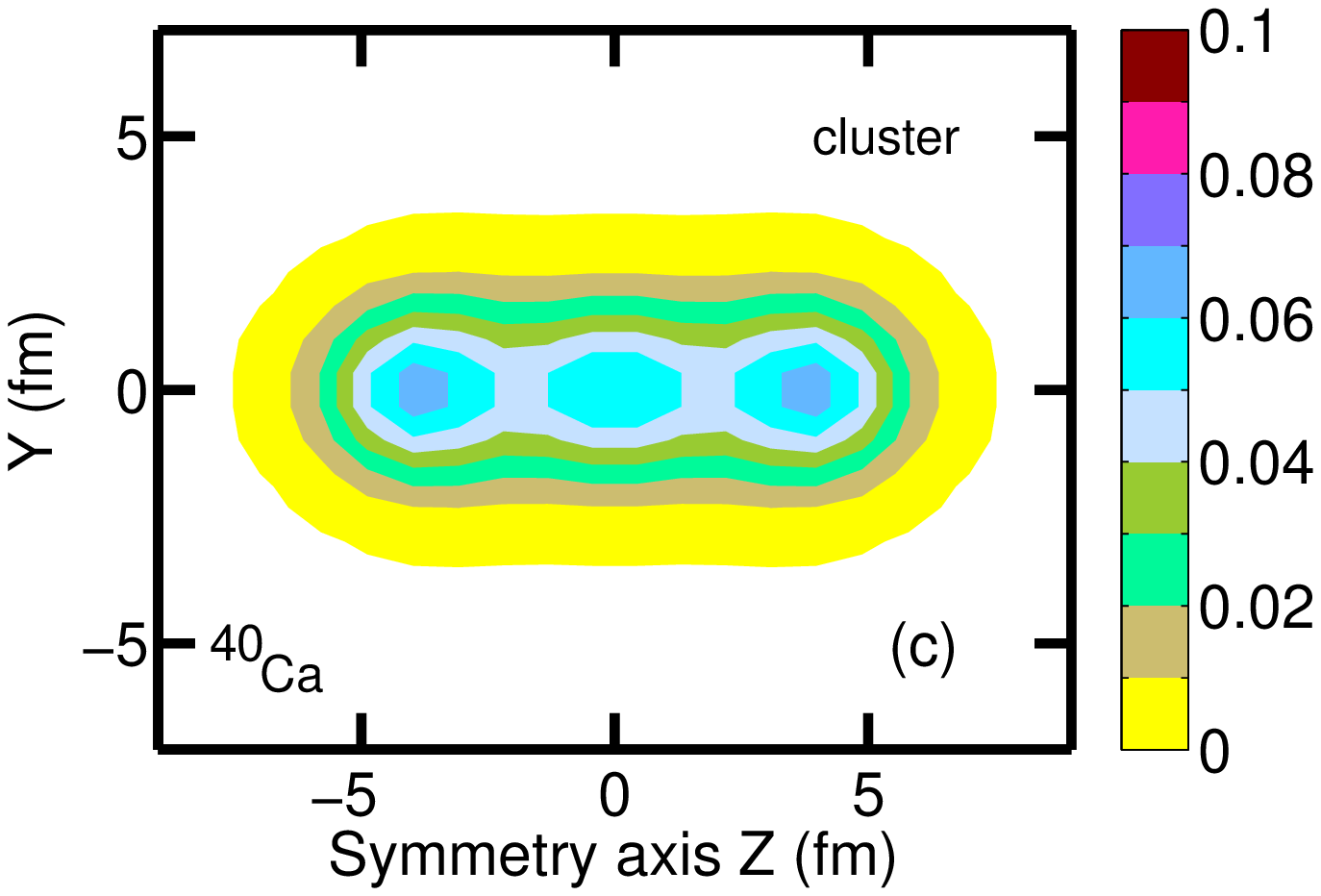}
\caption{(Color online)  Total neutron densities produced by
six neutrons occupying the lowest orbitals of the nucleonic
potential, namely, [000]1/2, [110]1/2 and [220]1/2 in considered
configurations of $^{12}$C, $^{28}$Si and $^{40}$Ca. The plotting
of the densities starts with yellow color at $0.001$ fm$^{-3}$.}
\label{Total-densities-rod}
\end{figure*}

 As illustrated in Table \ref{table-wf} the rotation leads to the
modification of the structure of the wave function typically reducing 
the weight of its dominant component. Its impact 
depends on the state. For some states these modifications are rather 
small, for others they may be substantial. 
In general, the impact of the rotation on the single-particle
densities can be characterized (i) by the change of their nodal 
structure, (ii) by the degree of delocalization of the
wave function and (iii) by the change of their azimuthal dependence.
For the states in the HD [2,2] configuration of $^{28}$Si, the nodal
structure of the single-particle density distributions is not 
affected by rotation. The rotation leads to some delocalization of
the wave function which reflects itself in some increase of the
space of the nucleus occupied by the particle [with related
decrease of average density] and by some decrease
of the maximum density in the density cluster (compare Figs.\ 
\ref{Si28-33012}(a-c) and Figs.\ \ref{Si28-33012}(d-f), compare upper, 
middle and bottom rows of Fig.\ \ref{Si28-10112} and of Fig. 4 in the
Supplemental Material \cite{Suppl-node}, compare upper and bottom rows 
in Fig.\ \ref{Si28-21132} and in Figs. 1, 2 and 3 of the Supplemental 
Material \cite{Suppl-node}). The changes in the azimuthal distribution 
of the densities induced by rotation are rather small for the $[NN0]1/2$ 
Nilsson states (see Fig. \ref{Si28-33012} and Figs. 1, 2 and 3 in the 
Supplemental Material \cite{Suppl-node}). On the other hand, they are 
quite substantial for the [101]1/2 (Fig.\ \ref{Si28-10112}) and [101]3/2 
(Fig. 4 in the Supplemental Material \cite{Suppl-node}) states. This is 
a consequence of the fact that the rotation leads to a different 
redistribution of the 
neutron matter for the $r=\pm i$ branches of the single-particle 
orbital resulting in an asymmetric doughnut density distributions 
in which the density depends on azimuthal angle. For example, the 
matter is moved away from the $xz$ plane in the $\pm y$ directions 
for the $[101]1/2(r=-i)$ orbital [compare upper and middle panels 
of Fig.\ \ref{Si28-10112}]. For the $[101]1/2(r=+i)$, this redistribution 
proceeds from the $yz$ plane in the $\pm x$ direction [compare upper 
and bottom panels of Fig.\ \ref{Si28-10112}]. Similar effect is also 
seen for the $[101]3/2(r=\pm i)$ states (Fig. 4 in the Supplemental 
Material \cite{Suppl-node}), but here it is inverted for $(r=\pm i)$ 
signatures as compared with the case of the $[101]1/2(r=\pm i)$ states. It is 
interesting that for the [101]1/2 and [101]3/2  states the rotation 
leads to the increase of maximum density in density cluster.

  The observed features of the single-particle density distributions allow 
to understand in a simple way the transition from the rod-shape total 
neutron density in $^{12}$C to the ellipsoid-like density distribution 
with two-pronounced clusters in $^{28}$Si (Fig.\ \ref{Total-densities}). 
In $^{28}$Si, six neutrons in the [000]1/2, [110]1/2 and [220]1/2 orbitals 
build the density distribution which is quite similar (slightly shorter 
in axial direction and slightly wider in radial direction) to the one 
seen in the rod-shape structure of $^{12}$C (compare Fig.\ 
\ref{Total-densities-rod}(b) with Fig.\ \ref{Total-densities-rod}(a)).
Thus, the 'rod-shape' cluster structure of $^{12}$C (but with less pronounced 
central cluster) is still present in the HD [2,2] configuration of 
$^{28}$Si. The addition of two neutrons 
into the [330]1/2 orbital will lead to some increase of the elongation of this substructure.
However, the addition of six neutrons into the [101]1/2, [101]3/2 and [221]3/2 
orbitals will lead to build up of the density at the radial coordinate 
away from the axis of symmetry located not far away from the equatorial plane.
 While the orbitals of the $[NN0]1/2$ type 
show the clusterization of the wavefunction with substantial concentration of the 
density in at least one density cluster, the occupation of the [101]1/2, [101]3/2 and 
[221]3/2 orbitals with single radial node, leading to either doughnut or double axial
ring structures, acts against $\alpha$-clusterization. This is because they occupy 
substantial space of nucleus and are characterized by low density which even at its 
maximum is only slightly more than half of the maximum of the density seen in the 
density clusters produced by the orbitals of the $[NN0]1/2$ type. In addition, there 
is no a center of the density distribution in the [101]1/2, [101]3/2 and [221]3/2 orbitals 
which could be associated with $\alpha$-particle.

\section{Megadeformed [42,42] configuration in $^{40}$Ca.} 
\label{Sect-40Ca}

  Megadeformed [42,42] configuration in $^{40}$Ca becomes yrast in the 
CRMF calculations at spin $I=23\hbar$ \cite{RA.16}. It is more elongated 
with narrower neck than the HD [2,2] configuration in $^{28}$Si (compare 
Fig.\ \ref{Total-densities}(e) with Figs.\ \ref{Total-densities}(b,c)). 
Despite these differences, the nodal structure of the densities of the 
single-particle states occupied below the $N=14$ shell gap and their 
pattern of density distribution is the same for 
these two configurations in two nuclei (compare Figs.\ \ref{Si28-33012}, 
\ref{Si28-10112}, and \ref{Si28-21132} in the paper and Fig. 4 in the 
Supplemental Material (Ref.\ \cite{Suppl-node}) with Figs. 5 and 6 
in the Supplemental Material \cite{Suppl-node}). Thus, 
we focus in this Section on the [211]1/2, [321]3/2 and [440]1/2 states 
which are located above the $N=14$ shell gap and which are occupied in 
the MD [42,42] configuration of $^{40}$Ca (Fig.\ \ref{Fig-routh}).

  In analogy to the case of the $[101]1/2$ and $[101]3/2$ states (see 
discussion in Sec.\ \ref{28Si-sect}), the density distributions of the 
$[211]3/2$ and $[211]1/2$ states are very similar at no rotation (compare 
Fig.\ \ref{Si28-21132} with Fig.\ 7 in the Supplemental Material 
\cite{Suppl-node}). The rotation affects the wave functions of the 
$(r=\pm i)$ branches of the [211]1/2 state in different way (see Table 
\ref{table-wf}). The wave function of the $r=+i$ orbital is only weakly 
affected by rotation so apart of the modification of azimuthal dependence 
the nodal structure of its density distribution is the same as the one 
at no rotation (compare upper and bottom rows of Fig.\ 7 in the Supplemental 
Material \cite{Suppl-node}). On the contrary, at $\Omega_x = 1.8$ MeV the 
$r=-i$ orbital is strongly mixed with substantial admixture of the $|440,1/2>$ 
basis state (Table \ref{table-wf}). This leads to the emergence of the 
additional density cluster at axis of symmetry near $z\sim \pm 6$ fm 
(see middle row of Fig.\ 7 in the Supplemental Material \cite{Suppl-node}).

\begin{figure*}[htb]
\includegraphics[angle=0,width=5.5cm]{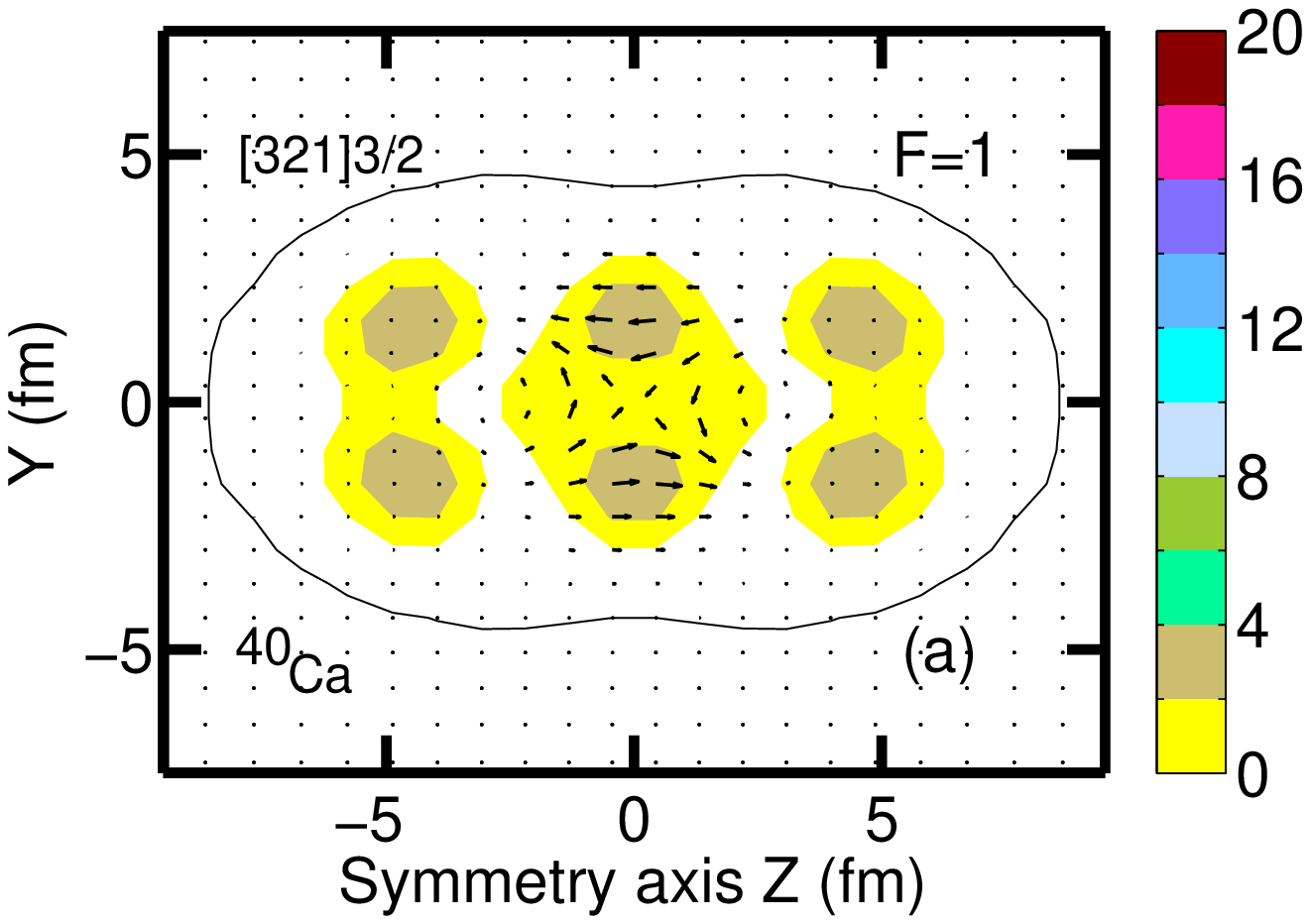}
\includegraphics[angle=0,width=5.5cm]{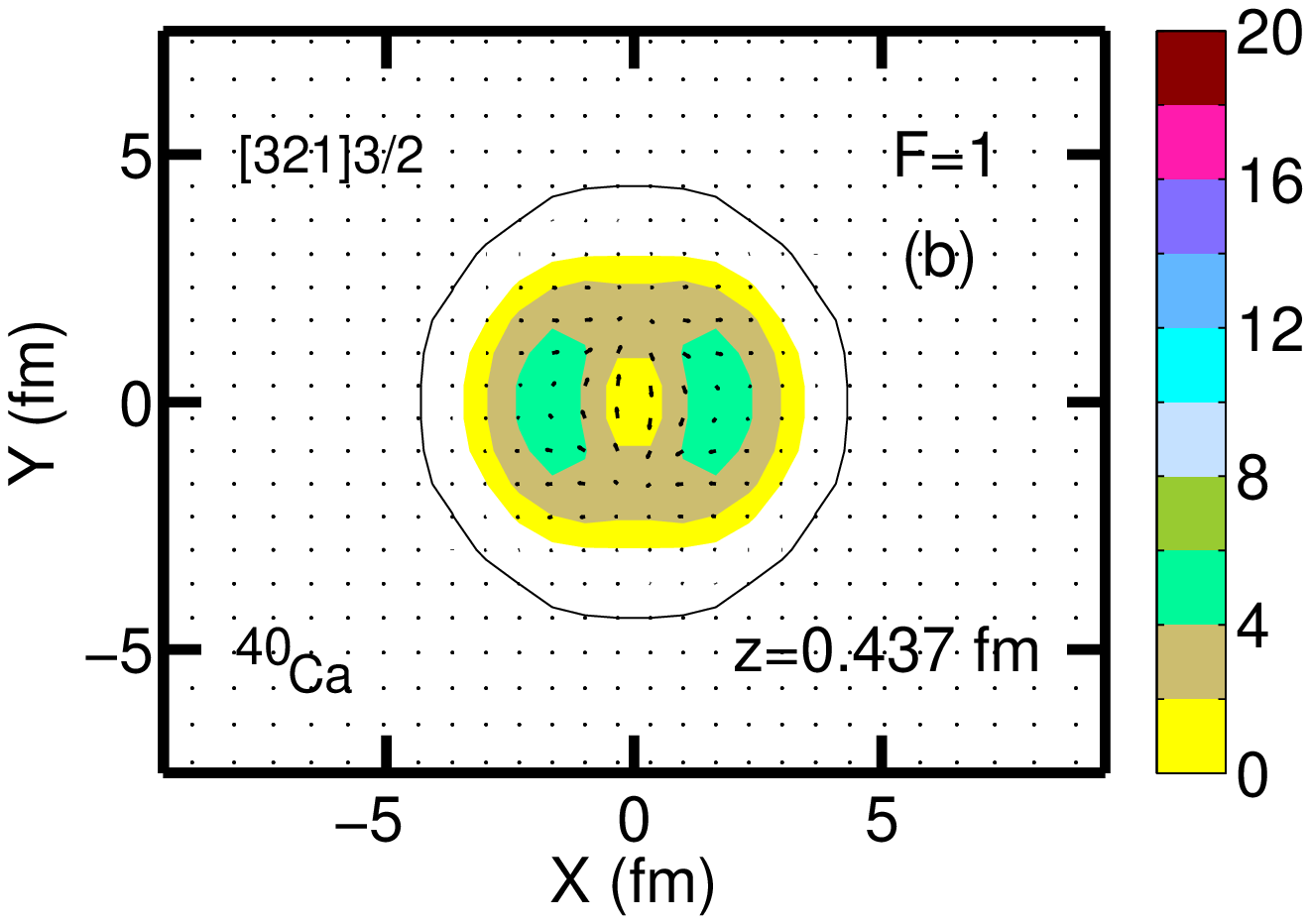}
\includegraphics[angle=0,width=5.5cm]{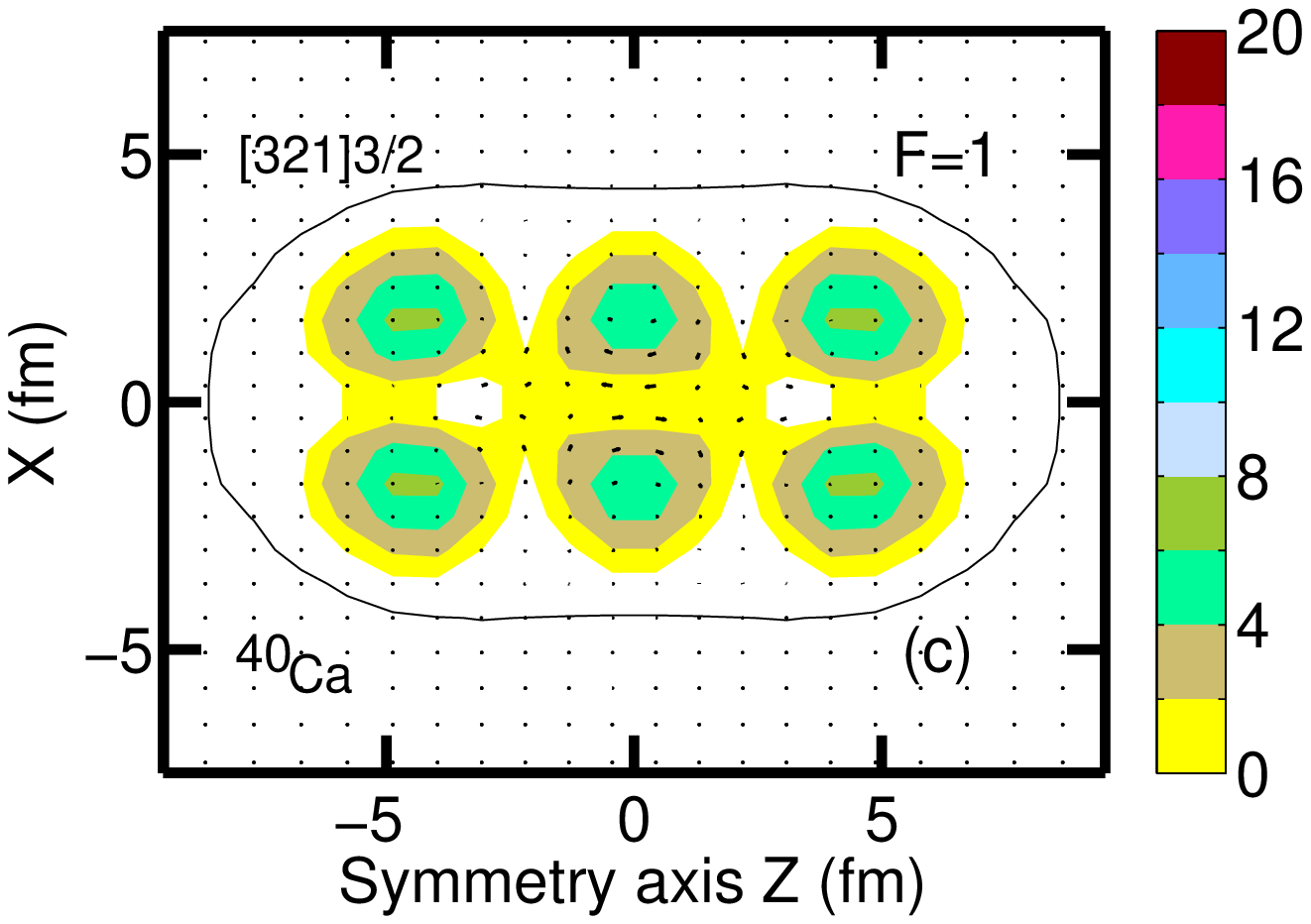} 
\caption{(Color online) The same as Fig.\ \ref{C12-NL3*} but for the [312]3/2$(r=-i)$ 
orbital in the MD [42,42] configuration of $^{40}$Ca at rotational frequency 
$\Omega_x=1.8$ MeV corresponding to spin $I=25\hbar$. The densities in the $yz$ and 
$xz$ planes are taken at $x=0.329$ fm and $z=0.329$ fm, respectively. }
\label{Ca40-31232}
\end{figure*}

  The wave function of the [321]3/2 state is dominated by the $|321,3/2>$ 
basis state which has one node in radial direction and two nodes in axial
direction. As a result, the density distribution of this state is given 
by three density rings (Fig.\ \ref{Ca40-31232}); one is located in the 
equatorial plane and other two symmetrically with respect of this plane. 
At no rotation, these rings are almost axially symmetric. The rotation 
induces the dependence of the density on the azimuthal angle (azimuthal 
asymmetry); this is clearly seen in Fig.\ \ref{Ca40-31232}. For the 
$r=-i$ branch of this state, the density is mostly localized around 
the $xz$-plane and its vicinity (Fig. \ref{Ca40-31232}). The situation 
becomes reversed for the $r=+i$ branch, for which most of the density 
becomes localized around the $yz$-plane and its vicinity.

\begin{figure*}[htb]
\includegraphics[angle=0,width=5.5cm]{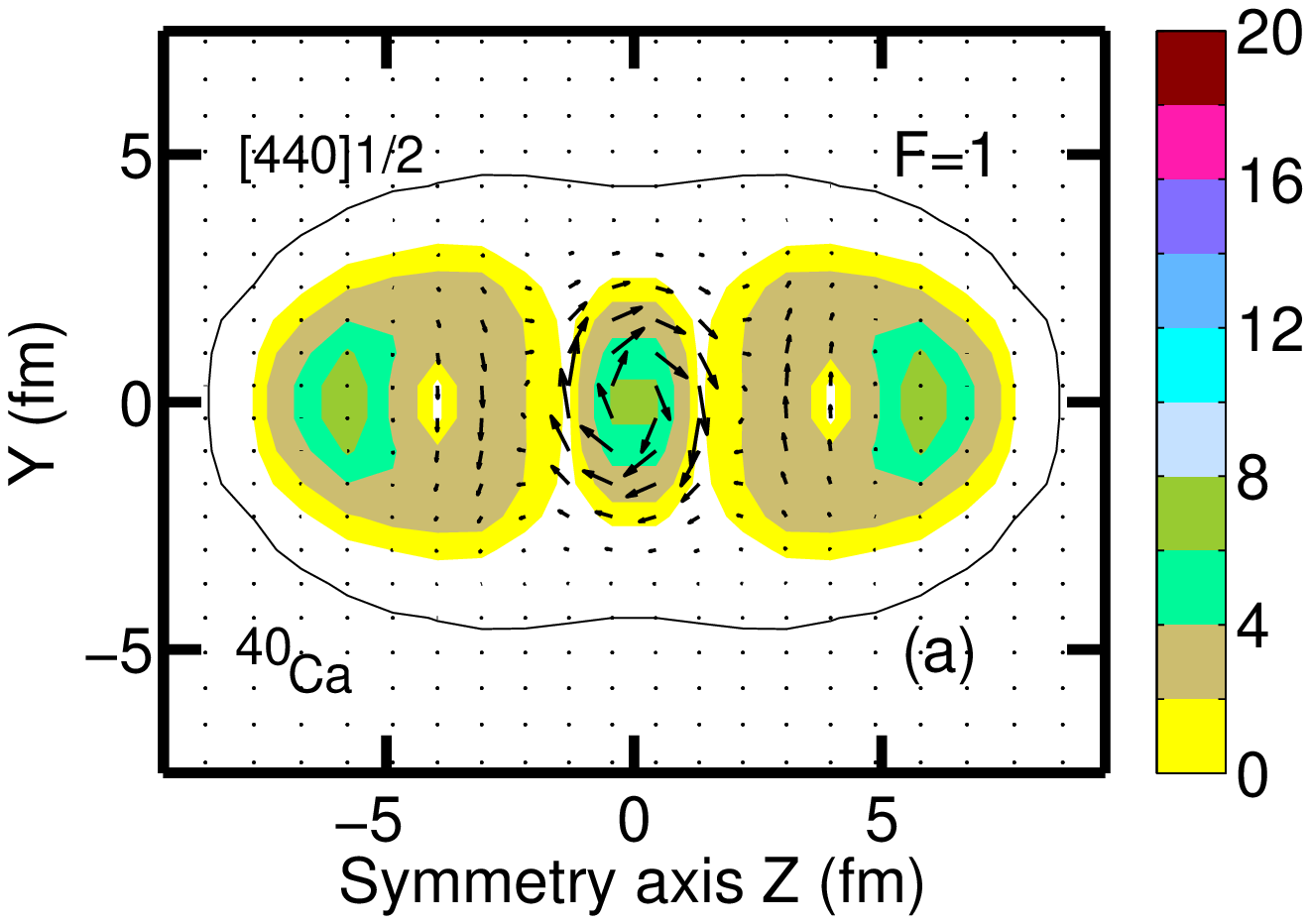}
\includegraphics[angle=0,width=5.5cm]{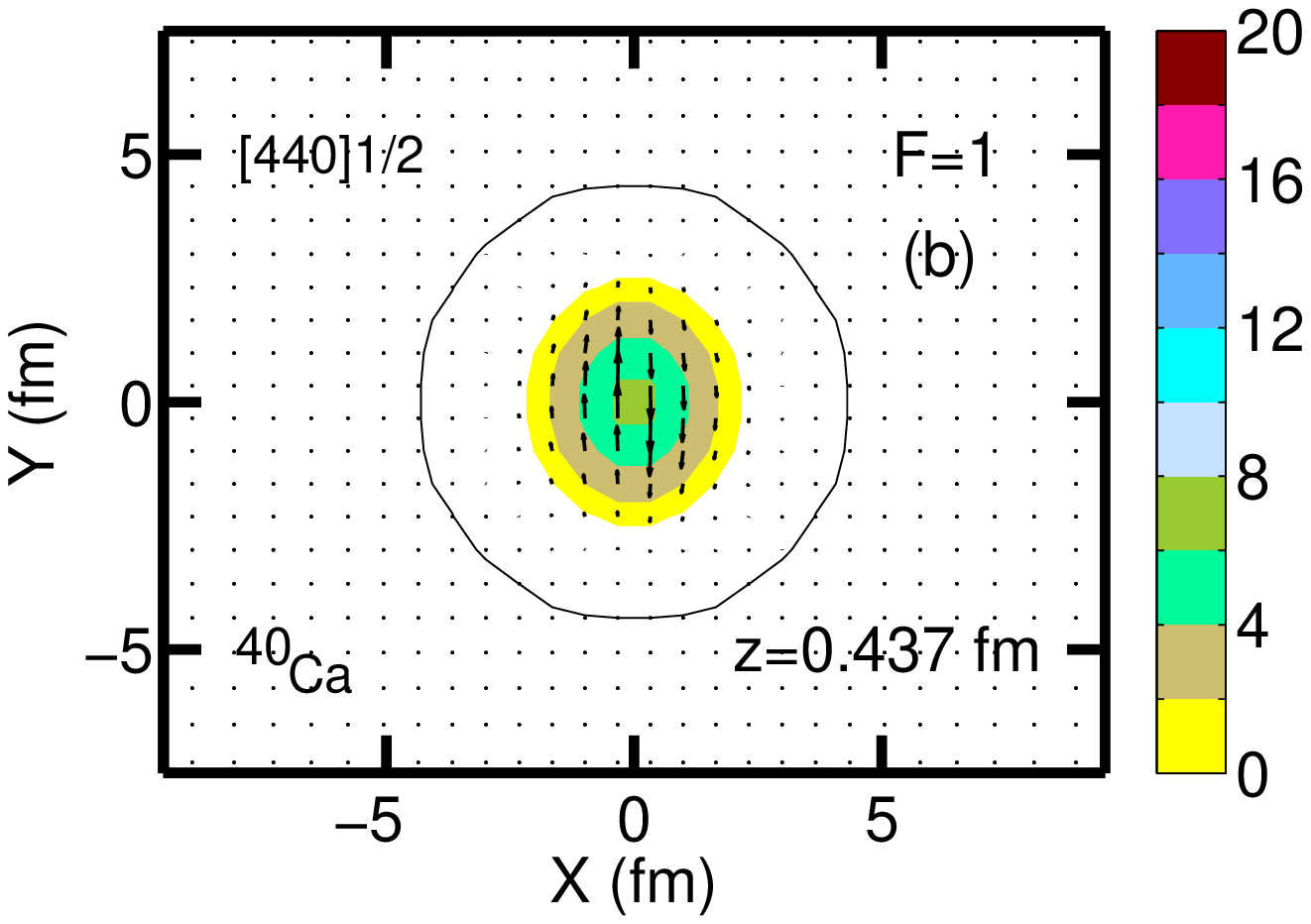}
\includegraphics[angle=0,width=5.5cm]{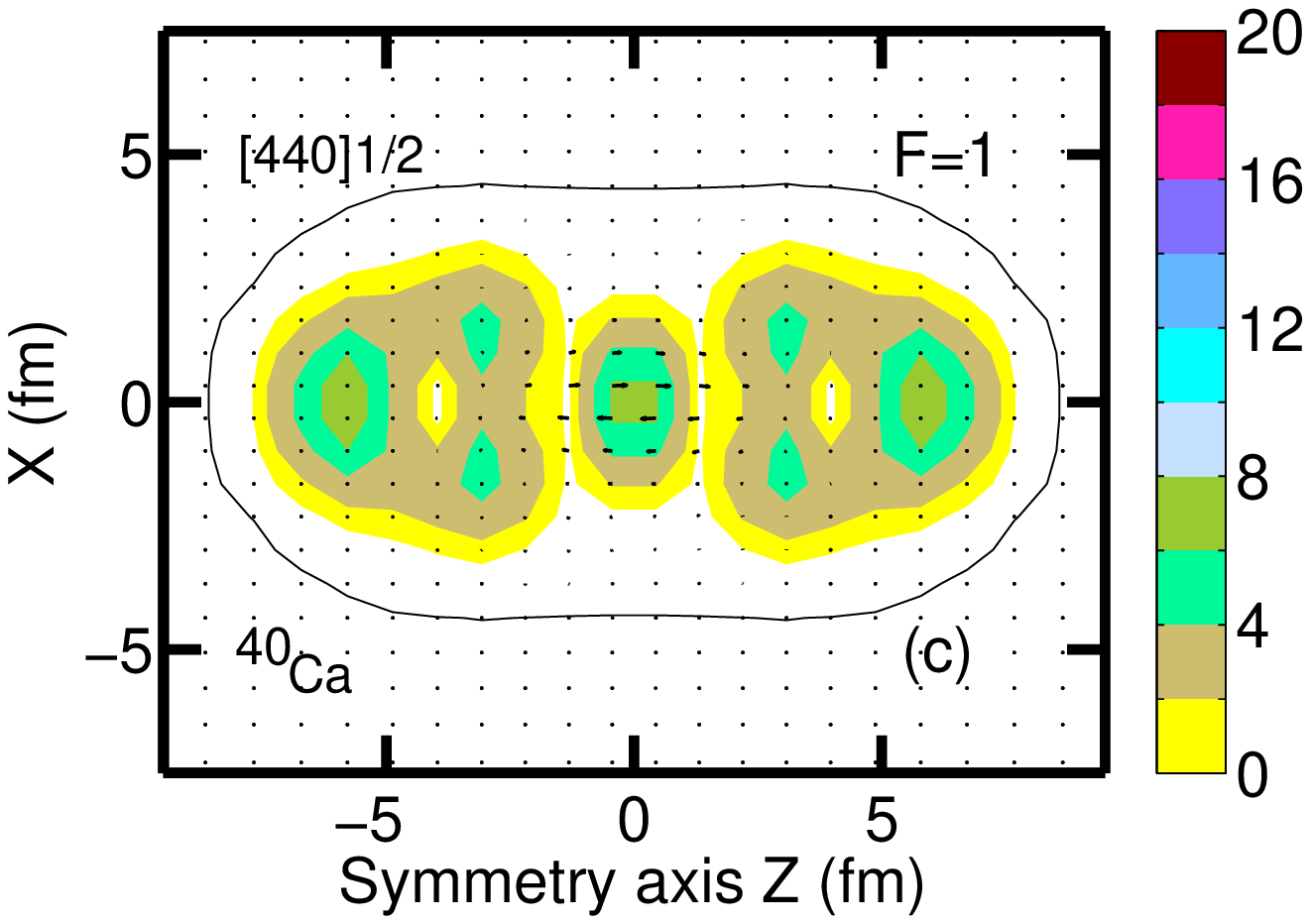}
\includegraphics[angle=0,width=5.5cm]{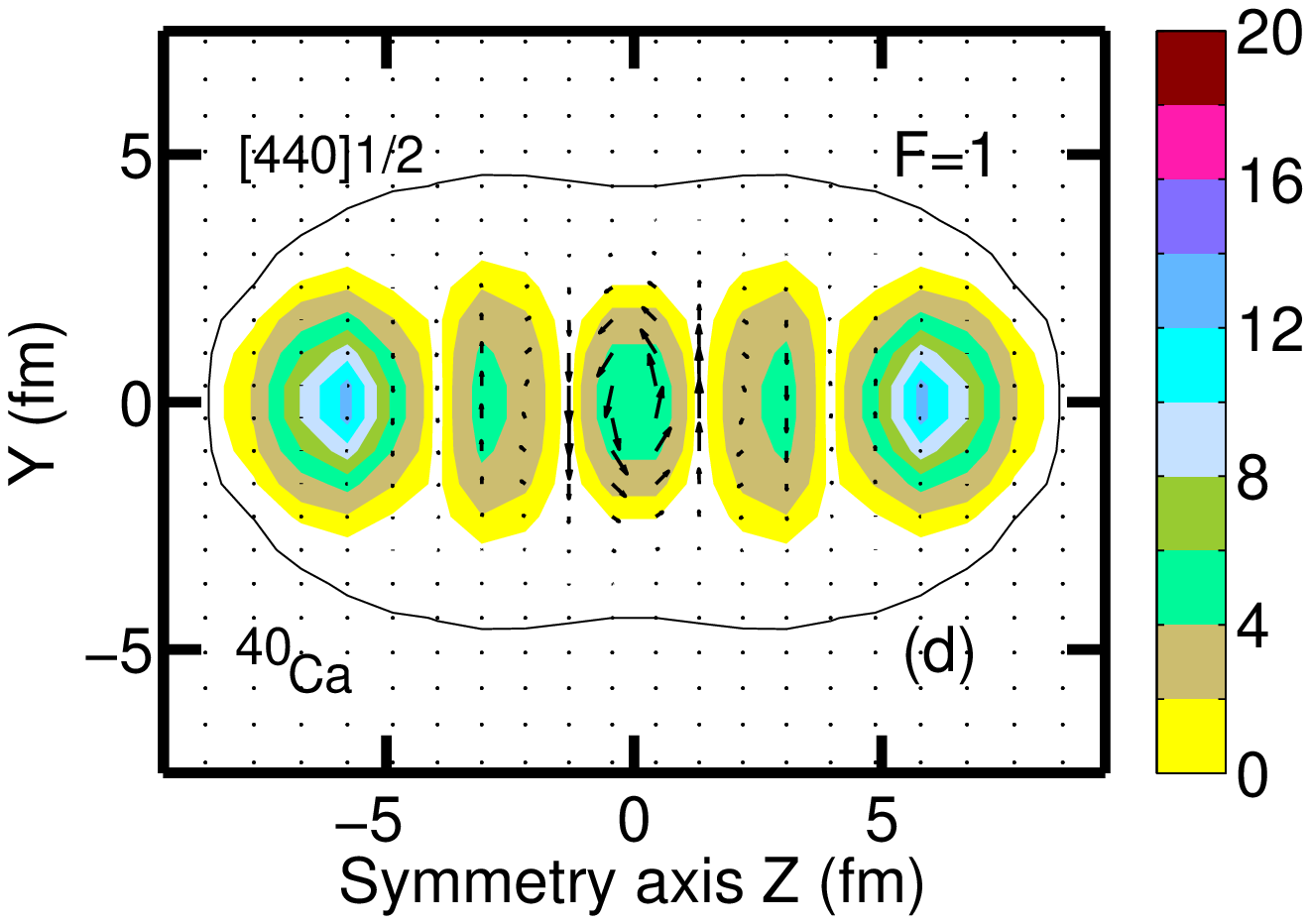}
\includegraphics[angle=0,width=5.5cm]{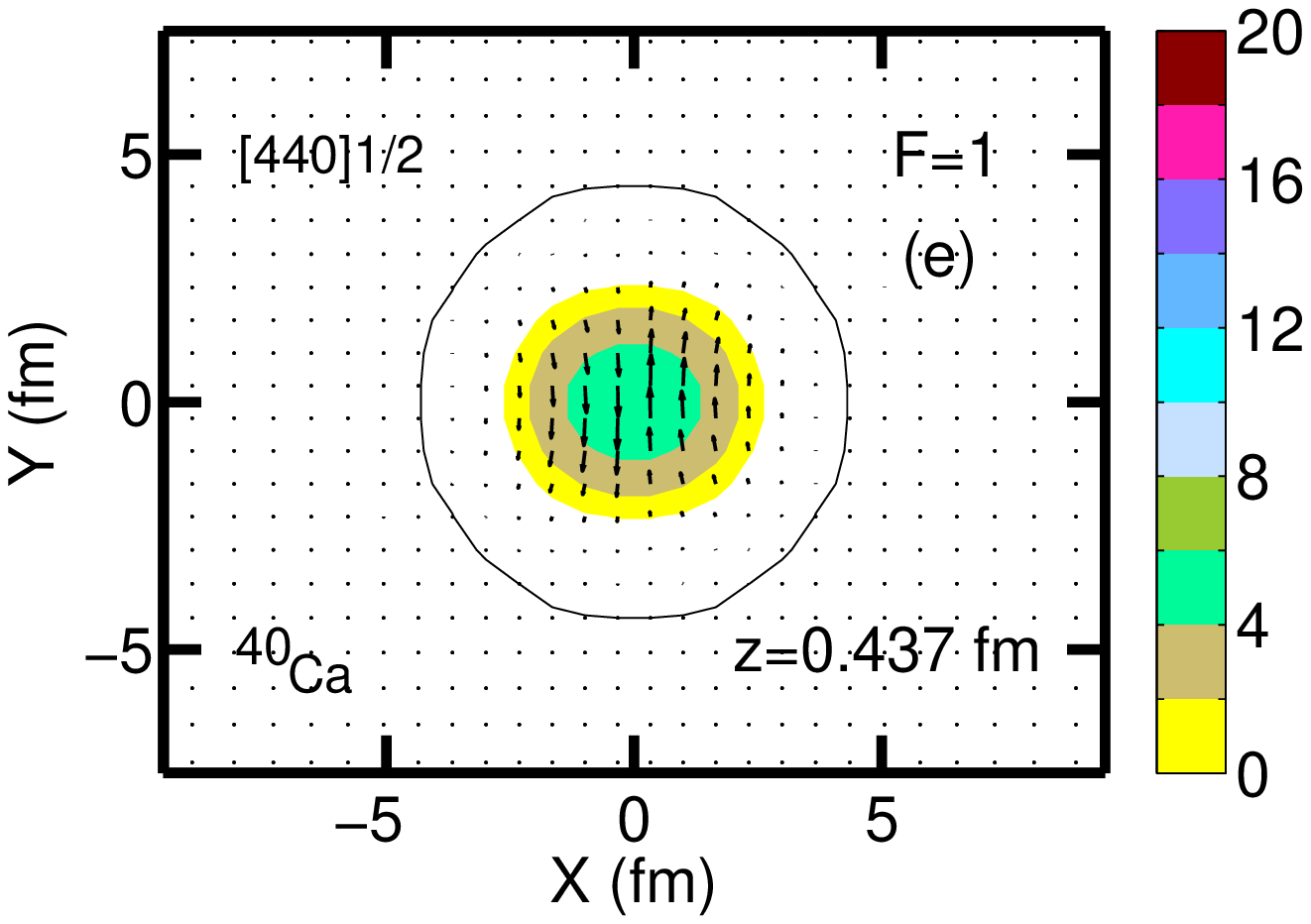}
\includegraphics[angle=0,width=5.5cm]{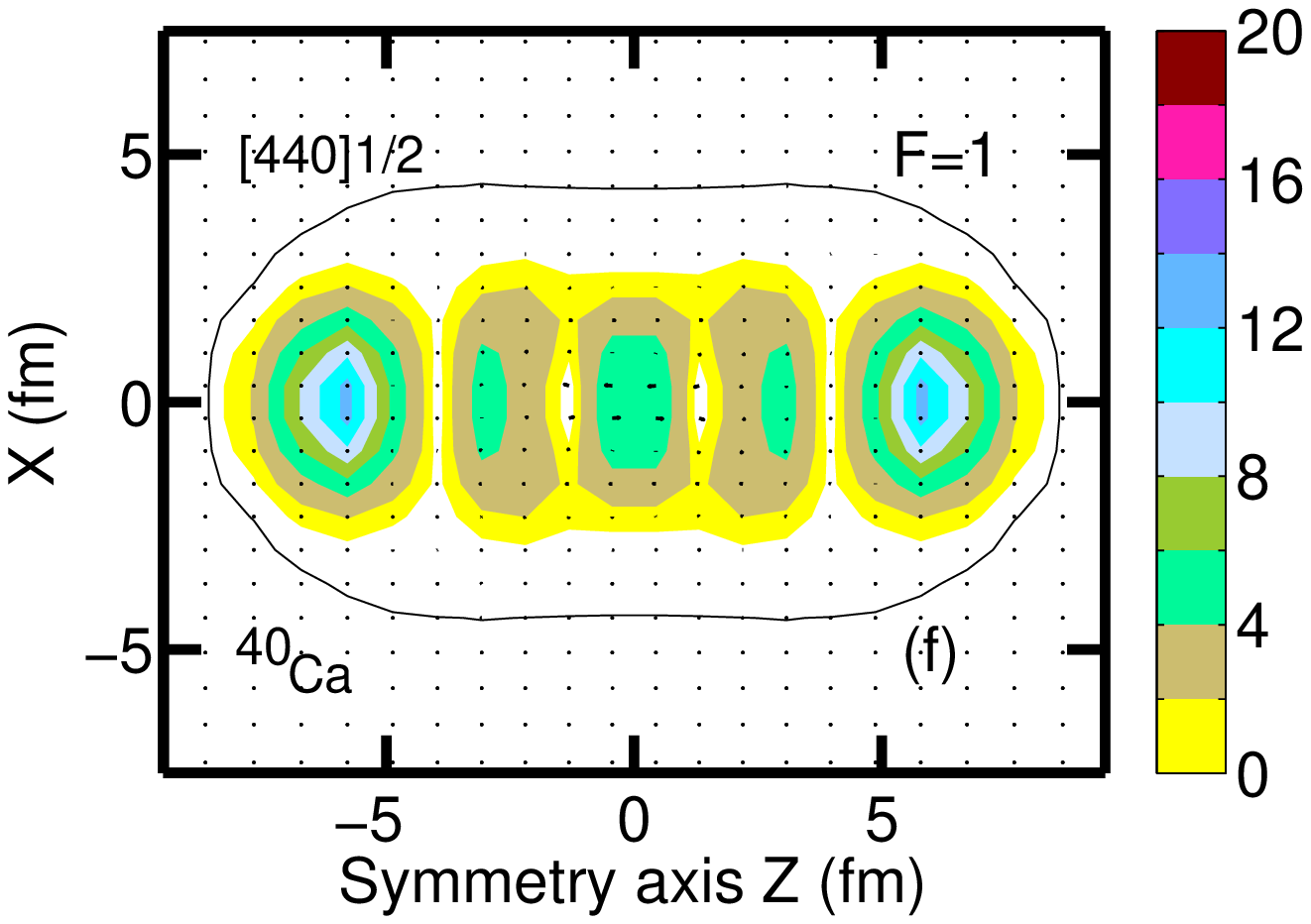}
\caption{(Color online) The same as Fig.\ \ref{Ca40-31232} but for the
[440]1/2$(r=\pm i)$ orbitals at $\Omega_x=1.8$ MeV. Top and bottom panels 
show the results for the ($r=-i)$ and $(r=+i)$ branches, respectively. Note that 
the density distribution at $\Omega_x=0.0$ MeV is very similar to what is 
seen in bottom panels.}
\label{Ca40-44012}
\end{figure*}

  Fig.\ \ref{Ca40-44012} displays the single-particle densities of the
$[440]1/2(r=\pm i)$ orbitals and illustrates the impact of state mixing 
on the single-particle densities. The wave function of the $(r=+i)$ branch 
is dominated by the basis $|440,1/2>$ state (Table \ref{table-wf}). As a 
consequence, its density distribution closely follows to that expected 
for the $[NN0]1/2$ states, namely, five density clusters (which is a 
consequence of four ($n_z=4$) nodes in axial direction for the $|440,1/2>$
basis state) with the maximum of the density in each of them at the axis 
of symmetry (the consequence of no node in radial direction). The density 
distribution of the $(r=-i)$ branch is different since it is built from 
three spheroidal clusters (one at center and two in the polar regions) 
separated by the ring structure. The later comes from substantial admixture 
of the $|221,1/2>$ basis state which has this kind of density distribution 
(see, for example, Fig.\ \ref{Si28-21132} and bottom row in Fig. 5 of the
Supplemental Material (Ref.\ \cite{Suppl-node}) as well as the discussion 
of the $[221]3/2$ state in $^{28}$Si (Sec.\ \ref{28Si-sect})).

\section{Main consequences of the nodal structure of 
the single-particle wave functions.}
\label{sect-general}

  The analysis of the results of the CRMF calculations performed
for extremely elongated shapes including hyperdeformed ones in $^{28}$Si, 
megadeformed shapes in $^{40}$Ca and rod-shape structures in $^{12}$C 
reveals the following general features

\begin{itemize}

\item
  The buildup of such shapes from individual contributions of 
particles is defined by two groups of the single-particle states
in the light nuclei with mass number $A$ up to around 50. The states 
with the $[NN0]1/2$ structure belong to the first group\footnote{
Low energy structures in relatively light nuclei can be described
in terms of molecular bonding \cite{OFE.06}. For covalent bonding, a negative 
parity orbital perpendicular to the $\alpha$-$\alpha$ axis is called 
a $\pi$-orbital, whereas a $\sigma$-orbital denotes a positive
parity orbital parallel to the $\alpha$-$\alpha$ direction. 
Thus, the [101]3/2 and [220]1/2 Nilsson states are the examples of
the $\pi$- and $\sigma$-orbitals, respectively.} The second 
group is represented by the states with the $[N,N-1,1]$1/2 and 
$[N,N-1,1]$3/2 structures; note that the spatial distribution of 
the wave function (density) almost does not depend on $\Omega$ 
and is  almost entirely defined by $[N,N-1,1]$.

\item 
  With relatively few exceptions, the wave functions of the 
single-particle states occupied in such extremely deformed 
shapes are dominated by a single basis state. This is because 
the mixing of the basis states is suppressed at the bottom of 
nucleonic potential since the energy distances between the
basis states which could mix are large and the number of 
possible counterparts with which appreciable mixing could 
take place is limited. Indeed, the fragmentation of the wave 
function (with related decrease of the dominant component of 
the wave function) typically increases with the raise of the 
position of the single-particle state with respect of the 
lowest state in the nucleonic potential.

\item
  As a consequence, the nodal structure of the wave function (and 
thus of density distribution) of the single-particle state is defined 
solely by the nodal structure of the dominant basis state.  However, 
the nodal structure of the density distribution of these basis states 
is determined by their quantum numbers. As a result, three basic types of 
single-particle density distributions, namely, spheroidal/elipsoidal 
shapes, doughnut and ring shapes, play an important role in forming the 
nuclear shapes at large elongation.

\begin{itemize}
\item
  The density distributions of the Nilsson states with $[NN0]1/2$ 
quantum numbers are nearly axially symmetric with $N+1$ spheroidal/elipsoidal 
like density clusters the maximum of the density of which is located 
at the axis of symmetry. With the exception of the $N=0$ case, the clusters 
with highest densities are located in polar regions of the nuclei. 
This structure of the density distribution is the consequence of the 
nodal structure of the dominant basis state: no nodes in radial 
direction and $n_z=N$  nodes in axial direction. 

\item
  The doughnut shapes are formed by the $[N01]\Omega$ states since 
the densities of their dominant basis states have one node in radial 
direction and no nodes in axial direction.

\item
  Finally, the states with the structure $[N,N-1,1]\Omega$ form 
multiply (two for $n_z=1$ and three for $n_z=2$) ring shapes for 
$N=2$ and 3.

\end{itemize}
   
\item 
  Another source of increased fragmentation of the single-particle 
wave function is Coriolis interaction. It leads to some reduction
of the weight of the dominant basis state in the single-particle 
wave function and to some delocalization of single-particle density. 
However, even with these effects accounted the single-particle states 
of interest for the rotational frequencies under study are dominated 
by a single basis state and their density distributions have the same 
nodal structure as the one at no rotation. Note that the rotation 
introduces some azimuthal dependence of the density distribution; this 
effect is especially pronounced for the states of the $[N,N-1,1]$ 
type. 

\item 
  The localization of the single-particle wave function strongly 
depends on its nodal structure. Only the states with low number of nodes 
in axial direction 
and with no nodes in radial direction could be well localized. The highest 
localization of the wave function is reached for the lowest states of 
the $[NN0]1/2$ type with $N=0, 1$ and 2; that is a reason why these states 
are so important in $\alpha$-clusterization. Subsequent increase of $N$ 
and/or the number of the nodes in radial direction substantially decreases 
the level of the localization of the wave function. The rotation also reduces 
somewhat the localization of the wave function.

\end{itemize} 

 The structure of the clusters forming the nucleus is frequently 
defined in the DFT framework by comparing the density distributions 
of possible clusters with total density of the nucleus. Although 
some useful information can be obtained in that way, especially, 
with the  use of the localization functions defined in Refs.\ 
\cite{RMUO.11,ZSN.16} such an approach has its own limitations.
This measure of localization is inherently defined for total 
quantities because the nucleon localization function considers the 
conditional probability of finding a nucleon within a distance 
$\delta$ from a given nucleon at point ${\bf r}$  with the same spin 
$\sigma$ and isospin $\tau$ \cite{BE.90,RMUO.11,ZSN.16}. Thus,
it cannot be applied to the single-particle quantities. As a result,
the present analysis suggests an alternative way in which the 
single-particle densities forming the total density of the nucleus 
and its constituent clusters are compared. Since single-particle 
densities bear a clear mark of the underlying single-particle wave 
functions, such way of the comparison can provide a microscopic 
understanding on how the nucleus is formed from the clusters. The 
work on that type of the analysis is in the progress and the results 
will be presented in a forthcoming publication.

\subsection{Nodal structure of the single-particle densities 
and the transition to liquid phase}

 The analysis of Ref.\ \cite{EKNV.13} based on the consideration
of total nucleonic densities and harmonic oscillator potential has 
suggested that the nuclei heavier than $A\sim 30$ consist of 
largely delocalized nucleons.  As a result, the transition from 
coexisting cluster and mean-field states to a Fermi liquid state 
should occur for nuclei with $A\approx 20-30$ \cite{EKNV.13}. Note 
that for solid phase, the nuclear configurations are characterized 
by the situation in which each particle is localized with respect 
of its neighbours \cite{M.99}. On the contrary, the individual 
particles are delocalized in a quantum liquid \cite{PZ-book,M.99}.  
By definition the quantum liquid is many-body system whose behaviuor 
is defined by the effects of both quantum mechanics and quantum 
statistics \cite{QL-book}. The latter enters into the game through 
the requirement of the indistinguishability of the constituent
particles (\cite{QL-book}) which defines the type of quantum 
statistics (Fermi or Bose). This requirement cannot be satisfied 
in finite nuclei if the occupied states are localized and 
have  different spatial distributions.
 
  The analysis of the single-particle densities performed in the 
present paper suggests that the transition to quantum liquid does 
not happen in the considered nuclei. Although with the increase of 
particle number  the occupation of the single-particle orbitals 
with lower level of localization becomes dominant, none of these 
states can be described as delocalized. They still preserve their 
nodal structure and typically occupy less than half of the volume 
of the nucleus.

\subsection{$\alpha$-clusterization and its evolution with particle number}

 The observed features of the single-particle density distributions emerging 
from the nodal structure of the wavefunctions allow to understand in a relatively 
simple way the necessary conditions for $\alpha$-clusterization. Two factors 
play an important role here: the degree of the localization of the wavefunction 
and the type of the density clusters formed by single-particle orbital. It 
is clear that for the $\alpha$-clusterization the single-particle density clusters 
should be compact (well localized), should have spheroidal or slightly elipsoidal
density distribution and overlap in space. These conditions are satisfied only for 
the lowest states of the $[NN0]1/2$ type with $N=0, 1$ and 2 which are active in 
the $\alpha$-cluster structures of very light nuclei \cite{AJ.94,FBW.95,OFE.06,EKNV.12,ZIM.15}.  
With increasing particle number the orbitals with doughnut and multiply ring type 
density distributions become occupied. These states are substantially less localized; 
the maximum of the density in such structures is typically much smaller than the 
maximum of the density in the lowest $[NN0]1/2$ orbitals. In addition, such density 
distributions (doughnuts and rings) are incompatible with $\alpha$-clusters.

  Based on these considerations it is clear that $\alpha$-clustering in highly 
elongated nuclear structures for typical deformations considered here should be 
an important mode only in very light nuclei in which the states of the $[NN0]1/2$ 
type are occupied. Although the $\alpha$-cluster substructures still survive in 
heavier nuclei (Fig.\ \ref{Total-densities-rod}), their contribution to the total 
wave function of the nucleus is expected to decrease as compared with light 
nuclei with increasing mass number because of increased contribution of the 
single-particle structures with ring- and doughnut-type density distributions.
This trend is similar to the one obtained in antisymmetrized molecular dynamics 
calculations of Ref.\ \cite{MKKRHT.06}.
%
%
%

\subsection{Building nuclear molecules by means of particle-hole
                             excitations}

  The coexistence of ellipsoidally shaped structures and nuclear 
molecules in the same nucleus has been seen in the CRMF calculations 
of Ref.\ \cite{RA.16} for similar elongations of nuclear shape. It 
turns out that the configurations of these two types of the shapes 
are connected by characteristic particle-hole excitations. A specific 
feature of the nuclear molecules is the existence of two fragments 
connected by the neck. The MD configurations [31,31] in $^{36}$Ar, 
[42,42] in $^{40}$Ca and $[421,421]$ in $^{42}$Sc  (Fig.\ 
\ref{Total-densities}d-f) are the examples of nuclear molecules 
(Ref.\ \cite{RA.16}). To build nuclear molecules from typical
elipsoidal density distributions one has to move the matter from the 
neck (equatorial) region into the polar regions of the 
nucleus. This can be achieved by specific particle-hole 
excitations\footnote{Note that particle-hole excitations are 
a powerful tool of the modification of the density distribution in 
finite nuclei. For example, they can substantially modify the radial 
dependence of matter distribution in spherical nuclei \cite{AF.05} or
introduce a deformation into nuclear system \cite{NilRag-book}.}
removing particles 
from (preferentially) doughnut type orbitals or from the orbitals which 
have a density ring in a equatorial plane into the orbitals (preferentially 
of the $[NN0]1/2$ type) which build the density mostly in the  polar 
regions of the nucleus.

  The results presented in Ref.\ \cite{RA.16} give a number of 
examples of such particle-hole excitations leading to the transitions 
from elipsoidal nuclear shapes to nuclear molecules. One such an example is 
the transition from the HD [4,4] configuration in $^{36}$Ar, which has 
ellipsoidal density distribution [see Fig. 24b in Ref.\ \cite{RA.16}]), 
to the MD [31,31] configuration which is an example of nuclear molecule 
[see Fig. 24c in Ref.\ \cite{RA.16} and Fig.\ \ref{Total-densities}d in 
the present paper].  This transition involves the proton and neutron 
particle-hole excitations from the 3/2[321] orbital (which has triple 
ring density distribution) into the [440]1/2$(r=-i)$ orbital. Another 
example is the transition from the [41,41] configuration in $^{42}$Sc, 
which has ellipsoidal density distribution (see Fig. 8a in Ref.\ \cite{RA.16}), 
to the MD [421,421] configuration which is a very good example of nuclear 
molecule (Fig.\ \ref{Total-densities}f). This transition is achieved in 
proton and neutron subsystems by the particle-hole excitations from the 
$[202]5/2$ (which has doughnut type density distribution) and $[321]3/2$ 
(which has triple ring density distribution) orbitals into the 
$[440]1/2(r=+i)$ and $[550]1/2(r=+i)$ orbitals. The latter orbitals have 
the largest and most dense density clusters in the polar regions of the 
nucleus (see Fig.\ \ref{Ca40-44012} and Fig. 8 in the Supplemental Material 
\cite{Suppl-node}).

\subsection{The currents and rigid rotation of the system}

\begin{figure}[htb]
\includegraphics[angle=0,width=8.5cm]{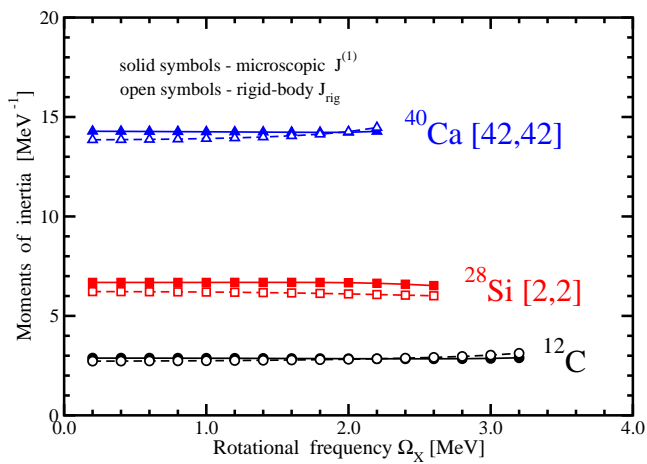}
\caption{(Color online) The calculated moments of inertia for the
indicated configurations.}
\label{Fig-moments}
\end{figure}

 Fig.\ \ref{Fig-moments} compares microscopically calculated 
kinematic moments of inertia $J^{(1)}$ with the rigid body moments of 
inertia $J_{rig}$ for the configurations under study. $J^{(1)}$ is 
calculated fully self-consistently via
\begin{eqnarray}
J^{(1)}(\Omega_x)=\frac{J}{\Omega_x},
\end{eqnarray}
where $J$ is the expectation value of the total angular momentum 
along the $x$-axis and $\Omega_x$ is rotational frequency along 
the same axis. In the 
CRMF framework, $J$ is defined as a sum of the expectation 
values of the single-particle angular momentum operators 
$\hat{\jmath}_{x}$ of the occupied states
\begin{eqnarray}
J=\sum_{i}\langle i|\hat{\jmath}_{x}|i\rangle .
\label{Jmicro}
\end{eqnarray}
Note that the effects of the time-odd mean fields, which are
extremely important for the moments of inertia (see Refs.\ 
\cite{KR.93,TO-rot}), are included fully self-consistently
in the CRMF calculations.

  The rigid body moment of inertia $J_{rig}$ is obtained in 
one-dimensional cranking approximation with the rotation 
defined around the $x$-axis from the calculated density 
distribution  $\rho({\bf r})$ by
\begin{eqnarray}
J_{rig}=\int \rho({\bf r}) (y^2+z^2)d^3r
\label{Jrigid}
\end{eqnarray}

 An interesting feature of the rotating bands in the nuclei under 
consideration is the fact that their microscopic kinematic moment 
of inertia $J^{(1)}$ changes very little with the increase of 
rotational frequency. Indeed, the variation of $J^{(1)}$ over
calculated rotational frequency is only 0.47\%, 2.4\% and 0.4\%
of its total value in rotational bands of $^{12}$C, $^{28}$Si and 
$^{40}$Ca, respectively\footnote{This is not general feature since
the kinematic moments of inertia show pronounced variations in a number
of configurations (see Fig. 35 in Ref.\ \cite{RA.16}).}. The rigid 
body moment of inertia is rather close to the microscopic one; it 
deviates from $J^{(1)}$ by $-5.3$\% ($+7.8$\%), $-6.9$\% ($-7.8$\%) 
and $-3.0$\% ($+1.3$\%) of the $J^{(1)}$ value at the lowest (highest) 
calculated frequencies for the rotational bands of $^{12}$C, 
$^{28}$Si and $^{40}$Ca, respectively. Note that similar analysis 
for the hyperdeformed bands in the $Z=40-58$ region shows that 
microscopic and rigid body moments of inertia differ typically by 
less than 5\% (Ref.\ \cite{TO-rot}). This difference is bigger in 
$^{12}$C and $^{28}$Si most likely due to smaller number of
the single-particle orbitals involved as a result of which their 
individual features still play a prominent role in the definition
of the total properties of the configuration. In any case, these 
differences between  microscopic and rigid body moments of inertia 
are significantly smaller than those  expected in normal-deformed 
bands (see Ref.\ \cite{TO-rot}). Thus, the bands under 
study behave in a first approximation like rigid rotors.

 The microscopic origin of these features can be traced back to 
underlying shell structure. Indeed, the analysis within the periodic 
orbit theory \cite{DFPCU.04} for superdeformed rotational bands shows 
that the single-particle orbitals that cause shell structure of 
prolate superdeformed nuclei do  not carry rotational flux if the axis 
of rotation is perpendicular to the symmetry axis. Therefore, the moments 
of inertia of such rotational bands should be equal to the rigid-body 
value \cite{DFPCU.04}. Based on general arguments this conclusion has 
to be valid also for prolate hyperdeformed and megadeformed bands. Such 
conclusion is supported by our microscopic calculations which show that the 
calculated moments of inertia of extremely deformed rotational bands are 
typically within 5\% of the rigid-body value in the medium mass nuclei 
(Ref.\ \cite{TO-rot}) and close to this value in light nuclei (as 
defined in the present manuscript).

  The distributions of the total neutron currents in the $yz$ plane 
are shown in Fig.\ \ref{Fig-currents}. {They are defined as \cite{TO-rot}
\begin{eqnarray}
{\bff j}^{n}({\bff r}) &=& \sum_{i=1}^{N} (\psi_i({\bff r}))^{\dagger}
\hat{{\bff \alpha}} \psi_i({\bff r}). \label{current-eq}
\end{eqnarray}
where $\psi_i({\bff r})$ are the single-neutron wave functions.
Thus, total current is built as a sum of the individual currents 
$\psi_i({\bff r}))^{\dagger} \hat{{\bff \alpha}} \psi_i({\bff r})$
of the occupied orbitals. These individual currents are shown in 
the figures with single-particle density distributions
(see Figs.\ \ref{C12-NL3*}, \ref{Si28-33012}, \ref{Si28-10112},  
\ref{Si28-21132}, \ref{Ca40-31232}, \ref{Ca40-44012}, and the figures 
in the Supplemental Material \cite{Suppl-node}). Note that the $yz$ plane 
is perpendicular to the axis of the rotation. As a result, in general the 
currents in this plane are substantially larger than the ones
in the $xz$ and $xy$ planes and they show the vortices. Note 
that the localization, the strength, and the structure of the current 
vortices created by a particle in a specific single-particle 
state depend on its nodal structure (for more details
see Sec. V in Ref.\ \cite{TO-rot}, Ref.\ \cite{GR.78} and Sec. III C
in Ref.\ \cite{AA.10}). All single-particle states are characterized
by the weak current in the surface area and neither of them 
shows the current distribution expected for rigid rotation.

  It is well known that there are no currents in the intrinsic 
frame if the rigid nonspherical body rotates uniformly (rigid
rotation)  (see Sec. IV A-V in Ref.\ \cite{Bohr1975}). The presence
of strong current vortices in Fig.\ \ref{Fig-currents} demonstrates 
the dramatic deviation of the currents from rigid rotation. This is 
despite the fact that the moments of inertia of considered configurations are
close to the rigid-body value. 
This fact underlines
the importance of quantum mechanical treatment of the currents.

 The experimental data on the moments of inertia can be easily extracted 
from the rotational sequences of observed states. However, the discussion in 
this section as well as the results obtained within periodic orbit and 
cranked relativistic mean field theories in Refs.\ 
\cite{DFPCU.04,TO-rot} show that the closeness of experimental 
moment of inertia to the rigid body value is not sufficient 
indicator of the rigid-body rotation. To confirm or reject such an 
interpretation one should measure the currents but they are not
experimentally accessible quantities.

\begin{figure*}[htb]
\includegraphics[angle=0,width=5.9cm]{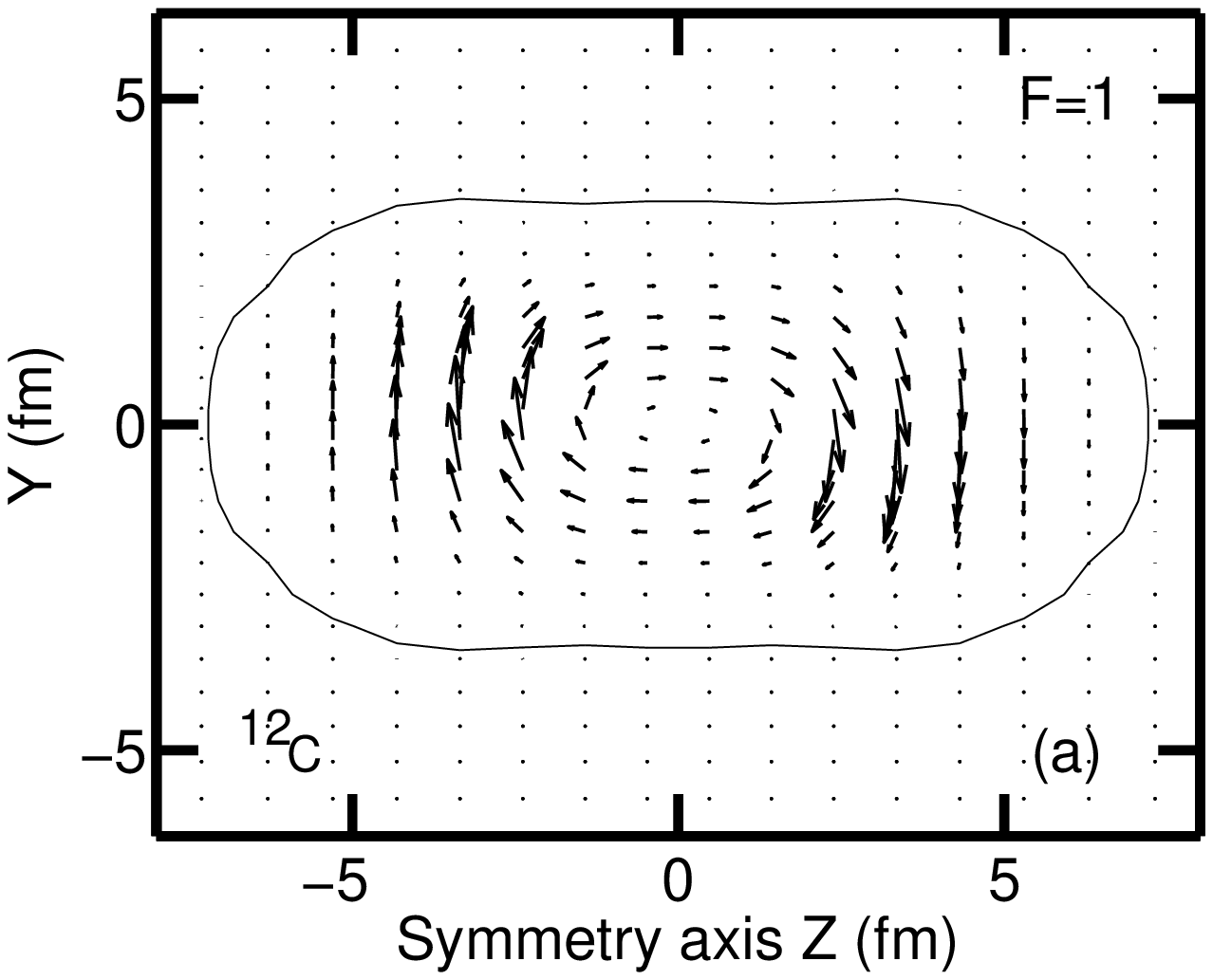}
\includegraphics[angle=0,width=5.9cm]{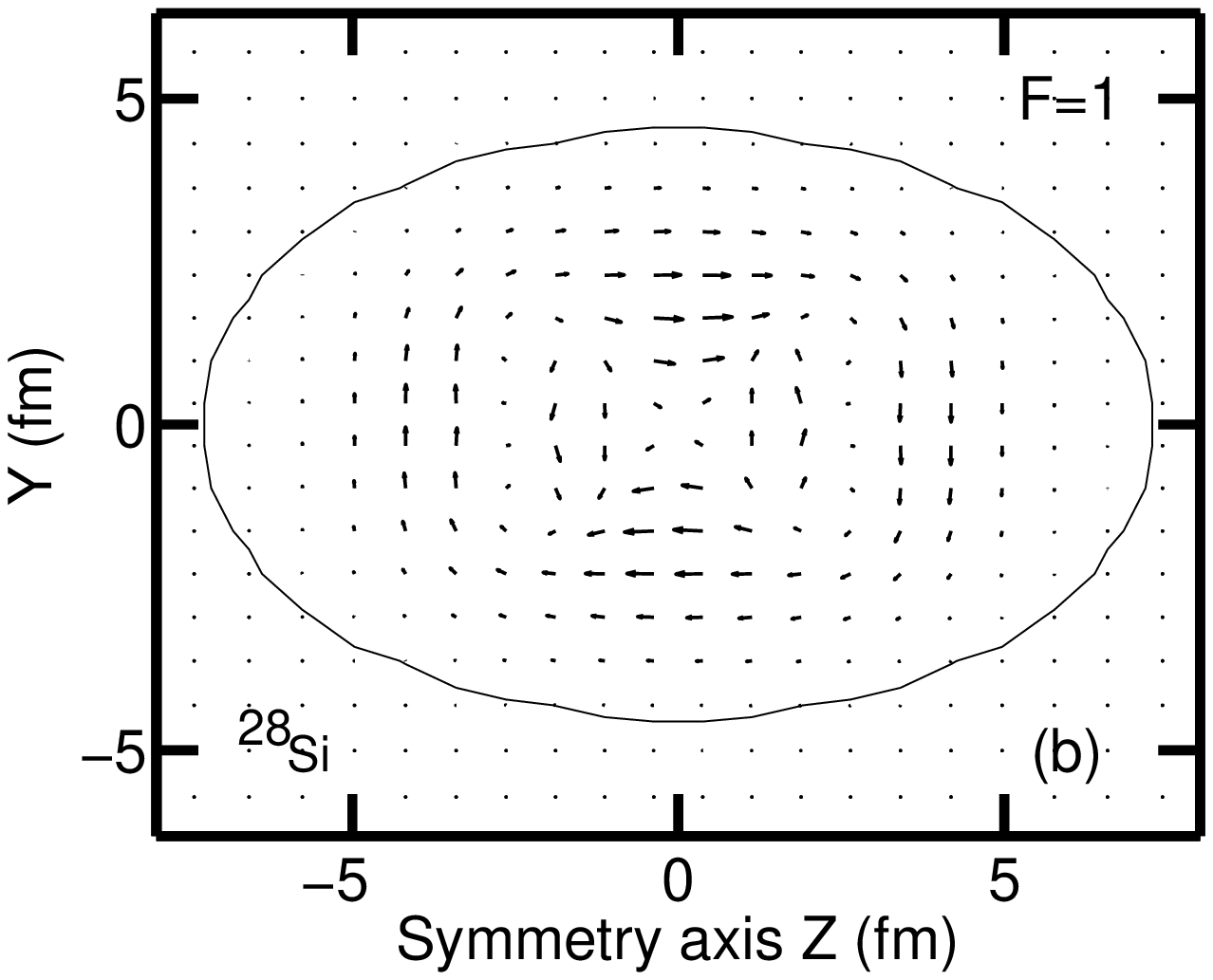}
\includegraphics[angle=0,width=5.9cm]{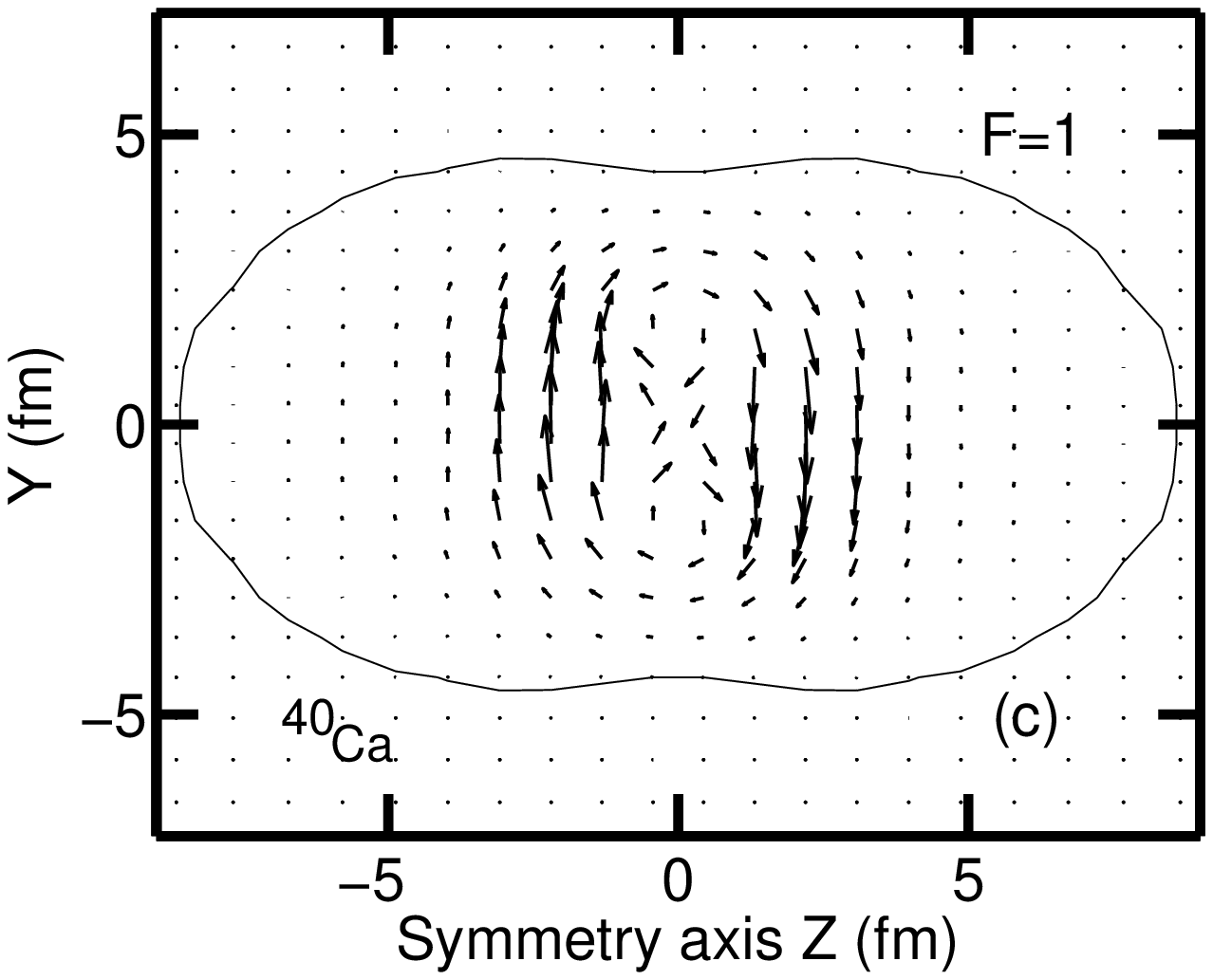}
\caption{(Color online) Total neutron currents {\bff j}$^n$({\bff r})
in the $yz$ plane plotted at $x=0.234$ fm, $x=0.326$ fm and $x=0.329$ fm for 
the considered configurations in $^{12}$C, $^{28}$Si and $^{40}$Ca, 
respectively. They are given at spin values indicated in Fig.\ \ref{Total-densities}.
The currents in panel (a) are plotted at arbitrary units 
for better visualization. The currents in other panels are normalized to 
the currents in above mentioned panels by using factor F.
}
\label{Fig-currents}
\end{figure*}

\section{Conclusions}
\label{concl}

  In conclusion, the nodal structure of the density distributions 
of the single-particle states occupied in extremely deformed structures
(such as rod-shaped, hyper- and megadeformed ones) of non-rotating
and rotating $N\sim Z$ nuclei has been investigated in detail. Such 
structures are either axial or nearly axial in the CRMF calculations and 
they are present in light nuclei with $Z=4-24$ \cite{RA.16,ZIM.15,YIM.14,AR.16}. This 
simplifies the situation and with relatively few exceptions the wave 
functions of the  single-particle states occupied in such extremely 
deformed shapes are dominated by a single basis state. As a consequence, 
the nodal structure of the wave function (and thus of the density 
distribution) of the single-particle state is defined solely by the 
nodal structure of this dominant basis state, the structure of which 
is given by the Nilsson label $[Nn_z\Lambda]\Omega$. Two types of the 
states, namely, $[NN0]1/2$ (with $N$ from 0 to 5) and $[N,N-1,1]1/2$ 
($N,N-1,1]3/2$) (with $N$ from 1 to 3) define the features of 
extremely deformed configurations in the nuclei under study.

 For extremely deformed shapes in the $A \leq 50$ nuclei considered 
here, the nodal structure of the single-particle states does not depend 
on the nucleonic configuration (occupation of the single-particle states)
or rotation. The only exception is the case of strong interaction of 
two single-particle states with the same parity and signature
which leads to the mixing of the wavefunctions of interacting
orbitals and thus of their single-particle densities. However, this happens 
rarely and in limited rotational frequency range.

  The observed features of the single-particle density distributions 
emerging from the nodal structure of the wavefunctions allow to understand 
in a relatively simple way the necessary conditions for 
$\alpha$-clusterization and the suppression of the $\alpha$-clusterization 
with the increase of mass number. In addition, it allows to understand the
coexistence of ellipsoidal mean-field type structures and nuclear molecules
at similar excitation energies and the features of particle-hole excitations
connecting these two types of the structures.

  Our investigation shows that although with increasing the particle 
number the occupation of the single-particle  orbitals with low level of 
localization of the single-particle densities becomes dominant, the states 
sitting deep in the nucleonic potential still remain well localized. In 
addition, neither of the occupied states lose their nodal structure and 
become delocalized. Thus, for the deformations and nuclei under 
study no transition to quantum liquid phase has been observed.

\section{Acknowledgements}

This material is based upon work supported by the U.S. Department
of Energy, Office of Science, Office of Nuclear Physics under Award
No. DE-SC0013037.

\bibliography{references18}
\end{document}